\documentclass[preprint,3p,number]{elsarticle}
\biboptions{sort&compress}
\usepackage{mathtools, nccmath}
\usepackage{amsfonts}

\usepackage{hypernat}
\usepackage[pdfencoding=auto, psdextra,hidelinks]{hyperref}
\usepackage[boxruled,vlined,linesnumbered]{algorithm2e}
\usepackage{algorithmic}

\makeatletter
\newcommand{\removelatexerror}{\let\@latex@error\@gobble}
\makeatother

\usepackage{multicol}
\usepackage{listings}

\usepackage[dvipsnames,table]{xcolor}
\usepackage{array}
\newcolumntype{M}[1]{>{\centering\arraybackslash}m{#1}}
\usepackage{afterpage}
\usepackage{placeins}
\usepackage{stackengine}
\usepackage{anyfontsize}
\usepackage{esint}

\makeatletter
\AtBeginEnvironment{algorithm}{\let\c@figure\c@nalgo}
\makeatother

\usepackage[nameinlink,capitalize]{cleveref}

\setstackgap{L}{0.75\normalbaselineskip}
\stackMath

\SetKwInput{KwTemp}{Temporary Variable}
\SetKwInput{KwTId}{\texttt{MPI} rank}
\SetKwInput{KwBIdx}{\texttt{blockIdx.x}}
\SetKwInput{KwBIdy}{\texttt{blockIdx.y}}
\SetKwInput{KwTIdx}{\texttt{threadIdx.x}}
\SetKwInput{KwTIdy}{\texttt{threadIdx.y}}

\SetCommentSty{mycommfont}

\definecolor{codegreen}{rgb}{0,0.6,0}
\definecolor{codegray}{rgb}{0.5,0.5,0.5}
\definecolor{codepurple}{rgb}{0.58,0,0.82}
\definecolor{backcolour}{rgb}{0.95,0.95,0.92}
\usepackage[many]{tcolorbox}
\tcbuselibrary{listings}
\tcbuselibrary{breakable}

\tcolorboxenvironment{lstlisting}{
  spartan,boxrule=0pt,
  colframe=Gray!15,
  boxsep=0mm,breakable,
  left=3mm,right=1.5mm,top=0mm,bottom=0mm,
  colback=Gray!15,
}
\lstdefinestyle{mystyle}{
    language=C++,
    backgroundcolor=\color{Gray!15},
    commentstyle=\color{Maroon},
    keywordstyle=\color{Blue},
    directivestyle={\color{green}},
    numberstyle=\color{black}\scriptsize\textbf,
    stringstyle=\color{codepurple},
    basicstyle=\ttfamily,
    breakatwhitespace=false,
    breaklines=true,
    captionpos=b,
    keepspaces=false,
    numbers=left,
    numbersep=1pt,
    showspaces=false,
    showstringspaces=false,
    showtabs=false,
    tabsize=1,
    otherkeywords={constexpr,__m512d},
    escapeinside={||}
}

\usepackage{amssymb}
\usepackage[utf8]{inputenc}
\usepackage{graphicx}
\usepackage{float}
\usepackage{stfloats}
\usepackage{microtype}
\graphicspath{{images/}}
\usepackage{amsmath}
\usepackage{bm}
\usepackage{wrapfig}
\usepackage{caption}
\usepackage{subcaption}
\usepackage{scalerel}
\usepackage{xcolor}

\newcommand{\bzero}{\mathbf{0}}
\newcommand{\del}{\nabla}

\newcommand{\imag}{\mathrm{i}}
\newcommand{\bk}{\boldsymbol{k}}

\newcommand{\bn}{\boldsymbol{n}}

\newcommand{\bu}{\boldsymbol{u}}

\newcommand{\bx}{\boldsymbol{\textbf{x}}}
\newcommand{\by}{\boldsymbol{\textbf{y}}}

\newcommand{\bB}{\boldsymbol{\textbf{B}}}
\newcommand{\bC}{\normalfont\boldsymbol{\textbf{C}}}
\newcommand{\bD}{\boldsymbol{\textbf{D}}}
\newcommand{\bE}{\normalfont\boldsymbol{\textbf{E}}}
\newcommand{\bF}{\boldsymbol{\textbf{F}}}
\newcommand{\bG}{\boldsymbol{\mathcal{G}}}
\newcommand{\bnG}{\normalfont\boldsymbol{\textbf{G}}}
\newcommand{\bH}{\boldsymbol{\textbf{H}}}
\newcommand{\bI}{\boldsymbol{\textbf{I}}}
\newcommand{\bJ}{\boldsymbol{\textbf{J}}}
\newcommand{\bK}{\boldsymbol{\textbf{K}}}

\newcommand{\bX}{\normalfont\boldsymbol{\textbf{X}}}
\newcommand{\bY}{\boldsymbol{\textbf{Y}}}
\newcommand{\bYt}{\boldsymbol{\mathcal{Y}}}

\newcommand{\bL}{\normalfont\boldsymbol{\textbf{L}}}
\newcommand{\bM}{\boldsymbol{\textbf{M}}}
\newcommand{\bN}{\normalfont\boldsymbol{\textbf{N}}}

\newcommand{\bP}{\normalfont\boldsymbol{\textbf{P}}}
\newcommand{\bQ}{\normalfont\boldsymbol{\textbf{Q}}}
\newcommand{\bR}{\normalfont\boldsymbol{\textbf{R}}}

\newcommand{\bT}{\normalfont\boldsymbol{\textbf{T}}}

\newcommand{\bXt}{\boldsymbol{\mathcal{X}}}
\newcommand{\bTt}{\boldsymbol{\mathcal{T}}}

\newcommand{\bV}{\normalfont\boldsymbol{\textbf{V}}}
\newcommand{\bnV}{\normalfont\boldsymbol{V}}

\newcommand{\bxi}{\boldsymbol{\xi}}

\newcommand{\bLam}{\boldsymbol{\Lambda}}

\newcommand{\dx}{\,d\bx}
\newcommand{\dy}{\,d\by}

\newcommand{\Rthree}{ {\mathbb{R}^{3}} }

\newcommand{\ket}[1]{\left| #1 \right>} 

\newcommand{\cblue}{\color{black}}
\newcommand{\cn}{\color{black}}

\newcommand\norm[1]{\lVert#1\rVert}

\def\appendixname{}

\journal{Computer Physics Communications}
\hypersetup{
  pdftitle={Matrix-free algorithms for fast ab initio calculations on distributed CPU architectures using finite-element discretization},
  pdfauthor={Gourab Panigrahi and Phani Motamarri}
}

\begin{document}

\begin{frontmatter}

\title{\LARGE Matrix-free algorithms for fast \emph{ab initio} calculations on distributed CPU architectures using finite-element discretization
}

\author[inst1]{Gourab Panigrahi}
\author[inst1]{Phani Motamarri}
\ead{phanim@iisc.ac.in}
\affiliation[inst1]{organization={Department of Computational and Data Sciences, Indian Institute of Science},
            addressline={CV Raman Road},
            city={Bengaluru},
            postcode={560012},
            state={Karnataka},
            country={India}}

\begin{abstract}
Finite-element (FE) discretizations have emerged as a powerful real-space alternative for efficient large-scale Kohn–Sham density functional theory (DFT) calculations, but their performance is often limited by the repeated application of the discretized Hamiltonian to large multivectors within iterative eigensolvers. Conventional global sparse matrix–vector and FE cell-matrix approaches increasingly encounter memory limitations and high data-movement overheads at the high FE polynomial orders commonly used in DFT. This work develops matrix-free algorithms for finite-element based DFT that significantly accelerate Hamiltonian application on modern distributed CPU architectures. In particular, matrix-free expressions are derived for all components of the FE-discretized Kohn–Sham operator, including kinetic, local potential, GGA gradient, periodic Bloch-shift terms, in conjunction with separable nonlocal pseudopotential terms, formulated using structured tensor contractions over one-dimensional basis functions and quadrature data. A unified multilevel batched data layout is introduced to efficiently handle both real and complex-valued operators and to maximize cache reuse and SIMD utilization on AVX2-, AVX-512-, and SVE-based systems. Additional strategies, including even–odd decomposition and mixed-precision intrinsics, further reduce floating-point operations and data movement costs. Extensive benchmarks on Frontier, Param Pravega, and Fugaku demonstrate {\cblue speedups of up to 3.6$\times$\cn} for pseudopotential DFT calculations involving large multivectors and up to 5.8$\times$ speedups for all-electron DFT calculations over cell-matrix baselines. When coupled with Chebyshev-filtered subspace iteration eigensolver tolerant to approximations in matrix-vector multiplications, the matrix-free approach delivers substantial end-to-end performance gains while maintaining desired accuracy in energies and forces. {\cblue The results show that matrix-free algorithms exploiting the tensor-product structure of the FE basis provide a practical means of accelerating small to medium-scale \emph{ab initio} simulations based on finite-element discretization of DFT on present-day CPU supercomputers.\cn}

\end{abstract}

\end{frontmatter}
\clearpage
\section{Introduction}
\emph{Ab initio} calculations based on quantum mechanical theories have long been central to the theoretical investigation of material properties, offering fundamental insights that guide both experimental and computational studies. Among these, the Kohn--Sham formulation of density functional theory (DFT) \cite{Kohn1965Self-consistentEffects,Hohenberg1964InhomogeneousGas} has emerged as one of the most widely used and impactful methods in computational materials science.  Over the past several decades, DFT calculations have enabled predictive modeling of structural, thermal, electronic, optical, and mechanical properties across diverse classes of materials{\cblue , and are among the largest consumers of compute time on many scientific high-performance computing facilities\cn}. Despite the remarkable success of DFT in reducing the complexity of the many-electron problem, DFT simulations remain computationally intensive, particularly for systems involving more than a few hundred atoms, complex chemical environments, or extended spatiotemporal scales. {\cblue There is, therefore, a continued need for numerical algorithms and implementation strategies that make DFT calculations faster on modern parallel architectures without compromising robustness or accuracy, since such developments directly extend the system sizes and time scales accessible to DFT-based simulations.\cn}

Over the past few decades, numerous codes have been developed for \emph{ab initio} materials simulations using DFT, with plane-wave (PW) \cite{Giannozzi2017AdvancedESPRESSO,Gulans_2014,b9d092bd3dd842e3ae06a5b3791921be,GONZE2002478,PhysRevB.54.11169} and atomic-orbital (AO) basis sets \cite{Pople,jensen2002,cp2k2014,blum2009,nwchem} being the most widely used. While PW methods offer systematic convergence for periodic systems, they are restricted to periodic boundary conditions and require artificial periodicity for non-periodic or semi-periodic systems, which can increase the computational cost. AO basis sets, though efficient for localized systems, lack systematic convergence and can struggle to achieve accuracy across diverse chemical environments. Moreover, both the PW and AO approaches suffer from scalability limitations on modern high-performance computing (HPC) architectures, which require massive parallelism and distributed memory efficiency. These limitations have motivated the development of real-space discretization techniques such as finite-difference (FD) \cite{Kronik2006PARSECNano-structures,Ghosh2017SPARC:Clusters,andradeRealspaceGridsOctopus2015,enkovaaraElectronicStructureCalculations2010}, finite-element (FE) \cite{Tsuchida1996AdaptiveCalculations,Pask2005FiniteCalculations,Motamarri2013Higher-orderTheory,Das2019FastSystem,kanungoLargescaleAllelectronDensity2017,Motamarri2020DFT-FECalculations,ramakrishnanFastScalableFiniteelement2025,kodali2025}, wavelets \cite{Genovese2008}, psinc basis \cite{Skylaris2005} and other reduced-order or adaptive basis formulations \cite{Hu2015DGDFT:Calculations,linTensorstructuredAlgorithmReducedorder2021}, all of which are inherently flexible and scalable. Among these, the finite-element method represents a relatively recent but rapidly growing entrant in the field of real-space DFT formulations. The FE basis offers several compelling advantages over other widely used discretization schemes for DFT calculations. Particularly, it provides a systematically improvable basis for convergence, while naturally accommodating generic boundary conditions including fully periodic, semi-periodic, and open boundaries  within a unified framework. The local support of FE basis functions also makes the method exceptionally well-suited for large-scale parallel computations. An additional and powerful feature of the FE framework lies in its ability to incorporate adaptive spatial resolution, allowing the basis to be locally refined in regions of rapid variation in the electronic structure, such as near atomic cores or bonding regions. This flexibility enables both all-electron and pseudopotential DFT simulations within the same finite-element framework. Recent studies have demonstrated that higher-order finite-element formulations (degree $\geq$ 5) can significantly outperform plane-wave (PW) approaches in norm-conserving pseudopotential DFT calculations \cite{Motamarri2020DFT-FECalculations,kodali2025} and projector augmented wave DFT calculations \cite{ramakrishnanFastScalableFiniteelement2025} for systems sizes beyond 5,000 electrons, achieving substantial reductions in computational cost. The open-source code DFT-FE, which served as the computational engine behind the ACM Gordon Bell Prize 2023 \cite{DasGB2023}, embodies these advantages through its systematically improvable FE discretization coupled with highly scalable and efficient solvers for the Kohn--Sham equations.

The finite-element discretization of the Kohn--Sham DFT problem results in a large sparse generalized eigenvalue problem, and the central computational challenge is to solve the $N_v$ smallest eigenpairs of this sparse eigenproblem where $N_v$ is proportional to number of electrons in the system. For practical problems involving multi-atomic species, $N_v$ typically ranges from 10 to beyond 10,000, depending on the size of the material system being simulated. These eigenproblems benefit from fully iterative eigensolvers \cite{saadNumericalMethodsLarge2011} that rely on iterative orthogonal projection methods. Such methods involve orthogonal projection of the underlying sparse matrix onto a carefully constructed smaller subspace rich in the wanted eigenvectors (Rayleigh-Ritz step), followed by subspace diagonalization of the projected matrix and a subspace rotation step to recover the desired orthogonal eigenvector estimates of the original sparse Hermitian matrix. Popular iterative approaches used for DFT eigenvalue problems include Davidson \cite{crouzeixDavidsonMethod1994,davidsonIterativeCalculationFew1975}, Generalised Davidson \cite{morganGeneralizationsDavidsonsMethod1986}, Chebyshev filtered subspace iteration (ChFSI) \cite{Zhou2006ParallelAcceleration}, LOBPCG \cite{knyazevOptimalPreconditionedEigensolver2001}, PPCG \cite{vecharynskiProjectedPreconditionedConjugate2015}, Lanczos \cite{sorensenImplicitApplicationPolynomial1992}, and block-Krylov methods \cite{saadNumericalMethodsLarge2011}. Recently, R-ChFSI \cite{KODALI2026110239}, a variation of ChFSI that is tolerant to approximations in matrix-vector multiplications, was proposed for accelerating large sparse eigenproblems arising in DFT calculations. Most of these iterative approaches rely on blocked approaches, and the key computational bottleneck till $10,000-15,000$ electrons is the construction of the subspace sufficiently enriched in the desired eigenvectors. Achieving this requires repeated evaluation of finite-element discretized sparse matrix-multivector products in a distributed setting, and doing so efficiently is critical for overall computationally performance.

Traditionally, finite-element discretized matrix-vector products are performed using sparse matrix-vector multiplication (SpMV) algorithms. However, for high-order FE basis functions, SpMV is penalised by irregular memory access and high data movement costs and prior studies \cite{Hughes1987Large-scaleGradients,Carey1988Element-by-elementComputations} have shown that superior performance on multithreaded architectures can be achieved by replacing global SpMV operations with FE cell-level dense matrix-vector multiplications, followed by the assembly of the resulting element(cell)-level contributions. This strategy, as employed by the DFT-FE \cite{Motamarri2020DFT-FECalculations, Das2022DFT-FEDiscretization} code on multi-node CPU and GPU systems, has demonstrated excellent throughput in evaluating FE discretized matrix-vector products involving the action on hundreds of vectors, thereby enabling efficient solutions to large-scale nonlinear eigenvalue problems arising in DFT-based quantum simulations of materials.

More recently, matrix-free finite-element algorithms have gained traction as a promising alternative in the areas of computational solid \cite{hyperelastic} and fluid mechanics \cite{Fischer2020ScalabilitySolvers,Kronbichler2012AApplication, Ljungkvist2017Matrix-FreeMeshesb}, wherein matrix-vector products are computed on the fly without explicitly forming or storing the element-level matrices. By exploiting the tensor-product structure of the FE basis, these approaches recast the three-dimensional integrals involved in the FE discretized operator action on a trial vector as a sequence of tensor contractions, thereby significantly reducing the arithmetic complexity for higher-order finite-elements, improving cache locality, and eliminating the memory footprint of explicitly stored cell-matrices or global assembled sparse matrices. Despite their promise, existing open-source implementations of such matrix-free techniques are not yet optimised or directly extensible for efficiently evaluating the action of FE discretized operators on a large number of dense vectors primarily arising in solving FE discretized large sparse eigenproblems. While recent developments, such as the use of the libCEED library for matrix-free operator evaluations on multi-component vectors \cite{Beams2020High-OrderGPUs} and efforts toward sparse multivector extensions \cite{Davydov2020AlgorithmsMulti-Vectors}, mark progress in this direction, an efficient and scalable algorithm for matrix-free FE matrix-multivector multiplication still remains a gap. Only very recently \cite{PANIGRAHI2024104925} has a batched matrix-free strategy been proposed for this purpose, but its demonstration has thus far been limited to finite-element discretizations of Helmholtz-type operators and not trivially extensible to FE discretized DFT operators that can be real and complex valued comprising both local and non-local contributions.

Motivated by these challenges, this work develops matrix-free algorithms for the Kohn--Sham DFT problem discretized using finite-element basis employing hexahedral elements. The implementation procedures are designed for efficient matrix-multivector products on modern distributed CPU architectures, building upon our recent developments \cite{PANIGRAHI2024104925}. We first derive expressions to formulate the matrix-free method for the DFT problem, wherein the local part of the FE discretized Kohn--Sham operator, including the kinetic energy term (Laplacian), exchange-correlation term (gradient), local potential, and periodic Bloch-shift terms, is expressed entirely through structured tensor contractions over one-dimensional FE shape functions and their gradients. By exploiting the tensor-product nature of high-order hexahedral finite elements, the FE cell-level local operator action is evaluated on the fly, utilising quadrature points to eliminate the need for constructing and storing element matrices, thereby reducing the memory footprint and substantially improving data locality. We further outline the treatment of real versus complex-valued operators, combining various terms with the same quadrature and reusing data structures for each step of tensor contractions.

We subsequently describe the numerical implementation of this methodology, in which a multilevel batched layout is proposed to handle both real and complex-valued matrix-free actions in an efficient single framework. We tune the batchsizes of the proposed layout in a way that tensor contractions are mapped efficiently to SIMD lanes on AVX2, AVX-512, and SVE architectures, while utilising even-odd decomposition to reduce floating-point operations by exploiting symmetry in FE basis functions. This layout, together with careful management of temporary tensors, ensures that most data remains resident in cache through the sequence of contractions. Our proposed matrix-free algorithm efficiently handles both local and non-local actions on trial multivectors by reusing data in cache with appropriate reorganization of loops that reduces data movement while switching between different layouts. We also discuss the algorithmic strategy for exchanging processor boundary degrees of freedom through non-blocking MPI communication within the context of a multilevel batched layout, as well as the selective use of uniform quadrature rules and lower-precision intrinsics for compute-intensive contractions, when coupled with an eigensolver that is tolerant to approximations in matrix-vector multiplications such as R-ChFSI.

We finally present an extensive performance evaluation of the resulting implementation across representative pseudopotential and all-electron material systems, testing the matrix-free action of both real and complex-valued DFT operators. This evaluation encompasses cases involving both the local and non-local components of the DFT operator, as well as the local component alone. The benchmarks span three representative CPU-based supercomputing platforms: Frontier (AVX2), Param Pravega (AVX-512), and Fugaku (SVE). We first evaluate the speedups obtained using the matrix-free approach compared to the cell-matrix baseline for the action of the FE discretized DFT operator on multivectors for various orders of FE interpolating polynomial. A roofline analysis is also presented, offering an architecture perspective on achieved arithmetic intensity, sustained performance, and expected scalability. We further discuss end-to-end improvements of the matrix-free approach within a Chebyshev-filtered subspace iteration eigensolver for the benchmark systems considered in this work. Across all systems, the matrix-free approach delivers clear and consistent gains over the state-of-the-art cell-matrix methodology of DFT-FE, achieving {\cblue up to 3.6$\times$\cn} acceleration for real and complex pseudopotential operators and up to 5.8$\times$ acceleration for all-electron operators, while maintaining chemical accuracy in total energies and forces. Together, these developments deliver an efficient and scalable matrix-free engine for the finite-element discretized Kohn--Sham DFT problem, significantly reducing time-to-solution across diverse materials systems and offering a general strategy that can extend to other tensor-product finite-element discretized PDE applications on modern supercomputing architectures.

The remainder of this manuscript is organized to progressively develop, analyze, and validate the proposed matrix-free finite-element framework for the Kohn--Sham DFT problem. \cref{sec:sec2} reviews the governing equations of real-space Kohn--Sham DFT and outlines the FE discretization strategy employed in this work. \cref{sec:sec3} and \cref{sec:sec4} introduces the proposed matrix-free methodology, detailing the tensor-product-based formulation of local and nonlocal operator actions, the multilevel batched data layout, and the associated hardware-aware optimizations. \cref{sec:sec5} presents a comprehensive suite of benchmark studies, evaluating performance and scalability across pseudopotential and all-electron systems on three distinct CPU architectures, along with a roofline analysis to interpret the achieved performance relative to hardware limits. Finally, \cref{sec:sec6} summarizes the key findings and discusses future directions for extending the proposed techniques to heterogeneous and GPU-accelerated platforms.

\section{Real-space Kohn--Sham density functional theory}\label{sec:sec2}
\subsection{\textbf{Governing equations}}\label{ssec:ksdfteqn}

In Kohn--Sham density functional theory, the ground-state properties of a materials system comprising $N_{a}$ nuclei and $N_e$ electrons are computed by solving $N_v(>N_e/2)$ smallest eigenpairs of a nonlinear eigenvalue problem \cite{Kohn1965Self-consistentEffects}. For completeness, we discuss here the DFT governing equations that are obtained by recasting the associated electrostatic energy to a local variational form amenable to finite-element discretization (see \cite{Das2022DFT-FEDiscretization}) which can accommodate generic boundary conditions in a unified framework. The underlying Kohn--Sham DFT governing equations within the framework of norm-conserving pseudopotentials for a periodic unit-cell $\Omega_p$ are given by:
\begin{align}
    &\left[-\frac{1}{2}\nabla^2 - \imag \bk\cdot\nabla + \frac{1}{2}|\bk|^2 + V_{\text{eff},\text{loc}} (\rho, \bR)\right]u_{n, \bk}(\bx) \nonumber \\ &+ e^{-\imag \bk \cdot \bx} \sum_a\sum_r \int_{\mathbb{R}^3} V^{a}_\text{psp,nloc}(\bx, \by, \bR_a + \bL_r) e^{\imag \bk \cdot \by}u_{n, \bk}(\by)\,d\by \nonumber \\ &\quad = \epsilon_{n, \bk} u_{n, \bk}(\bx)\;\; \text{where}\;\; n = 1,2, \cdots, N_v   \quad \forall {\ } \bk \in BZ \label{eqn:ksdftpde}\\
 &\quad \quad \quad V_{\text{eff},\text{loc}} (\rho, \bR) = V_\text{xc}(\bx) + \varphi(\bx) + \sum_a\sum_r (V_{\text{psp, loc}}^a(|\bx - (\bR_a + \bL_r)|) - V^a_{\text{sm}}(\bx, \bR_a + \bL_r)) \label{eqn:Veff}
\end{align}
\begin{equation}
    -\frac{1}{4\pi}\nabla^2\varphi(\bx) = \rho(\bx) + b(\bx, \bR) ,\quad \rho(\bx) = 2\sum_n \fint_{BZ} f(\epsilon_{n, \bk}, \mu)|u_{n, \bk}(\bx)|^2d\bk, \quad 2\sum_n \fint_{BZ} f(\epsilon_{n, \bk}, \mu) d\bk = N_e. \label{eqn:poissonrho}
\end{equation}
where $\imag = \sqrt{-1}$ and $u_{n, \bk}(\bx)$ is a complex-valued function that is periodic on the unit-cell satisfying $u_{n, \bk}(\bx + \bL_r) = u_{n, \bk}(\bx)$ for all lattice vectors $\bL_r$ and $\bk$ denotes a point in the first Brillouin zone (BZ) of the reciprocal space. Further  $\bR =
\{\bR_1, \bR_2, \cdots \}$ denotes the positions of the $N_a$ nuclei and the summations over $a$ in \cref{eqn:ksdftpde} runs over all these atoms in the unit-cell. Here, $\rho(\bx)$ denotes the electron density while $f_{n,\bk} = f(\epsilon_{n,\bk}, \mu)  \in [0, 1]$ denotes the orbital occupancy function and $\fint_{BZ}$ denotes the volume average of the integral over the first BZ corresponding to the periodic unit cell $\Omega_p$, and $\mu$ represents the Fermi-energy, determined by the constraint on the total number of electrons $N_e$ in the system as indicated in \cref{eqn:poissonrho}. In addition, $V_{\text{xc}}$ in the above governing equations denotes the exchange-correlation potential, the functional derivative of the exchange-correlation energy $E_{\text{xc}}(\rho)$, which accounts for the many-body quantum mechanical interactions between electrons. In this work, we adopt the widely used generalized gradient approximation (GGA) \cite{rmartin,gga1} given by $E_{xc}(\rho) = \int_{\Omega_p}\varepsilon_\text{xc}(\rho, \nabla\rho) d\bx$. Further, $V_{\text{psp,loc}}^a$ and $V^{a}_\text{psp,nloc}$ in \cref{eqn:ksdftpde} denote the local and non-local contributions of the pseudopotential operator associated with an atom $a$. Accordingly, the action of the non-local pseudopotential operator $V^{a}_\text{psp,nloc}$ on a wavefunction can be written in a separable form \cite{rmartin,kb82} as follows:
\begin{align}\label{eq:nlps}
 &e^{-\imag \bk\cdot\bx}\int\displaylimits_{\Rthree} \sum_{a} \sum_r V_{\text{psp,nloc}}^a(\bx,\by,\bR_a+\bL_r) e^{\imag \bk\cdot\by}\,u_{n,k}(\by) \dy \notag
 = \sum_{a} \sum_{lpm} \sum_{r} e^{-\imag\bk\cdot(\bx - \bL_r)} F_{lpm}^{a,n\bk} h^{a}_{lp}\,\chi_{lpm}^{a}(\bx,\bR_a+\bL_r)\,,\notag\\
&\text{with}\; F_{lpm}^{a,n\bk} =  \int_{\Omega_p} \sum_{r^{\prime}} \chi_{lpm}^{a}(\by,\bR_a+\bL_{r^{\prime}}) e^{\imag\bk\cdot(\by - \bL_{r^{\prime}})}  u_{n,\bk}(\by)\,\dy, \quad \frac{1}{h^{a}_{lp}} = \langle\phi_{lm}^a|\chi_{lpm}^a\rangle \,,
\end{align}
with $l$, $m$ denoting the azimuthal quantum number and the magnetic quantum number respectively and $\ket{\chi_{lpm}}$ denotes the pseudopotential projector while $p$ is the index corresponding to the projector component for a given $l$. Further, $h_{lp}$ denotes the pseudopotential constant, and $\ket{\phi_{lm}^a}$ denotes the single atom pseudo-wavefunction.
Additionally, $V^a_{\text{sm}}$ in \cref{eqn:Veff} denotes the self-potential arising from introducing the atom-centered smeared charges $b^{a}_{\text{sm}}(\bx-\bR^a)$ in the local real-space formulation of the electrostatics and is obtained by solving a Poisson equation with $b^{a}_{\text{sm}}(\bx-\bR^a)$ as the forcing term subject to Dirichlet boundary conditions on a spherical ball around the atomic charge (refer to \cite{Das2022DFT-FEDiscretization} for more details). Here, $b(\bx,\bR) = \sum_a b_{\text{sm}}^a(\bx-\bR^a)$ in \cref{eqn:poissonrho} and $\varphi(\bx)$ in this equation denotes the electrostatic potential arising due to $\rho(\bx)$ and $b(\bx, \bR)$ which is obtained by solving the Poisson problem with appropriate boundary conditions. Finally, \cref{eqn:ksdftpde} and \cref{eq:nlps} can be easily modified to the non-periodic or semi-periodic setting by considering the components of $\bk$ and $\bL_r$ to be 0 in the non-periodic directions. Further, we consider a large non-periodic domain with a vacuum layer in the non-periodic directions surrounding the material system with appropriate Dirichlet or Neumann boundary conditions on $\varphi(\bx)$.

\subsection{\textbf{Finite element discretization}}\label{ssec:ksdftfe}
We consider here the discretization of the governing equations corresponding to the Kohn-Sham DFT problem in \cref{eqn:ksdftpde} using a FE basis. The FE basis functions chosen in this work comprise piecewise polynomials that are strictly local \cite{Brenner2008TheMethods}. Specifically, these are $C^0$ continuous Lagrange interpolants constructed using Gauss-Lobatto-Legendre (GLL) nodal points, commonly referred to as spectral finite elements. For more details on the advantages of higher-order spectral FE basis discretization of the Kohn--Sham equations, we refer the reader to \cite{Das2022DFT-FEDiscretization,Motamarri2020DFT-FECalculations,Motamarri2013Higher-orderTheory}. The electronic fields in the Kohn--Sham equations, namely the single-particle Kohn--Sham wavefunctions and the electrostatic potentials are expanded in the FE basis (referred as shape functions) as follows:
\begin{align}\label{eqn:FEdiscr}
u^h_{n,\bk}(\bx) = \sum^m_{J=1}N^h_J(\bx)u^J_{n,\bk}, \quad\quad \varphi^h(\bx) = \sum^m_{J=1}N^h_J(\bx)\varphi^J 
\end{align}
where $N^h_J : 1 \leq J \leq m$ denote the localised 3D Lagrange polynomials generated using the $m$ nodes of the FE triangulation $\mathcal{T}^h$, where $h$ is the characteristic mesh size. The FE interpolating polynomial order within each element is denoted by $\texttt{FEOrder}=n_p-1$, with $n_p$ denoting the number of nodal points in each spatial direction for a given FE cell. The coefficients $u^J_{n, \bk}$, $\varphi^J$ represent the nodal values of the discretized wavefunction corresponding to the $n^{th}$ wavefunction and electrostatic potential respectively.

Using \cref{eqn:FEdiscr}, the FE discretized Kohn--Sham eigenvalue problem in \cref{eqn:ksdftpde} yields a generalized Hermitian eigenvalue problem (GHEP) to be solved for every $\bk$ point expressed as:
\begin{gather}
\bH^{\bk}\hat{\bu}_{n, \bk} = \epsilon^h_{n, \bk} \bM \hat{\bu}_{n, \bk}; \quad\quad \forall {\ } \bk \in BZ \label{eqn:FEeigen}
\end{gather}
where $\bH^{\bk}$ is the discrete Hamiltonian matrix, $\bM$ is the finite-element overlap matrix (or commonly referred to as the mass matrix in finite element literature), and $\epsilon^h_{n, \bk}$ is the $n^{th}$ eigenvalue corresponding to the discrete eigenvector $\hat{\bu}_{n, \bk}$. The discrete Hamiltonian matrix $\bH^{\bk}$ has two contributions i.e., $\bH^{\bk} = \bH_{\text{loc}}^{\bk} + \bH_{\text{nloc}}^{\bk}$ with $\bH_{\text{loc}}^{\bk} = \bT + \bL + \bnG -\imag \bK$ where the matrix components of $\bT$, $\bL$, $\bnG$ and $\bK$ are given below:
\begin{gather}
\text{T}_{IJ} = \frac{1}{2}\int_{\Omega_p}\nabla N^h_I(\bx) \cdot \nabla N^h_J(\bx) d\bx, \;\;\; \text{L}_{IJ} = \int_{\Omega_p}V^{L}(\bx,\bk) N^h_I(\bx) N^h_J(\bx) d\bx \nonumber\\
\text{G}_{IJ} = \int_{\Omega_p}  \bnV^{G}(\bx) \cdot \left(\nabla N^h_I(\bx) N^h_J(\bx) + N^h_I(\bx) \nabla N^h_J(\bx) \right) d\bx,\;\;\; \text{K}_{IJ} = {\cblue\int_{\Omega_p} \bk \cdot \left( N^h_I(\bx) \nabla N^h_J(\bx) \right) d\bx \cn}\label{eq:Hloc}
\end{gather}
where $V^L(\bx,\bk)$ and $\bnV^{G}(\bx)$ are given by:
\begin{align}\label{eq:Vloc}
V^L(\bx,\bk) &= \frac{\partial\varepsilon_{xc}(\rho, \nabla\rho)}{\partial\rho}\Bigg|_{\rho = \rho^h} + \varphi^h(\bx) + \sum_a\sum_r\left( V^a_{\text{ext, loc}} (|\bx - (\bR_a + \bL_r)|) - V^{a^h}_{\text{sm}} (\bx, \bR_a + \bL_r) \right) + \frac{1}{2}|\bk|^2\\
\bnV^{G}(\bx)  &= \frac{\partial\varepsilon_\text{xc}(\rho, \nabla\rho)}{\partial\nabla\rho}\Bigg|_{\rho = \rho^h}, \quad\quad \rho^h (\bx) = 2 \sum_{n=1}^{N_v} \fint_{BZ} f(\epsilon^h_{n, \bk}, \mu) |\bu^h_{n, \bk}(\bx)|^2 d\bk \label{eqn:den}
\end{align}
where the electrostatic potentials $V^{a^h}_{\text{sm}} (\bx, \bR_a + \bL_r)$ and $\varphi^h(\bx)$ are obtained by solving the discretized Poisson equations as discussed in \cite{Das2022DFT-FEDiscretization}. Importantly, the term $\bH_{\text{nloc}}^{\bk}$ is zero in all-electron calculations and $V^a_{\text{ext, loc}} (|\bx - (\bR_a + \bL_r)|)$ in \cref{eq:Vloc} equals $Z_I/|\bx-\bR_I|$, whereas in the case of norm-conserving pseudopotential calculations, we have $V^a_{\text{ext, loc}} (|\bx - (\bR_a + \bL_r)|) = V^a_{\text{psp, loc}} (|\bx - (\bR_a + \bL_r)|)$. Additionally, the matrix components for $\bH_{\text{nloc}}^{\bk}$ is given in terms of compactly supported projector functions $\chi_{lpm}^{a}$ as 
\begin{align}\label{eq:Vnonloc}
\text{H}^{\bk}_{\text{nloc},IJ} = \sum_a\sum_{lpm} F^{a, \bk}_{lpm, I} h^a_{lp} F^{a, \bk *}_{lpm, J}, \quad \text{where} {\ } F^{a, \bk}_{lpm, I} = \int_{\Omega_p} \sum_r e^{-i\bk\cdot (\bx - \bL_r)} \chi^a_{lpm} (\bx, \bR_a + \bL_r) N^h_I(\bx) d\bx
\end{align}
Further, the matrix $\bM$ arising in the GHEP has entries given by $M_{IJ} = \int_{\Omega_p}N^h_I(\bx) N^h_J (\bx) d\bx$. In addition, we also note that the matrices $\bH_{\text{loc}}^{\bk}$ and $\bM$ are sparse due to the compact support of the FE basis functions in real space. We also note that $\bH_{\text{nloc}}^{\bk}$ is sparse since the projectors $\chi^a_{lpm} (\bx, \bR_a + \bL_r)$ are localized and have a compact support, rendering sparsity to the matrix $\bF$ with components $F^{a, \bk}_{lpm, I}$, resulting in an overall sparse representation of the discrete Hamiltonian $\bH^{\bk}$. {\cblue As is evident from \cref{eq:Vnonloc}, $\bH_{\text{nloc}}^{\bk}$ is a sum of rank-one matrices, one for each projector per atom. In the methodology developed in this work, $\bH_{\text{nloc}}^{\bk}$ is never assembled and its action on trial vectors is evaluated directly through this separable structure (see \cref{sec:sec3}), and hence it is this rank-one structure rather than the sparsity of the assembled $\bH_{\text{nloc}}^{\bk}$ that the proposed strategy relies on. The compact support of the projectors nevertheless plays an important role in keeping the cost of this evaluation low, since only the FE cells overlapping with the support of $\chi^a_{lpm}$ contribute for a given atom $a$.\cn} Finally, obtaining the Kohn--Sham DFT electronic ground-state involves the self-consistent solution (SCF) of the discrete governing equations, where a nonlinear generalised algebraic eigenvalue problem is solved for $N_v$ smallest eigenpairs coupled with a linear system arising from the Poisson equation that computes the total electrostatic potential $\varphi^h$ with constraint on number of electrons $N_e$ in the system.

\subsection{\textbf{Fully iterative eigensolvers for DFT problem}}
The finite-element discretized Hermitian sparse generalised eigenproblem in \cref{eqn:FEeigen} arising in Kohn--Sham DFT benefits from fully iterative eigensolvers that rely on iterative orthogonal projection methods as discussed before. Current state-of-the-art finite-element implementations for electronic structure calculations using DFT employ the cell-matrix approach (CM) \cite{Motamarri2020DFT-FECalculations,Das2022DFT-FEDiscretization}, wherein the underlying sparse matrix-vector multiplications to construct the desired eigensubspace are performed using the FE cell-level dense matrix-vector multiplications followed by the assembly of FE cell-level product vectors. 

We now describe the proposed matrix-free (MF) methodology within the context of the finite-element discretized DFT problem, leading to the development of multilevel batched implementation strategies for on-the-fly matrix-vector products in a distributed setting, a departure from the state-of-the-art approaches. As demonstrated later, these matrix-free approaches can significantly improve computational efficiency in the subspace construction step of an eigensolver without the necessity of storing the FE cell-level dense matrices, thereby reducing arithmetic complexity, data movement and memory footprint.

\section{Matrix-free methodology for the DFT problem}\label{sec:sec3}
To begin, we will establish the necessary notations to describe the proposed multilevel batched matrix-free algorithm for the Kohn--Sham DFT problem. The mathematical approach derived in this subsection for the DFT problem is in the spirit of \citet{Kronbichler2012AApplication}.
In this work, finite-element discretization of the computational domain $\Omega$ is carried out by decomposing it into $E$ non-overlapping hexahedral FE cells $\Omega^{(e)}$, such that $\Omega=\bigcup_{e=1}^E\Omega^{(e)}$. To enable efficient distributed parallelism, the domain $\Omega$ is further partitioned into $n_t$ subdomains $\Omega_{(t)} \; \forall {\ } t=1,2,\dots,n_t$, with each subdomain $\Omega_{(t)}$ assigned to a corresponding MPI task $t$. Within a subdomain $\Omega_{(t)}$, let $E_t$ denote the number of FE cells and $m_t$ the number of basis functions, so that $\Omega_{(t)}=\bigcup_{e=1}^{E_t}\Omega^{(e)}$. Let $N_v$ be the number of eigenvectors solved in the Kohn--Sham DFT problem \cref{eqn:FEeigen}, which the batched operator processes in multivectors (blocks) of $n_v$ vectors. Let each atom in the simulation domain be indexed by $a$ and define $\mathcal{T}(a)$ as the set of tasks with compact support of the projector function  $\chi^{a}$  corresponding to this atom $a$ associated with the non-local contribution of the FE discretized operator described in \cref{eq:Vnonloc}. Similarly, define $\mathcal{E}(a, t)$ as the set of cells containing the compact support of  $\chi^{a}$ for a given atom $a$ and task $t$. In addition, $e^a \in \mathcal{E}(a, t)$ and $t^a \in \mathcal{T}(a)$ are the indices used to represent the cell-level FE discretized projector matrices $\mathbf{F}^{(a,e^a,t^a)}$ described in \cref{eq:Vnonloc}. Further, let $\alpha_t$ denote the number of atoms in a given task $t$.

\subsection{\textbf{Action of FE discretized DFT operator on multivectors}}
As discussed in \cref{eqn:FEeigen}, $\bH$ denotes the global sparse matrix associated with the FE discretized Kohn--Sham operator (the superscript $\bk$ is dropped here and subsequently for notational convenience) and further $\bX$ denotes the multivector matrix whose columns comprise trial eigenvectors. Thus, the matrix-multivector product $\bH \bX$ arising during the course of any iterative eigensolver in solving the DFT problem can be written mathematically as follows:
\begin{align} \label{eqn:HX}
\bY=\bH \bX = \bH_\text{loc}\bX &+ \bH_\text{nloc}\bX=\left[\sum_{t}^{n_t}{\bP^{(t)}}^T{\bC^{(t)}}^T\left(\sum_e^{E_t}{\bQ^{(e,t)}}^T\bH_\text{loc}^{\left(e\right)}\bQ^{(e,t)}\right)\bC^{(t)}\bP^{(t)}\right]\bX\nonumber \\ &+ \left[\sum_{t}^{n_t}{\bP^{(t)}}^T{\bC^{(t)}}^T \left[\sum_e^{E_t}{\bQ^{(e,t)}}^T \left( \sum_a \bF^{(a,e,t)} \Delta^a\!\left( \sum_{t^a}\sum_{e^a} \bF^{(a, e^a, t^a) \dagger} \right)\right)\right.\right.\nonumber\\ &\quad\quad\left.\left.{\bQ^{(e^a,t^a)}} \right] \bC^{(t^a)}\bP^{(t^a)}\right]\bX
\end{align}

Here, the operator $\mathbf{H}$ is decomposed into a local part $\mathbf{H}_\text{loc}$ and a nonlocal part $\mathbf{H}_\text{nloc}$, such that $\mathbf{H} = \mathbf{H}_\text{loc} + \mathbf{H}_\text{nloc}$. The index pair $(e,t)$ identifies the FE cell index $e$ associated with an MPI task $t$, and $\mathbf{P}^{(t)}$ is a Boolean sparse matrix that acts as the partitioner. Applying $\mathbf{P}^{(t)}$ to the global multivector $\mathbf{X}$ yields the subdomain-level multivector $\mathbf{X}^{(t)}$. The matrix $\mathbf{P}^{(t)}$ enforces the continuity of the discretized multivector $\mathbf{X}$ across the partitioned subdomains. The sparse matrix $\mathbf{C}^{(t)}$ in \cref{eqn:HX} is a constraint matrix of size $m_t \times m_t$ used to constrain the values of $m_t \times n_v$ matrix $\mathbf{X}^{(t)}$ at specific nodal points. These constraints are applied to satisfy necessary boundary conditions (either periodic or semi-periodic or non-periodic) on the discretized electronic wavefunctions $u^{h}_{n, \bk}(\bx)$ or to handle constraints arising from non-conforming meshes \cite{Bangerth2009DataSoftware}. The continuity condition of the discretized electronic fields $u^{h}_{n, \bk}(\bx)$ within a partitioned subdomain $\Omega_{(t)}$ across its constituent FE cells is imposed by the action of the $n_p^3 \times m_t$ Boolean sparse matrix $\mathbf{Q}^{(e,t)}$ on the constrained subdomain-level multivector $\mathbf{C}^{(t)}\mathbf{P}^{(t)}\mathbf{X}$. In effect, $\mathbf{Q}^{(e,t)}$ represents the mapping from the subdomain-level to the FE cell-level within the subdomain $\Omega_{(t)}$. The computation of the FE cell-level matrix $\mathbf{H}_\text{loc}^{(e)}$, resulting from the local contributions to the finite-element discretized DFT problem, involves evaluating 3D integrals using a $n_q$-point quadrature rule over the reference domain $\widehat{\Omega} = [-1,1]^3$ as is usually done. In \cref{eqn:HX}, $\Delta^a$ are the pseudopotential coefficients, and $\mathbf{F}^{(a,e,t)}$ are the cell-level non-local projector matrices computed from projectors having a compact support for a given atom $a$, in a specific cell $e$ within a task $t$. These projector matrices are used to evaluate the action of the discretized nonlocal part of the underlying PDE on the multivector $\mathbf{X}$.

A standard method for computing the matrix-multivector product in \cref{eqn:HX} involves assembling the \emph{global FE discretized matrix} $\bH_\text{loc}$ from the cell-level matrices $\bH_\text{loc}^{(e)}$ and then performing the sparse-matrix dense-matrix multiplication in a distributed manner. However, evaluation of such sparse matrix-vector products can be significantly slower on modern architectures due to expensive data movement costs incurred in loading sparse matrix data onto CPU registers. In DFT calculations, evaluation of these products arising from the higher-order finite-element discretizations was found to be more efficient \cite{Motamarri2020DFT-FECalculations,Das2022DFT-FEDiscretization} on multithreaded architectures using FE cell-level dense matrix-vector multiplications followed by assembly of FE cell-level product vectors in each MPI task and then across all MPI tasks, thereby incurring minimal parallel communication costs in a distributed setting. Additionally in the case of norm-conserving pseudopotential case, to better exploit the sparsity of the matrix $\bH_\text{nloc}$,  the action of  $\bH_\text{nloc}$ on $\bX$ in \cref{eqn:HX} is evaluated by first computing the action of cell-level matrices $\mathbf{F}^{(a,e,t) \dagger}$ on the extracted cell-level vectors, followed by the addition of the individual atomic contributions from all FE cells for each atom, scaled with the diagonal matrix $\Delta^{a}$ and subsequently the action of cell-level $\mathbf{F}^{(a,e,t)}$ on these atomic contributions to obtain FE cell-level product vectors that are finally assembled and added to $\bH_{\text{loc}}\bX$. Inspired by other scientific domains involving finite-strain hyperelasticity \cite{hyperelastic}, fluid flow problems \cite{Fischer2020ScalabilitySolvers}, we now discuss the matrix-free methodology for the DFT problem in computing  $\bH_{\text{loc}}\bX$, the computationally dominant step in the action of FE discretized DFT operator on multivectors $\bX$. We therefore derive the relevant expressions for the action of the local FE discretized operator on multivectors within the matrix-free framework.

\subsubsection{\textbf{Reformulation of FE cell-level matrices for the DFT operator}}\label{sec:cellmatrix}
Using the expressions for the FE discretized local operator described in the DFT problem in  \cref{eq:Hloc}, the cell-level matrices $\bH_\text{loc}^{\left(e\right)}$ assume the following form,
\begin{align} \label{eqn:Hloccell}
&\bH_\text{loc}^{\left(e\right)} = \left(\bT^{\left(e\right)} + \bL^{\left(e\right)} + \bnG^{\left(e\right)} - \imag \bK^{\left(e\right)} \right)
\end{align}
with $\bT^{\left(e\right)}$ denoting the discretized kinetic energy operator, $\bL^{\left(e\right)}$ being the discretized operator corresponding to contributions from electrostatic potential and the density derivative part of the GGA exchange-correlation potential, $\bnG^{\left(e\right)}$ corresponds to the contribution from the gradient-density derivative part of the GGA exchange-correlation potential, while $\bK^{\left(e\right)}$ arises because of $\bk$-points sampling the Brillouin zone. The components associated with each of the four terms in \cref{eqn:Hloccell} are given below for completeness
\begin{subequations} \label{eqn:HlocIntegral}
\begin{align}
&\text{T}^{(e)}_{IJ}=\int_{\Omega_p^{(e)}}\frac{1}{2}\del N^{h}_I(\bx) \cdot \del N^{h}_J(\bx) \dx \quad\quad
\text{G}^{(e)}_{IJ}=\int_{\Omega_p^{(e)}} \bnV^{G}(\bx) \cdot( N^{h}_I(\bx) \, \del N^{h}_J(\bx) + \del N^{h}_I(\bx) \, N^{h}_J(\bx) ) \dx \\
&\text{L}^{(e)}_{IJ}=\int_{\Omega_p^{(e)}} V^{L}(\bx,\bk) N^{h}_I(\bx) \, N^{h}_J(\bx) \dx  \quad\quad
\text{K}^{(e)}_{IJ}=\int_{\Omega_p^{(e)}} \bk \cdot( N^{h}_I(\bx) \, \del N^{h}_J(\bx)) \dx
\end{align}
\end{subequations}
As is usually done in FEM, a reference domain $\hat{\Omega}$ with a coordinate system $\bxi$ is introduced with $\bJ^{(e)}$ denoting the cell-level Jacobian matrix of the map from $\Omega_{p}^{(e)}$ to $\widehat{\Omega}$ with $\widehat{N}_I(\bxi)$ representing the shape function in the reference cell. With this, the integrals over FE cell domains in \cref{eqn:HlocIntegral} can be transformed to $\hat{\Omega}$ and furthermore the spatial gradients of shape functions $\del N^{h}_I(\bx)$ can be expressed in terms of gradients with respect to reference coordinates $\nabla_{\bxi} \widehat N_I(\bxi)$ and $\bJ^{(e)}$. The integrals are then evaluated using Gauss-Legendre quadrature points, a tensor-structured $n_q$-point rule with quadrature points $\hat{\bxi}_{Q}$ and the weights $w_{Q}$ as shown below:
\begin{align}
{\cblue\int_{\Omega_p^{(e)}}(.) \dx\rightarrow \int_{\widehat{\Omega}}(.) \det(\bJ^{(e)}) d\bxi}\rightarrow\sum_{Q=1}^{n_q^{3}} (.)\: w_Q \det(\bJ^{(e)})\biggr\rvert_{\widehat{\bxi}_{Q}}\label{eqn:cellIntegrals}
\end{align}
Defining the $n_q^3 \times n_p^3$ matrices, $\text{N}_{QI}=\widehat N_I(\bxi_Q)$ and $\text{D}^{(s)}_{QI}=\nabla_{\bxi}\widehat N_I\left(\bxi_Q\right)\cdot \widehat\bn_s$  where $\widehat\bn_s, \; s=0,1,2$ represents the unit vector along the $s$ axis {\cblue (the rows of these matrices are indexed by the $n_q^3$ quadrature points and the columns by the $n_p^3$ nodal points of a given FE cell)\cn}, we can now rewrite the matrix expressions in \cref{eqn:HlocIntegral} corresponding to the four contributions in \cref{eqn:Hloccell}
 as follows:
\begin{align}\label{eqn:reformcellmat}
      &\bT^{\left(e\right)}=\begin{bmatrix}
        \bD^{(0)}\\
        \bD^{(1)}\\
        \bD^{(2)}
    \end{bmatrix}^T
    \begin{bmatrix}
        \bG^{(00)}&&\bG^{(01)} &&\bG^{(02)}\\
        \bG^{(10)}&&\bG^{(11)} &&\bG^{(12)}\\
        \bG^{(20)}&&\bG^{(21)} &&\bG^{(22)}\\
    \end{bmatrix}
    \begin{bmatrix}
        \bD^{(0)}\\
        \bD^{(1)}\\
        \bD^{(2)}
    \end{bmatrix},
    & & \bL^{\left(e\right)}=\bN^T\bV^{L}\bN, \nonumber \\
    &\bnG^{\left(e\right)} = \left( \begin{bmatrix}
        \bD^{(0)}\\
        \bD^{(1)}\\
        \bD^{(2)}
    \end{bmatrix}^T\begin{bmatrix}
        \bV^G_0\\
        \bV^G_1\\
        \bV^G_2
    \end{bmatrix}\bN + \bN^T\begin{bmatrix}
        \bV^G_0\\
        \bV^G_1\\
        \bV^G_2
    \end{bmatrix}^T\begin{bmatrix}
        \bD^{(0)}\\
        \bD^{(1)}\\
        \bD^{(2)}
    \end{bmatrix}\right),
     & & \bK^{\left(e\right)} = \bN^T\begin{bmatrix}
        \bV^K_0\\
        \bV^K_1\\
        \bV^K_2
    \end{bmatrix}^T\begin{bmatrix}
        \bD^{(0)}\\
        \bD^{(1)}\\
        \bD^{(2)}
    \end{bmatrix}
\end{align}
where $\bG^{\left(s,d\right)}$, $\bV^L$, $\bV^G_s$ and $\bV^K_s$ for $s,d = 0,1,2$ are $n_q^3 \times n_q^3$ diagonal matrices with the diagonal entries as below:
\begin{gather}
\mathcal{G}^{\left(s,d\right)}_{QQ}=\left[\left(\bJ^{\left(e\right)}\right)^{-1}\left(\bJ^{\left(e\right)}\right)^{-T}\right]_{sd}\det{\bJ^{\left(e\right)}} w_Q\biggl|_{\widehat\bxi_Q}, \;\;\;\;
\text{V}^{L}_{QQ} = V^{L}(\bx(\bxi),\bk) \det{\bJ^{\left(e\right)}}w_Q\biggl|_{\widehat\bxi_Q}\\
\text{V}^{G}_{s,QQ} = V_{s}^{G}(\bx(\bxi)) \det{\bJ^{\left(e\right)}}w_Q\biggl|_{\widehat\bxi_Q},\;\;\;
\text{V}_{s,QQ}^{K} = k_s \det{\bJ^{\left(e\right)}}w_Q\biggl|_{\widehat\bxi_Q}
\end{gather}
{\cblue The reformulation above, and the matrix-free procedure deduced from it subsequently, does not require the map from $\widehat{\Omega}$ to $\Omega_p^{(e)}$ to be affine. For a general (distorted) hexahedral cell, $\bJ^{(e)}$ varies within the cell, and the diagonal matrices $\bG^{(s,d)}$, $\bV^L$, $\bV^G_s$ and $\bV^K_s$ are simply populated with the corresponding values at each quadrature point, leaving the tensor-product structure exploited later unchanged. This is indeed the setting in general-purpose matrix-free frameworks \cite{Kronbichler2012AApplication}. In the present work, however, the FE meshes are generated by recursive subdivision of parallelepiped cells, as is the common practice in FE-based DFT calculations \cite{Motamarri2020DFT-FECalculations,Das2022DFT-FEDiscretization}, so that every cell is an affine image of $\widehat{\Omega}$ and $\bJ^{(e)}$ is constant within a cell. We exploit this fact in our implementation by storing a single Jacobian matrix and its determinant per cell, thereby reducing the memory traffic associated with the geometric data.\cn}


The cell-matrix approach alluded to before explicitly builds the above FE cell-level matrices for the action of FE discretized operator on trial multivectors at the cell-level. We now describe how the reformulated expressions for $\bH_\text{loc}^{\left(e\right)}$ described in \cref{eqn:reformcellmat} act as a segue for deriving the expressions for matrix-free action of the DFT operator on multivectors.

\subsection{\textbf{Matrix-free action of DFT operator}}
This strategy begins with the extraction of FE cell-level multivectors $\bX^{(e,t)}$
using the subdomain-level to FE cell-level maps, the constraint and the partitioner matrices as described in \cref{eqn:HX} i.e., $\bX^{(e,t)}=\bQ^{(e,t)}\bC^{(t)}\bP^{(t)}\bX \quad \forall {\ } e=1,2,\dots,E_t$.  As discussed before, in contrast to cell-matrix approach, the matrix-free approach circumvents the precomputation of the local FE cell-level matrices $\bH^{(e)}_{\text{loc}}$ and instead the action of  $\bH^{(e)}_{\text{loc}}$ on $\bX^{\left(e,t\right)}$ in \cref{eqn:HX} is computed on-the-fly. This is accomplished by leveraging the tensor-product structure of the 3D FE basis functions and the tensor-product Gauss quadrature rules used to evaluate the integrals within $\bH_{\text{loc}}^{(e)}$, with more details described subsequently. We now deduce the expressions for matrix-free action on $\bX^{\left(e,t\right)}$ from \cref{eqn:reformcellmat} as the starting point. Similar in spirit to matrix-free methods developed for FE discretized operators arising in other scientific domains \cite{Deville2002High-OrderFlow,Fischer2020ScalabilitySolvers}, we use the factorization $\bD^{(s)} = \widetilde{\bD}^{(s)} \bN$ to simplify \cref{eqn:reformcellmat}, where $\widetilde{\bD}^{(s)}$ is a $n_q^{3} \times n_q^{3}$ matrix with $\widetilde{\text{D}}^{(s)}_{Q\widehat{Q}}=\nabla_{\bxi}\widetilde{N}_{\widehat{Q}}\left(\bxi_Q\right)\cdot \widehat\bn_s$ and $\widetilde{N}_{\widehat{Q}}$ is the Lagrange polynomial centered over the quadrature point $\widehat{Q}$. {\cblue We remark that the functions $\widetilde{N}_{\widehat{Q}}(\bxi)$ are the Lagrange interpolating polynomials of degree $n_q-1$ in each coordinate direction constructed on the grid of $n_q^3$ quadrature points, satisfying $\widetilde{N}_{\widehat{Q}}(\bxi_Q)=\delta_{\widehat{Q}Q}$. The factorization itself is a consequence of polynomial exactness of this interpolation i.e., each shape function $\widehat{N}_I(\bxi)$ is a polynomial of degree $n_p - 1 \leq n_q - 1$ in each coordinate direction, and is therefore exactly reproduced by its interpolant on the quadrature grid, i.e., $\widehat{N}_I(\bxi)=\sum_{\widehat{Q}}\widehat{N}_I(\bxi_{\widehat{Q}})\widetilde{N}_{\widehat{Q}}(\bxi)$. Differentiating this identity along $\widehat\bn_s$ and evaluating at the quadrature points $\bxi_Q$ gives $\text{D}^{(s)}_{QI}=\sum_{\widehat{Q}}\widetilde{\text{D}}^{(s)}_{Q\widehat{Q}}\text{N}_{\widehat{Q}I}$, which is the stated factorization $\bD^{(s)} = \widetilde{\bD}^{(s)} \bN$. Its utility lies in the fact that the shape function data enters the operator action only through $\bN$, whose action (and that of its transpose) is evaluated just once and reused, while the gradients are handled entirely on the quadrature grid through $\widetilde{\bD}^{(s)}$, the collocation derivative matrices \cite{Kopriva2009ImplementingEquations,Solomonoff1992ADifferentiation}.\cn} Hence the four contributions in the expressions for $\bH_{\text{loc}}^{(e)} =\left(\bT^{\left(e\right)} + \bL^{\left(e\right)} + \bnG^{\left(e\right)} - \imag \bK^{\left(e\right)}\right)$ in the DFT problem can be written as
\begin{align}\label{eqn:refact}
    \bH_{\text{loc}}^{(e)} = \bN^{T}\Biggl(\begin{bmatrix}
        \widetilde{\bD}^{(0)}\\
        \widetilde{\bD}^{(1)}\\
        \widetilde{\bD}^{(2)}
    \end{bmatrix}^T
    \begin{bmatrix}
        \bG^{(00)}&&\bG^{(01)} &&\bG^{(02)}\\
        \bG^{(10)}&&\bG^{(11)} &&\bG^{(12)}\\
        \bG^{(20)}&&\bG^{(21)} &&\bG^{(22)}\\
    \end{bmatrix}
    \begin{bmatrix}
        \widetilde{\bD}^{(0)}\\
        \widetilde{\bD}^{(1)}\\
        \widetilde{\bD}^{(2)}
    \end{bmatrix} + \bV^{L} &+  \begin{bmatrix}
        \widetilde{\bD}^{(0)}\\
        \widetilde{\bD}^{(1)}\\
        \widetilde{\bD}^{(2)}
    \end{bmatrix}^T\begin{bmatrix}
        \bV^G_0\\
        \bV^G_1\\
        \bV^G_2
    \end{bmatrix} + \begin{bmatrix}
        \bV^G_0\\
        \bV^G_1\\
        \bV^G_2
    \end{bmatrix}^T\begin{bmatrix}
        \widetilde{\bD}^{(0)}\\
        \widetilde{\bD}^{(1)}\\
        \widetilde{\bD}^{(2)}
    \end{bmatrix}  \nonumber \\
    &-\imag \begin{bmatrix}
        \bV^K_0\\
        \bV^K_1\\
        \bV^K_2
    \end{bmatrix}^T
    \begin{bmatrix}
        \widetilde{\bD}^{(0)}\\
        \widetilde{\bD}^{(1)}\\
        \widetilde{\bD}^{(2)}
    \end{bmatrix}\Biggr)\bN
\end{align}

Consequently, on-the-fly matrix-free action of $\bH_{\text{loc}}^{(e)}$ on $\bX^{(e,t)}$ is accomplished by first considering the action of $\bN$ on $\bX^{(e,t)}$ represented as $\bN \bX^{(e,t)}\equiv\left(\bN^{1D}\otimes\bN^{1D}\otimes\bN^{1D}\otimes\bI\right) \bXt^{\left(e,t\right)} = \bYt$ where $\bN^{1D}$ is a $n_q\times n_p$ matrix corresponding to the one-dimensional FE basis function values at quadrature points, $\otimes$ represents the Kronecker product while $\bXt^{\left(e,t\right)}$ denotes the 4$^{th}$ order tensor corresponding to the FE cell multivector $\bX^{(e,t)}$ with its components $\bXt^{\left(e,t\right)}_{\beta, j_1, j_2, j_3}$ corresponding to the discretized $\beta^{th}$ field at $J^{th}$ finite-element node and $(j_1, j_2, j_3)$ denote spatial indices of the node $J$. Furthermore, the action of $\widetilde{\bD}^{(s)}$ on $\bYt$ arising in \cref{eqn:refact} involving shape function gradients also leverages the tensor-structured nature of 3D FE basis functions. To this end, we have
$\widetilde{\bD}^{(0)}\bYt = (\bI\otimes\bI\otimes\widetilde\bD^{1D}\otimes\bI) \bYt$; $\widetilde{\bD}^{(1)}\bYt = (\bI\otimes\widetilde\bD^{1D}\otimes\bI\otimes\bI)\bYt$; $\widetilde{\bD}^{(2)}\bYt = (\widetilde\bD^{1D}\otimes\bI\otimes\bI\otimes\bI)\bYt$. These multiplications can be evaluated as a series of tensor contractions \cite{PANIGRAHI2024104925} and since the tensor contractions with $\bI$ do not involve any operations, the floating point computations in $\widetilde{\bD}^{(s)}$ action on $\bYt$ reduce. Using these ingredients, we now describe the expressions of matrix-free action of FE discretized DFT operators (local contribution) in two different settings arising from \cref{eqn:refact}: $\left(\bT^{\left(e\right)} + \bL^{\left(e\right)} + \bnG^{\left(e\right)}\right)$ and $\left(\bT^{\left(e\right)} + \bL^{\left(e\right)} + \bnG^{\left(e\right)} - \imag \bK^{\left(e\right)}\right)$
with the former one corresponding to real arithmetic and the latter to complex arithmetic, arising in DFT calculations with non-periodic and periodic boundary conditions, respectively.

\subsubsection{\textbf{Action of \texorpdfstring{$\bT^{(e)} + \bL^{(e)} + \bnG^{(e)}$}{T(e)+L(e)+G(e)} on \texorpdfstring{$\bX^{(e,t)}$}{X(e,t)}}}\label{sec:realGGAOp}

The matrix-free action of $\bT^{(e)} + \bL^{(e)} + \bnG^{(e)}$ on $\bX^{(e,t)}$ can be recast as the following sequence of steps involving a series of tensor contractions exploiting the tensor-structured nature of the FE basis functions.

\begin{gather} \bYt_0=\left(\bN^{1D}\otimes\bN^{1D}\otimes\bN^{1D}\otimes\bI\right)\bXt^{\left(e,t\right)} \;\; \longrightarrow \;\; \bYt_1 =  \begin{bmatrix}
        \bI\otimes\bI\otimes\widetilde\bD^{1D}\otimes\bI\\
        \bI\otimes\widetilde\bD^{1D}\otimes\bI\otimes\bI\\
        \widetilde\bD^{1D}\otimes\bI\otimes\bI\otimes\bI
\end{bmatrix}\bYt_0 \;\; \longrightarrow \;\;
\bYt_2 = \begin{bmatrix}
        \bV^G_0\\
        \bV^G_1\\
        \bV^G_2
    \end{bmatrix}^T\bYt_1  \nonumber \\
 \bYt_3 = \begin{bmatrix}
        \bV^G_0\\
        \bV^G_1\\
        \bV^G_2
    \end{bmatrix}\bYt_0 +
    \begin{bmatrix}
        \bG_{00}&&\bG_{01} &&\bG_{02}\\
        \bG_{10}&&\bG_{11} &&\bG_{12}\\
        \bG_{20}&&\bG_{21} &&\bG_{22}\\
    \end{bmatrix} \bYt_1 \;\; \longrightarrow \;\;
    \bYt_4 = \begin{bmatrix}
       \bI\otimes\bI\otimes\widetilde\bD^{1D}\otimes\bI\\
        \bI\otimes\widetilde\bD^{1D}\otimes\bI\otimes\bI\\
        \widetilde\bD^{1D}\otimes\bI\otimes\bI\otimes\bI
    \end{bmatrix}^T \bYt_3 + \bYt_2 + \bV^L \bYt_0 \nonumber \\[0.1in]
  \bY^{\left(e,t\right)} =  (\bT^{(e)} + \bL^{(e)} + \bnG^{(e)})\bX^{\left(e,t\right)}  \equiv \left(\bN^{1D}\otimes\bN^{1D}\otimes\bN^{1D}\otimes\bI\right)^T\bYt_4 \label{eqn:ggaReal}
\end{gather}

The action of $\widetilde{\bD}^{(s)}$ on multivectors is encountered both in the action of operators $\bT^{(e)}$ and $\bG^{(e)}$, and is evaluated once in $\bYt_1$ and reused in $\bYt_2$ and $\bYt_3$. Similar is the case with the action of the transpose of $\widetilde{\bD}^{(s)}$ encountered in $\bT^{(e)}$ and $\bG^{(e)}$, where the evaluation is done only once in $\bYt_4$ for both terms in $\bYt_3$. In addition, the action of $\bN$ and $\bN^T$ is evaluated once in $\bYt_0$ and $ \bY^{\left(e,t\right)}$ respectively, thus getting reused for all intermediate steps. Finally, the intermediate multivectors, $\bYt_4, \bYt_3, \bYt_2$ and $\bYt_0$ are used together in evaluating $\bY^{(e,t)}$ as shown in \cref{eqn:ggaReal}.

\subsubsection{\textbf{Action of complex operator \texorpdfstring{$\bT^{(e)} + \bL^{(e)} + \bnG^{(e)} - \imag \bK^{(e)}$}{T(e)+L(e)+G(e) - iK(e)} on \texorpdfstring{$\bX^{(e,t)}$}{X(e,t)}}}\label{sec:complexGGAOp}
The matrix-free action of $\bT^{(e)} + \bL^{(e)} + \bnG^{(e)} - \imag \bK^{(e)}$ on $\bX^{(e,t)}$ can be reformulated as a sequence of tensor contractions that leverages the tensor-structured representation of the FE basis functions in the following way.
\begin{align}
\bYt_0 &= \left(\bN^{1D}\otimes\bN^{1D}\otimes\bN^{1D}\otimes\bI\right)\bXt^{\left(e,t\right)}, \quad
\bYt_1 = \begin{bmatrix}
        \bI\otimes\bI\otimes\widetilde\bD^{1D}\otimes\bI\\
        \bI\otimes\widetilde\bD^{1D}\otimes\bI\otimes\bI\\
        \widetilde\bD^{1D}\otimes\bI\otimes\bI\otimes\bI
    \end{bmatrix}\bYt_0 \nonumber \\
\bYt_2 &= \begin{bsmallmatrix}
        \bV^G_0\\ \bV^G_1\\ \bV^G_2
    \end{bsmallmatrix}^T\bYt_1, \quad
\bYt_3 = \begin{bsmallmatrix}
        \bV^K_0\\ \bV^K_1\\ \bV^K_2
    \end{bsmallmatrix}^T\bYt_1, \quad
\bYt_4 = \begin{bsmallmatrix}
        \bV^G_0\\ \bV^G_1\\ \bV^G_2
    \end{bsmallmatrix}\bYt_0 +
    \begin{bsmallmatrix}
        \bG_{00}&\bG_{01}&\bG_{02}\\
        \bG_{10}&\bG_{11}&\bG_{12}\\
        \bG_{20}&\bG_{21}&\bG_{22}
    \end{bsmallmatrix} \bYt_1 \nonumber \\
\bYt_5 &= \begin{bmatrix}
       \bI\otimes\bI\otimes\widetilde\bD^{1D}\otimes\bI\\
        \bI\otimes\widetilde\bD^{1D}\otimes\bI\otimes\bI\\
        \widetilde\bD^{1D}\otimes\bI\otimes\bI\otimes\bI
    \end{bmatrix}^T \!\bYt_4 + \bYt_2 + \bV^L \bYt_0 - \imag \bYt_3 \nonumber \\
\bY^{\left(e,t\right)} &\equiv (\bT^{(e)} + \bL^{(e)} + \bnG^{(e)} - \imag \bK^{(e)})\bX^{\left(e,t\right)}
    = \left(\bN^{1D}\otimes\bN^{1D}\otimes\bN^{1D}\otimes\bI\right)^T\bYt_5 \label{eqn:ggaCmplx}
\end{align}

The action of $\widetilde{\bD}^{(s)}$ on multivectors is encountered in the action of operators $\bT^{(e)}$, $\bG^{(e)}$ and $\bK^{(e)}$ and is evaluated once in $\bYt_1$ and reused in $\bYt_2$, $\bYt_3$ and $\bYt_4$. Similar is the case with the action of the transpose of $\widetilde{\bD}^{(s)}$ encountered in $\bT^{(e)}$ and $\bG^{(e)}$, where the evaluation is done only once in $\bYt_5$ for both terms in $\bYt_4$. In addition, the action of $\bN$ and $\bN^T$ is evaluated once in $\bYt_0$ and $ \bY^{\left(e,t\right)}$ respectively, thus getting reused for all intermediate steps. Finally, the intermediate multivectors, $\bYt_5, \bYt_4, \bYt_3, \bYt_2$ and $\bYt_0$ are used together in evaluating $\bY^{(e,t)}$ as shown in \cref{eqn:ggaCmplx}. Note, the operators $\bT^{(e)}$, $\bL^{(e)}$, $\bnG^{(e)}$ and $\bK^{(e)}$ are always real, and hence the matrix-free action on the complex multivectors involves real times complex arithmetic operations in all the sequence of steps shown in \cref{eqn:ggaCmplx}. The multivector $\bYt_5$, which has action of $-\imag$ on $\bYt_3$, is swapped, negating the real and imaginary parts of $\bYt_3$, and added to $\bYt_5$, which can be easily adapted in our matrix-free method due to the proposed multilevel batched layout as discussed in the subsequent section. This is in contrast to the cell-matrix approach, where the operator itself is complex, and hence the multiplication will involve complex times complex arithmetic. Thus, all arithmetic operations in matrix-free can be performed in real arithmetic, which helps reduce the overall number of floating-point operations performed.

\section{Numerical implementation of the matrix-free algorithm}\label{sec:sec4}
We now discuss the multilevel batched algorithmic strategies to implement the matrix-free action of finite-element discretized DFT operator on multivectors discussed in \cref{sec:sec3} on distributed multinode CPU architectures. The key steps are (i) the extraction phase, where the FE cell-level multivectors $\bX^{(e,t)}$ are extracted using the subdomain to FE cell-level map, (ii) FE cell-level matrix-multivector product evaluation within the matrix-free framework, carried out through tensor contractions and point-wise multiplications as discussed in \cref{eqn:ggaReal} and \cref{eqn:ggaCmplx} for local part of the FE discretized Hamiltonian in conjunction with the non-local part for the case of pseudopotential calculation, and finally (iii) the assembly of the resulting product FE cell-level matrices to form the output node-level multivector, using the same map employed in the extraction step.  A detailed description of this procedure is provided in the following subsections.

\subsection{\textbf{Proposed multilevel batched layout adapted for real and complex arithmetic}}\label{sec:multiBatch}

In this section, we propose a generalization of the batched algorithm developed in our prior work \citet{PANIGRAHI2024104925} in order to optimize the tensor contractions encountered in the matrix-free action of the FE discretized DFT operator on multivectors.  Levels of batches are introduced where the total number of vectors $n_v$ is divided into $n_l$ levels. The further levels of batching enable better data locality and parallelization for the operations involving real and complex arithmetic. In the proposed layout, a given batch level $j$ has $b^{\left(j\right)}$ vectors (batchsize) with $n_{b}^{\left(j\right)}$ batches and $i$-th batch is indexed with $i_b^{\left(j\right)}$.  We introduce a factor $c$ to extend the same idea for the complex arithmetic case. To this end, we define the following notations:
\begin{align} \label{eq:batchedlayout}
&b^{\left(j\right)}\prod_{k=1}^{j} n_{b}^{\left(k\right)} = b^{\left(0\right)}n_{b}^{\left(0\right)} = cn_v \quad\quad \,\,
    c = \left\{\begin{array}{@{}lr@{}}
        2 & \text{for complex case}\\
        1 & \text{for real case}\\
        \end{array}\right\}
   \nonumber \\
&n_{b}^{\left(j\right)} = \frac{b^{\left(j-1\right)}}{b^{\left(j\right)}}, \quad\quad n_{b}^{\left(0\right)} = 1 \quad\quad\quad \forall \quad j = 1, 2 \cdots n_l \nonumber \\
&\bB^{\left(i_b\right)} = \prod^{n_l}_{j=1} \bB^{\left(i_b^{\left(j\right)}\right)}
\end{align}
\begin{figure*}
    \centering
    \includegraphics[width=0.98\linewidth]{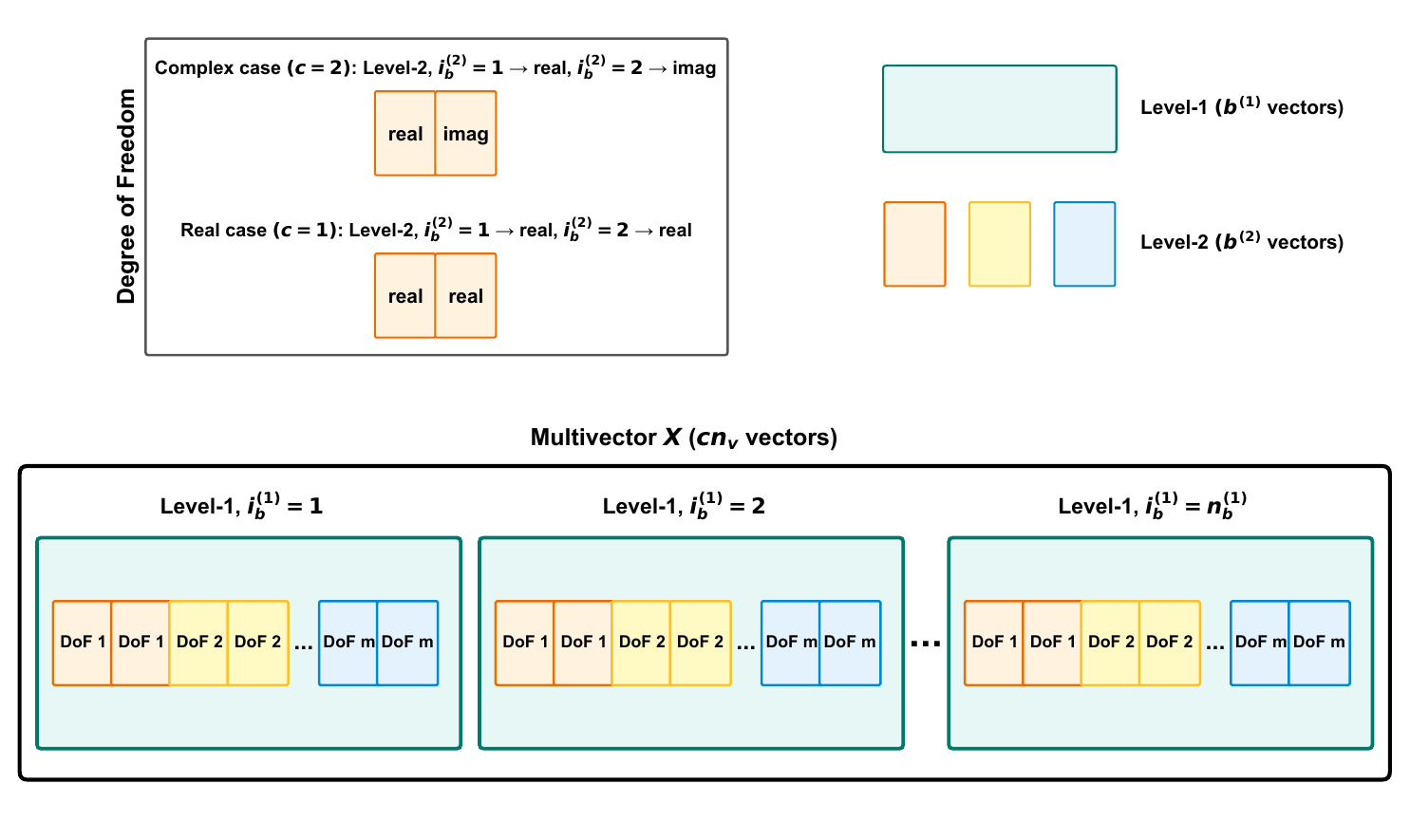}
    \caption{\small Pictorial depiction of the multilevel batched layout described in \cref{sec:multiBatch}}
\end{figure*}
We introduce Boolean sparse matrices $\bB^{\left(i_b^{\left(j\right)}\right)}$ whose action on the multivector $\bX$ results in the extraction of the multivector
batch $\bX^{(i_b)} = \bX\bB^{\left(i_b\right)} = \bX\prod^{n_l}_{j=1} \bB^{\left(i_b^{\left(j\right)}\right)}$, where $\bX^{(i_b)}$ is the multivector batch indexed by $i_b$ which is a multi-index with components ($i_b^{\left(1\right)}, \cdots, i_b^{\left(n_l\right)}$). For the DFT problem at hand, we introduce 2 levels of batches in total to accommodate both real and complex arithmetic. This allows for minimal changes to our implementation strategy when switching from real arithmetic to complex.
Increasing the number of levels involves a trade-off between cache locality and the number of vectors processed per loop iteration. Our numerical experiments indicate that the best performance in terms of time to solution is obtained with 2 levels and, therefore, this setting is used for all benchmarks. In the case when operators are real, we utilize the SIMD feature of CPUs and set $b^{\left(2\right)}$ as the SIMD width. The batchsize $b^{\left(1\right)}$ is set as $2b^{\left(2\right)}$ and found to give the best performance in terms of time to solution, out of various multiples of $b^{\left(2\right)}$. Similar to the real case, when operator is complex, we set $b^{\left(2\right)}$ as the SIMD width and $b^{\left(1\right)}$ as $cb^{\left(2\right)}$. The batch index $i_b^{\left(2\right)} = 1$ represents the real part and batch $i_b^{\left(2\right)} = 2$ represents the imaginary part of the multivector each for $b^{\left(2\right)}$ vectors, thus requiring minimal changes to our implementation strategy when executing the matrix-free action of FE discretized DFT operator on multivector for complex case.

\subsection{\textbf{Multilevel batched algorithm for matrix-free action of DFT-operator}}
The multilevel batched algorithm described here is built on the ideas in our previous work \citet{PANIGRAHI2024104925}, appropriately extending it to evaluate the matrix-free action on multivectors in \cref{eqn:HX} and is illustrated in \cref{eqn:AXmatrixfree}.

\begin{gather}
\bX^{(i_b,e,t)} = \bQ^{(i_b,e,t)}\bC^{(i_b,t)}\bP^{(i_b,t)}\bX^{\left(i_b\right)} \;\;\longrightarrow \;\;
\bY_1^{(t)} = \sum_{i_b}^{n_b} \left[\sum_e^{E_t} {\bQ^{(i_b,e,t)}}^T \bH_\text{loc}^{\left(e\right)}\bX^{(i_b,e,t)}\right]{\bB^{\left(i_b\right)}}^T \label{eqn:AXmatrixfree}\\
\bY_2^{(a)} = \sum_{t^a}\sum_{i_b}^{n_b}\left[\sum_{e^a} \bF^{(i_b,a, e^a, t^a) \dagger}\bX^{(i_b,e^a,t^a)} \right] {\bB^{\left(i_b\right)}}^T \nonumber \\ \longrightarrow\;\;
\bY = \sum_{t}^{n_t}{\bP^{(t)}}^T{\bC^{(t)}}^T \!\left( \bY_1^{(t)} +  \sum_e^{E_t} {\bQ^{(e,t)}}^T \sum_a \bF^{(a,e,t)} \Delta^a \bY_2^{(a)} \right) \label{eqn:AXmatrixfreenonloc}
\end{gather}

First, the FE cell-level multivector batch $\bX^{(i_b,e,t)}$ is extracted from $\bX$ by the action of $\bB^{\left(i_b\right)}$ on $\bX$ ($\bX^{(i_b)} = \bX\bB^{\left(i_b\right)}$), followed by the action of partitioner matrix $\bP^{(i_b,t)}$, constraint matrix $\bC^{(i_b,t)}$ (refer to \cite{PANIGRAHI2024104925} for the efficient action of this matrix) and finally the subdomain-level to FE cell-level map $\bQ^{(i_b,e,t)}$  as shown in \cref{eqn:AXmatrixfree}.
Then the matrix-free action of $\bH_\text{loc}^{\left(e\right)}$ on $\bX^{(i_b,e,t)}$ is performed at the cell-level followed by mapping the FE cell-level product to subdomain-level product vector -via- ${\bQ^{(i_b,e,t)}}^T$ and summing over the contributions from all the FE cells contained in the subdomain $\Omega^{(t)}$. The multivector $\bY_1^{(t)}$ is then evaluated by the action of ${\bB^{\left(i_b\right)}}^T$ followed by summing over batches as illustrated in \cref{eqn:AXmatrixfree}. Furthermore, in the case of pseudopotential DFT calculations, the multivector $\bY_2^{(a)}$ is evaluated reusing the already extracted $\bX^{(i_b,e,t)}$ for a given batch $i_b$ and cell $e$ in a processor $t$ as seen in \cref{eqn:AXmatrixfreenonloc}. This evaluation involves action of operators $\bF^{(i_b,a, e^a, t^a) \dagger}$ and ${\bB^{\left(i_b\right)}}^T$, and summed over cells and processors with local support for $\bF^{(i_b,a, e^a, t^a) \dagger}$ and over batches, as shown in \cref{eqn:AXmatrixfreenonloc}. Finally, the global product multivector $\bY$ is evaluated using ${\bP^{(t)}}^T$, ${\bC^{(t)}}^T$ acting on sum of $\bY_1^{(t)}$ and ${\bQ^{(e,t)}}^T$ acting on output of $\bF^{(a,e,t)} \Delta^a$ on $\bY_2^{(a)}$ as shown in \cref{eqn:AXmatrixfreenonloc}. \cref{alg:batchedAX} describes the implementation of \crefrange{eqn:AXmatrixfree}{eqn:AXmatrixfreenonloc}.

\begin{algorithm}[ht]
    \caption{Multilevel batched level evaluation of ${\bY}$}\label{alg:batchedAX}
    \KwIn{${\bX}$}
    \KwData{$\bB^{\left(i_b\right)},\bP^{\left(i_b,t\right)},\bC^{\left(i_b,t\right)},\bQ^{\left(i_b,e,t\right)}$}
    \KwTemp{$ \bT^{\left(0\right)},\bT^{\left(1\right)},\bT^{\left(2\right)},\bT^{\left(3\right)},\bT^{\left(4\right)}$}
    \KwResult{${\bY}$}

    \For{$t \gets1$ \KwTo $n_t$}{
        \For{$i_b^{\left(1\right)}\gets1$ \KwTo $n_{b}^{\left(1\right)}$}{
            $\bX^{(i_b^{\left(1\right)},e,t)} \gets \bC^{(i_b^{\left(1\right)},t)}\bP^{(i_b^{\left(1\right)},t)}\bX\bB^{\left(i_b^{\left(1\right)}\right)}$

            \For{$e\gets1$ \KwTo $E_t$}{
                $\bT^{\left(0\right)} \gets \bQ^{\left(i_b^{\left(1\right)},e,t\right)} \bX^{(i_b^{\left(1\right)},e,t)}$

                \For{$a\gets1$ \KwTo $\alpha_t$}{
                    \If{$e \in \mathcal{E}(a, t)$} {
                        $\bT^{\left(1\right)} \gets \bT^{\left(1\right)} + \bF^{(i_b,a, e, t) \dagger} \bT^{\left(0\right)}$
                    }
                }

                $\bT^{\left(0\right)} \gets \bH_\text{loc}^{\left(e\right)} \bT^{\left(0\right)}$

                $\bY^{(t)} \gets \bY^{(t)} + {\bQ^{(i_b^{\left(1\right)},e,t)}}^T \bT^{\left(0\right)}{\bB^{\left(i_b^{\left(1\right)}\right)}}^T$
            }
            $\bT^{\left(2\right)} \gets \bT^{\left(2\right)} + \bT^{\left(1\right)}{\bB^{\left(i_b^{\left(1\right)}\right)}}^T$
        }

        \For{$a\gets1$ \KwTo $\alpha_t$} {
            \If{$t \in \mathcal{T}(a)$} {
                $\bT^{\left(3\right)} \gets \bT^{\left(3\right)} + \bT^{\left(2\right)}$
            }
        }

        \For{$e\gets1$ \KwTo $E_t$} {
            \For{$a\gets1$ \KwTo $\alpha_t$} {
                \If{$e \in \mathcal{E}(a, t)$} {
                    $\bT^{\left(4\right)} \gets \bT^{\left(4\right)} + \bF^{(a,e,t)} \Delta^a\bT^{\left(3\right)} $
                }
            }

            $\bY^{(t)} \gets \bY^{(t)} + {\bQ^{(e,t)}}^T \bT^{\left(4\right)}$
        }

        $\bY \gets \bY + {\bP^{(t)}}^T{\bC^{(t)}}^T \bY^{(t)} $
    }

    \Return{${\bY}$}
\end{algorithm}
\begin{algorithm}[ht]
    \caption{Evaluation of $\bY^{(i_b^{\left(1\right)},e,t)} = \bH_\text{loc}^{\left(e\right)} \bX^{(i_b^{\left(1\right)},e,t)}$}\label{alg:GGA}
    \KwIn{$\bX^{(i_b^{\left(1\right)},e,t)}$}
    \KwData{$\bN^{1D},\widetilde\bD^{1D},\bG^{\left(s,d\right)}, \bV^L, \bV^G_s, \bV^K_s$ where $s,d=0,1,2$}
    \KwTemp{$\bTt, \bTt_0,\bTt_1,\bTt_2,\bTt_3$}
    \KwResult{$\bY^{(i_b^{\left(1\right)},e,t)}$}

     $\bTt \gets \left(\bI\otimes\bI\otimes\bN^{1D}\otimes\bI\right)\bX^{(i_b^{\left(1\right)},e,t)}$\

    $\bTt \gets \left(\bI\otimes\bN^{1D}\otimes\bI\otimes\bI\right)\bTt$\

    $\bTt \gets \left(\bN^{1D}\otimes\bI\otimes\bI\otimes\bI\right)\bTt$\

    $\bTt_0 \gets \left(\bI\otimes\bI\otimes\widetilde\bD^{1D}\otimes\bI\right)\bTt$

    $\bTt_1 \gets \left(\bI\otimes\widetilde\bD^{1D}\otimes\bI\otimes\bI\right)\bTt$

    $\bTt_2 \gets \left(\widetilde\bD^{1D}\otimes\bI\otimes\bI\otimes\bI\right)\bTt$

    $\bTt_3 \gets \displaystyle\sum_{d=0}^{2} \bV^G_d\bTt_d -\imag \displaystyle\sum_{d=0}^{2}\bV^K_d\bTt_d + \bV^{L}\bTt \quad\quad\quad \imag = \sqrt{-1}$

    $\bTt_s \gets \displaystyle\sum_{d=0}^{2} \bG^{\left(s,d\right)}\bTt_d + \bV^G_s\bTt$

    $\bTt_3 \gets \bTt_3 + \left(\bI\otimes\bI\otimes\widetilde\bD^{1D}\otimes\bI\right)^T\bTt_0$

    $\bTt_3 \gets \bTt_3 + \left(\bI\otimes\widetilde\bD^{1D}\otimes\bI\otimes\bI\right)^T\bTt_1$

    $\bTt_3 \gets \bTt_3 + \left(\widetilde\bD^{1D}\otimes\bI\otimes\bI\otimes\bI\right)^T\bTt_2$

    $\bTt_3 \gets \left(\bI\otimes\bI\otimes\bN^{1D}\otimes\bI\right)^T\bTt_3$\

    $\bTt_3 \gets \left(\bI\otimes\bN^{1D}\otimes\bI\otimes\bI\right)^T\bTt_3$\

    $\bY^{(i_b^{\left(1\right)},e,t)} \gets \left(\bN^{1D}\otimes\bI\otimes\bI\otimes\bI\right)^T\bTt_3$

    \Return{$\mathbf{Y}^{(i_b^{\left(1\right)},e,t)}$}
\end{algorithm}
As can be seen from \cref{alg:batchedAX}, the data structure $\bT^{\left(0\right)}$ is reused for action of $\bH_\text{loc}^{\left(e\right)}$ and also for action of $\bF^{(i_b,a, e, t) \dagger}$. The proposed multilevel batched layout with 2 levels of batches is used in lines 2 to 11, particularly the loop from lines 2 to 11 corresponds to the level-1 batch and the level-2 batch is evaluated in lines 8 and 9 when evaluating the action of operators $\bF^{(i_b,a, e, t) \dagger}$ and $\bH_\text{loc}^{\left(e\right)}$ on $\bT^{\left(0\right)}$. But for the rest of the algorithm from lines 12 to 19, we use the same proposed multilevel batched layout but with level-0 batch with batchsize $b^{\left(0\right)} = cn_v$ and $n_{b}^{\left(0\right)} = 1$. This choice for a different level and batchsize was done because the increased data movement for pseudopotential operators $\bF^{(a,e,t)}$ negated the speedup gained with parallelism of the  level-2 batch. But since the data $\bT^{\left(0\right)}$ is already extracted, the first step of nonlocal operation of $\bH_\text{nloc}$,  the action of $\bF^{(i_b,a, e, t) \dagger}$ is performed in level-2 batched layout. For performing the rest of operations from line 12 onwards appropriate reshaping of the underlying data was done. The novelty in the proposed algorithm is that it is applicable to both real and complex arithmetic and for both the forms of the FE discretized DFT operator $\bH_\text{loc}^{\left(e\right)}$ discussed in \cref{sec:realGGAOp} and \cref{sec:complexGGAOp}. In \cref{alg:GGA}, we illustrate the implementation of the action of a complex operator on multivectors, demonstrating how reusing data and combining terms can be achieved and further showcase the implementation of the 2nd level of the 2-level batched layout in the proposed method. The reuse of data and combining terms reduces data movement and the cache memory requirement. Another advantage of the proposed multilevel batched layout is that it makes the temporary variables used in \cref{alg:GGA} reside in cache, giving further speedups. We employ the even-odd decomposition \cite{PANIGRAHI2024104925} to reduce FLOP in various computations in \cref{alg:GGA} by exploiting the symmetry of the basis shape function $\bN^{1D}$ and shape function gradient $\widetilde\bD^{1D}$. We assume a single quadrature rule is used for all terms in \cref{sec:complexGGAOp} (and for \cref{sec:realGGAOp}), i.e. for $\bT^{(e)}$, $\bL^{(e)}$, $\bnG^{(e)}$ and $\bK^{(e)}$, usually the highest-order rule required by any of these operators for the material system at hand. Using one (sufficiently accurate) quadrature rule simplifies the matrix‑free implementation because it removes the need to store multiple sets of shape functions and shape function gradients for different rules, which would otherwise increase data movement and bookkeeping. When a high quadrature order is required (e.g. DFT with $\geq 7$), the gains are in the form of reduced memory footprint, lower FLOP and improved data locality due to a unified evaluation. The principal disadvantage of this one quadrature choice is that some terms will be evaluated with a higher‑order quadrature rule than strictly necessary, which can increase floating‑point operations for those terms. However, as later discussed in \cref{sec:rchfsi}, the effective computational cost can be further reduced by exploiting iterative eigensolvers, such as the residual-Chebyshev filtered subspace iteration procedure \cite{KODALI2026110239}, which is tolerant to approximations in matrix-vector multiplications.  This allows one to lower the quadrature order ($n_q = n_p$) in evaluating matrix-free action of DFT operator on multivectors without sacrificing the desired accuracy in energy and forces, thereby turning the quadrature dependence of matrix‑free FLOP into an additional avenue for performance gains. This is not possible in the baseline cell-matrix method since the integrals in evaluating $\bH_\text{loc}^{\left(e\right)}$ already use specified quadrature rules. Hence, the FLOP due to its action on $\bX^{(i_b,e,t)}$ is independent of quadrature rule used for any term.

The proposed matrix-free implementation does not form the extracted multivector $\mathbf{X}^{(i_b)}$ explicitly in memory. Instead, the action of $\mathbf{P}^{(i_b,t)}$ is applied directly by exchanging the necessary boundary portions of the multivectors between MPI ranks and then operating on the received data. The nested summations over elements $e$ and batch indices $i_b^{\left(1\right)}$ in \cref{alg:batchedAX} are performed as serial loops on each MPI task. For a given MPI task $t$, let $m_{\text{loc}}^{(t)}$ denote the number of locally owned degrees of freedom (DoFs) and $m_{ghost}^{(t)}$ the number of non-owned DoFs on shared subdomain boundaries (ghost DoFs). For the case with 2 levels of batches, each process stores its multivector data in two contiguous blocks: first a locally owned block of size $(b^{(2)} n_b^{(2)} \times m_{\text{loc}}^{(t)} \times n_b^{(1)})$, and immediately following it a ghost block of size $(b^{(2)} n_b^{(2)} \times m_{ghost}^{(t)} \times n_b^{(1)})$. Similarly, for the case with level-0 batch, each process stores its multivector data as: first a locally owned block of size $(b^{(0)} \times m_{\text{loc}}^{(t)} \times n_b^{(0)})$, and immediately following it a ghost block of size $(b^{(0)} \times m_{ghost}^{(t)} \times n_b^{(0)})$. The dimensions are listed in storage order so that the leftmost index is the fastest varying in memory. Nonblocking point-to-point MPI calls (\texttt{MPI\_ISend} and \texttt{MPI\_IRecv}) are used to exchange the boundary data, and the transfers are completed with \texttt{MPI\_Waitall}, following the common pattern used in libraries such as \texttt{deal.II} \cite{Arndt2019TheLibrary}, PETSc \cite{Balay2023PETSc/TAO3.20} and Trilinos \cite{trilinos}.

\section{Performance Benchmarks}\label{sec:sec5}
We now examine the performance characteristics of the proposed matrix-free algorithms to assess their numerical efficiency and scalability. The benchmark studies are carried out on representative material systems selected to capture three distinct categories of finite-element discretized DFT operators $\bH = \bH_\text{loc} + \bH_\text{nloc}$: (i) \textit{Real operators}, which arise in pseudopotential DFT simulations of non-periodic systems, where the FE operator is given by ($\bH^{(e)}_\text{loc} = \bT^{(e)} + \bL^{(e)} + \bnG^{(e)}$), (ii) \textit{Complex operators}, occurring in pseudopotential DFT simulations of periodic systems, represented as, ($\bH^{(e)}_\text{loc} = \bT^{(e)} + \bL^{(e)} + \bnG^{(e)} - \imag \bK^{(e)}$) (iii) \textit{Real and complex operators}, arising in all-electron DFT simulations of both periodic and non-periodic systems respectively where $\bH_\text{nloc} = 0$.
To investigate these operator types, we consider four representative material systems: (a) Pseudopotential calculations -- (i) aluminum nanoparticle (non-periodic boundary conditions) (ii) a body-centered cubic (BCC) molybdenum supercell with a vacancy (periodic boundary conditions), (b) All-electron DFT calculations -- (iii) benzamide (non-periodic) (iv) BCC lithium supercell with a vacancy (periodic). We first focus on analysing the performance of the matrix-free implementation for the action of the FE discretized DFT operator $\bH = \bH_\text{loc} + \bH_\text{nloc}$ on trial multivectors. The number of trial vectors is chosen based on the number of electrons in each system and varies between 64 and 8448. The finite-element interpolation orders (\texttt{FEOrder}) used in these benchmarks are 7, 8, and 9, the values typically employed in FE-based DFT simulations to achieve the desired accuracy in ground-state energies and atomic forces \cite{Motamarri2013Higher-orderTheory, Motamarri2020DFT-FECalculations, Das2019FastSystem}. In all benchmarks, mesh sizes are selected to ensure ground-state energy errors below  O($10^{-4}$) Ha/atom and force errors below  O($10^{-4}$) Ha/Bohr for each \texttt{FEOrder}. These choices of \texttt{FEOrder} and mesh size provide a balanced trade-off between accuracy and computational efficiency, reducing the DoFs required to achieve a given accuracy while controlling the per-DoF computational cost. For each system, we report speedups in the computational time of the proposed matrix-free method relative to the state-of-the-art cell-matrix approach \cite{Motamarri2013Higher-orderTheory, Motamarri2020DFT-FECalculations, Das2019FastSystem} for applying the FE discretized DFT operator to trial multivectors. We then identify the \texttt{FEOrder} that yields the smallest total solution time with low per-DoF computational cost (in core-hours) and employ this \texttt{FEOrder} for further analysis. We further present an analytical performance model for our implementation. A roofline analysis of the various FE discretized DFT operators then quantifies the achievable performance bounds and compares the expected against the actual speedups. Finally, we demonstrate the utility of the proposed matrix-free method in the context of an iterative eigensolver, applying it within a Chebyshev-filtered subspace iteration (ChFSI) algorithm, a technique widely used in real-space DFT calculations~\cite{Zhou2006ParallelAcceleration,Motamarri2013Higher-orderTheory}. To this end, we analyze the performance improvements achieved during the subspace construction phase of ChFSI over the chosen baseline, performing strong-scaling studies on multi-node CPU architectures.

We use \texttt{deal.II} library version 9.5.2 \cite{Arndt2019TheLibrary, 2023:arndt.bangerth.ea:deal} with \texttt{p4est} \cite{BursteddeWilcoxGhattas11} backend to perform MPI-parallel finite-element meshing and domain decomposition. The benchmarks are carried out on three supercomputers: Frontier, Param Pravega, and Fugaku. For Fugaku, we utilize 4 MPI tasks per node to accommodate the problem within the available memory. But we utilise 56 and 48 MPI tasks per node on Frontier and Pravega, respectively for all the material systems considered here.  On Frontier, AMD CPUs with AVX2 vectorization are used; on Pravega, Intel CPUs with AVX-512 vectorization; and on Fugaku, Fujitsu CPUs with SVE vectorization. These systems represent some of the most widely used CPU architectures for large-scale scientific computing. {\cblue The scope of the present study is restricted to CPU architectures. The performance of batched matrix-free strategies relative to the corresponding GPU cell-matrix baselines of \texttt{DFT-FE} has been examined in our earlier work \cite{PANIGRAHI2024104925} in the context of FE discretized Helmholtz operators, and the extension of the full DFT operator implementation developed here to GPU and heterogeneous architectures is deferred to future work.\cn} Detailed hardware configurations, compiler specifications, and MPI libraries are summarized in \cref{tbl:sysconf} and \cref{tbl:libconf}.

\begin{table}[ht]
\rowcolors{2}{gray!25}{white}
\begin{tabular}{M{0.19\linewidth} M{0.23\linewidth} M{0.23\linewidth} M{0.23\linewidth}}
\rowcolor{gray!50}
\textbf{System Config} & \small\textbf{Frontier} & \small\textbf{Param Pravega} & \small\textbf{Fugaku} \\
Processor         & \small AMD EPYC 7A53 & \small Intel Xeon Platinum 8268 & \small Fujitsu A64FX  \\
Nodes             & \small 9856 & \small 428 + 156 (High Memory)  & \small 158976 \\
\small CPU cores/Node    & 64 (56 + 8 reserved) & 48 & 48  \\
\small Node Performance  & \small 2.51 TFLOP/s (AVX2 FP64) & \small 4.45 TFLOP/s (AVX-512 FP64) & \small 3.38 TFLOP/s (SVE FP64)\\
\small Memory/Node       & \small 512 GB DDR4 & \small 192 GB or 768 GB (High Memory) DDR4 & \small 32 GB HBM2\\
Interconnect      &\footnotesize HPE Slingshot & \footnotesize Mellanox ConnectX-6 MT28908 &\footnotesize Tofu Interconnect D\\
OS & \small SLES 15.6 & \small CentOS 7 & \small RHEL 8
\end{tabular}
\caption{System configurations for the benchmark architectures.}\label{tbl:sysconf}
\end{table}

The compilers, MPI and BLAS libraries used are listed in \cref{tbl:libconf}.
\begin{table}[ht]
    \rowcolors{2}{gray!25}{white}
    \begin{tabular}{M{0.19\linewidth} M{0.23\linewidth} M{0.23\linewidth} M{0.23\linewidth}}
    \rowcolor{gray!50}
    \textbf{Library} & \small\textbf{Frontier} & \small\textbf{Param Pravega} & \small\textbf{Fugaku} \\
    Compiler         &\;\;\;\;\; gcc 12.3.0 & gcc 12.2.0 & Fujitsu compiler  \\
    Compiler Flags & \texttt{-O3 -march=znver3} &\texttt{-O3 -march=native} &\texttt{-Nclang -std=c++17 -O3 -msve-vector-bits=512 --linkfortran} \\
    MPI    & Cray MPICH 8.1.31 & Intel oneAPI MPI 2021.9.0  & Fujitsu MPI\\
    BLAS  & \small OpenBLAS 0.3.28 & \small Intel oneAPI MKL 2023.1.0 & \small Fujitsu BLAS\\
    \end{tabular}
    \caption{External libraries and compiler flags used for compilation.}\label{tbl:libconf}
\end{table}

\subsection{\textbf{DFT Operator Action}}\label{sec:DFToperator}

To evaluate the performance of our matrix-free implementation, we benchmark in this subsection four representative types of the finite-element discretized Kohn--Sham DFT operator involving pseudopotential (real and complex) and all-electron (real and complex) cases on realistic material systems commonly encountered in \emph{ab initio} modeling of materials. In all cases, we compare the proposed matrix-free algorithm against the baseline cell-matrix approach.

\subsubsection{\textbf{Pseudopotential DFT operator for non-periodic systems}}\label{sec:NonPeriodic}
\begin{figure}[ht]
    \centering
    \begin{subfigure}[t]{.49\textwidth}
        \centering
        \includegraphics[width=\textwidth]{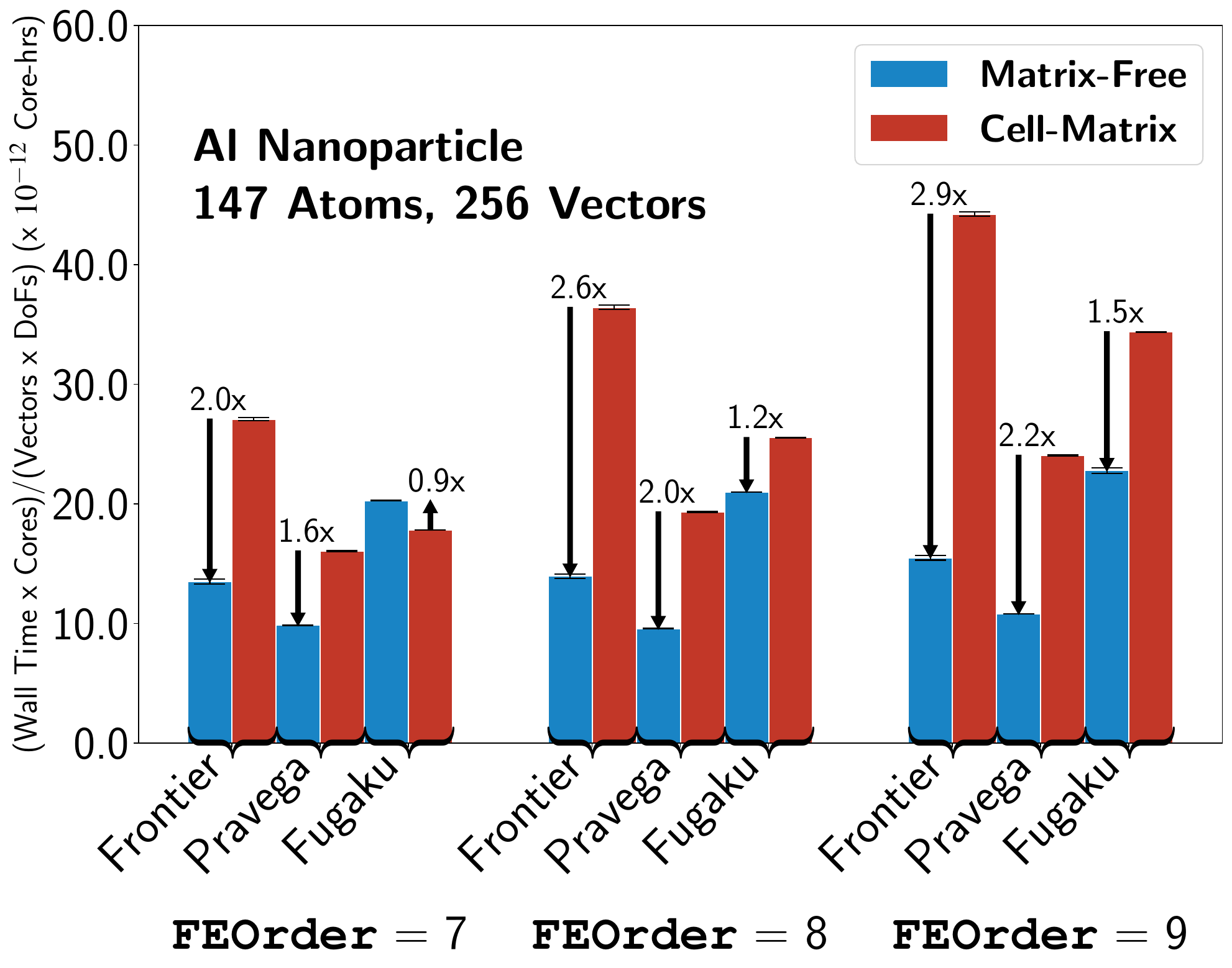}
    \end{subfigure}\hfill
    \begin{subfigure}[t]{.49\textwidth}
        \centering
        \includegraphics[width=\textwidth]{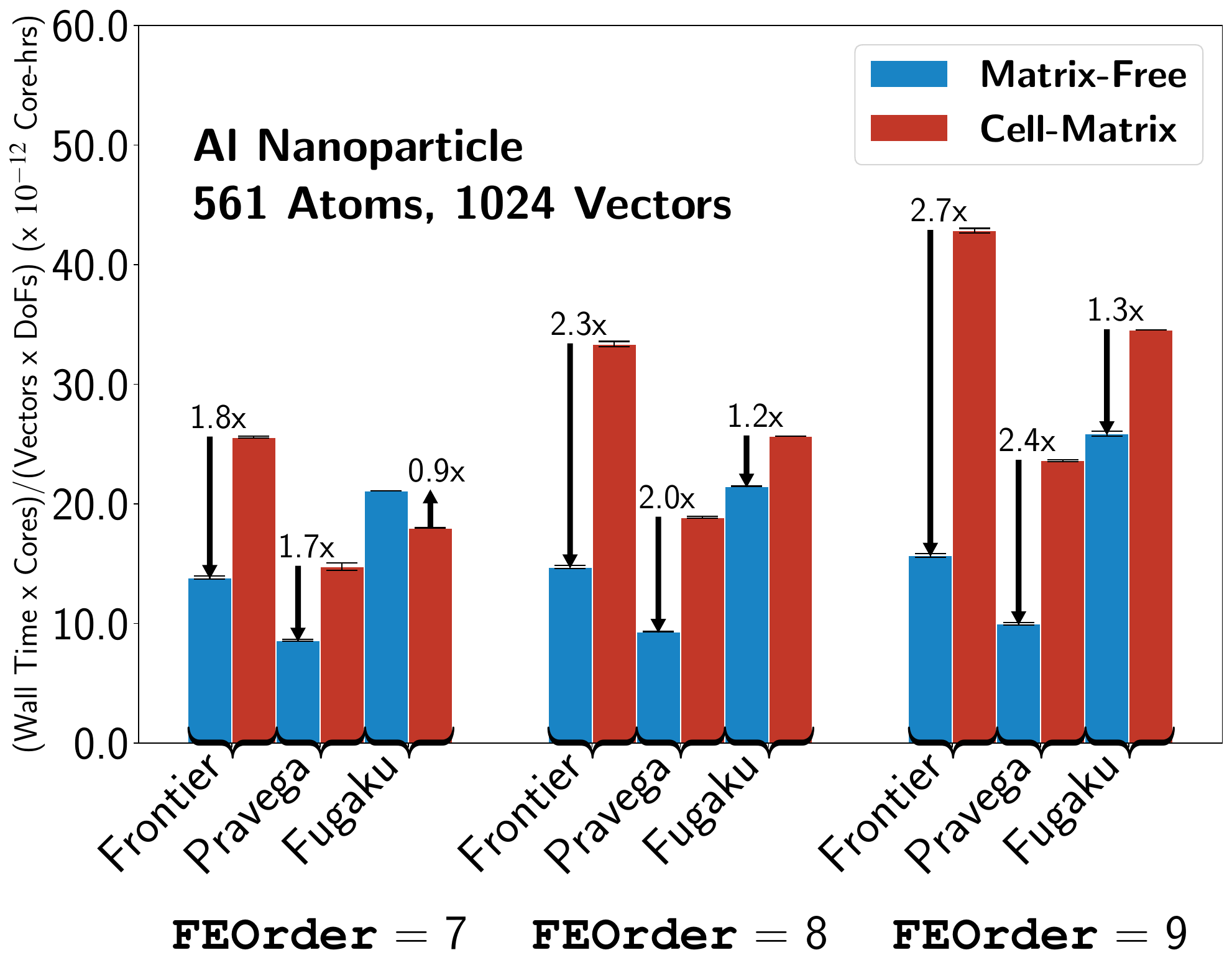}
    \end{subfigure}

    \caption{Speedups of our matrix-free implementation strategies for Al nanoparticle (147 and 561 atoms with ${\sim}10k$ DoFs/atom and ${\sim}7k$ DoFs/atom respectively) non-periodic system with pseudopotential DFT operator on
    Frontier (Left: 2 nodes, Right: 4 nodes), Param Pravega (Left: 2 nodes, Right: 4 nodes) and Fugaku (Left: 20 nodes, Right: 70 nodes) supercomputers.} \label{fig:speedup_alnp}
\end{figure}

For non-periodic material systems, the pseudopotential DFT operator is given by $\bH = \bH_\text{loc} + \bH_\text{nloc}$ and the key computational kernel of evaluating  $\bY=\bH\bX$ involves the following matrix-free action of the local part of the DFT operator at the finite-element cell-level, in addition to the non-local action of the DFT operator
\begin{align}
\bH_\text{loc}^{\left(e\right)} \bX^{\left(e,t\right)} = \left(\bT^{\left(e\right)} + \bL^{\left(e\right)} + \bnG^{\left(e\right)}\right)\bX^{\left(e,t\right)}
\end{align}
The benchmarks are conducted on aluminum nanoparticles containing 147 and 561 atoms, using finite-element interpolation orders of $\texttt{FEOrder}=7,8$ and $9$. For these systems, we employ 256 and 1024 trial vectors, respectively, chosen to represent the number of valence electrons associated with each nanoparticle (half that number for spin-unpolarized calculations). The performance of the matrix-free implementation is compared across Frontier, Param Pravega, and Fugaku supercomputers by running on the minimum number of compute nodes needed to fit the given problem in memory for each system. The smallest Al nanoparticle (147 atoms) utilises 2 nodes on Frontier and Pravega, and 20 nodes on Fugaku, while the larger system (561 atoms) requires 4 nodes on Frontier and Pravega, and 70 nodes on Fugaku. The higher node count on Fugaku is dictated by its lower per-node memory capacity. {\cblue We note that the comparisons reported in this subsection (\cref{fig:speedup_alnp,fig:speedup_bccmo,fig:speedup_benzLi}) are not scaling studies and each material system is benchmarked at a fixed node count on a given machine, chosen as the minimum number of nodes that accommodates the problem in memory, and the timings are normalized by the DoFs and number of vectors only to allow comparison across machines and discretizations. The strong-scaling behaviour of the proposed implementation, where the node count is increased for a fixed problem, is reported separately in \cref{fig:scalingAlBCCMo}.\cn}

\cref{fig:speedup_alnp} presents the scaled performance metric, defined as the time normalized by the number of vectors and the number of finite-element mesh DoFs. For this analysis, the y-axis reports the \textit{scaled time = (wall time $\times$ CPU cores) / (vectors $\times$ DoFs)}. Across all systems, we observe substantial speedups of the matrix-free implementation relative to the cell-matrix baseline. For the 147-atom nanoparticle, the matrix-free approach achieves 2.0$\times$--2.9$\times$ speedups on Frontier, 1.6$\times$--2.2$\times$ on Pravega, and {\cblue 0.9$\times$--1.5$\times$ on Fugaku, the value below unity corresponding to the mild slowdown observed at $\texttt{FEOrder}=7$ discussed below. Comparable improvements are obtained for the 561-atom system (1.8$\times$--2.7$\times$ on Frontier, 1.7$\times$--2.4$\times$ on Pravega and 0.9$\times$--1.3$\times$ on Fugaku, with the $\texttt{FEOrder}=7$ slowdown on Fugaku recurring here as well).\cn} Although the computational cost of the nonlocal operator increases with system size, the matrix-free formulation continues to provide clear benefits over the baseline for both aluminum systems. A mild slowdown is observed on Fugaku for $\texttt{FEOrder}=7$, primarily because of the action of $\bH_{\text{nloc}}$ computed using \cref{alg:batchedAX}. This arises from the higher cost of data reshaping required to match the data layout for various operations on Fugaku, particularly pronounced at lower polynomial orders. This effect is absent for all-electron systems, as is evident from \cref{fig:speedup_benzLi} (where $\bH_{\text{nloc}}=\bzero$), confirming that the overhead is associated with the nonlocal action rather than the action of the local part of the Hamiltonian. Notably, this slowdown does not persist for higher $\texttt{FEOrder}$ values because the {\cblue reduced floating-point cost of the matrix-free kernels relative to the cell-matrix approach grows with the polynomial order and compensates\cn} for the additional overhead. The performance gains on Fugaku are lower overall compared to Frontier and Pravega, largely due to Fugaku’s lower memory-per-core ratio, which restricts in-cache data reuse. Additionally, we note that our matrix-free implementation is fundamentally memory-bound (see the roofline analysis in \cref{fig:rooflineReal}), and we employ four MPI tasks per node on Fugaku. As a result, the effective memory bandwidth available per task reduces in this configuration, with the node's memory-access pathways underutilised, exacerbating the bandwidth-bound behaviour and thereby impacting overall performance on Fugaku.

Finally, we note that the scaled time is consistently minimal for $\texttt{FEOrder}=8$, indicating that this degree of finite-element interpolating polynomial provides the optimal trade-off between accuracy and computational cost. In particular, we also note that the total solution time and the number of DoFs required to achieve the desired accuracy ($10^{-4}$ Ha/atom in energies and $10^{-4}$ Ha/Bohr in forces) are also the lowest for this order.

\subsubsection{\textbf{Pseudopotential DFT complex-valued operator for periodic systems}}\label{sec:Periodic}

For periodic material systems, the pseudopotential DFT operator $\bH = \bH_\text{loc} + \bH_\text{nloc}$ includes a complex-valued contribution. The key computational kernel of evaluating $\bY = \bH \bX$ then involves the following action of the local part of the DFT operator at the finite-element cell-level,

\begin{align}
\bH_\text{loc}^{\left(e\right)} \bX^{\left(e,t\right)} = \left(\bT^{\left(e\right)} + \bL^{\left(e\right)} + \bnG^{\left(e\right)} - \imag \bK^{\left(e\right)} \right)\bX^{\left(e,t\right)}
\end{align}

We benchmark body-centered cubic (BCC) molybdenum $2\times 2\times 2$ supercell with a monovacancy containing 127 and 1023 atoms using a $2\times2\times2$ Monkhorst–Pack k-point rule \cite{PhysRevB.13.5188} to sample the Brillouin zone. The same range of polynomial orders ($\texttt{FEOrder}=7$, 8 and 9) as in the previous study has been employed for benchmarking these periodic systems. We employ 1280 and 8448 trial vectors for 127-atom and 1023-atom material systems respectively, chosen to be representative of the number of valence electrons (half the number of electrons for spin unpolarized calculations) associated with each BCC molybdenum system. The 127-atom case is benchmarked on 2 nodes on both Frontier and Pravega, and 20 nodes on Fugaku. In contrast, the 1023-atom system is benchmarked on 12, 10, and 200 nodes on the respective supercomputing systems, chosen as the smallest node counts capable of fitting the given material systems in memory.

\begin{figure}[ht]
    \centering
    \begin{subfigure}[t]{.49\textwidth}
        \centering
        \includegraphics[width=\textwidth]{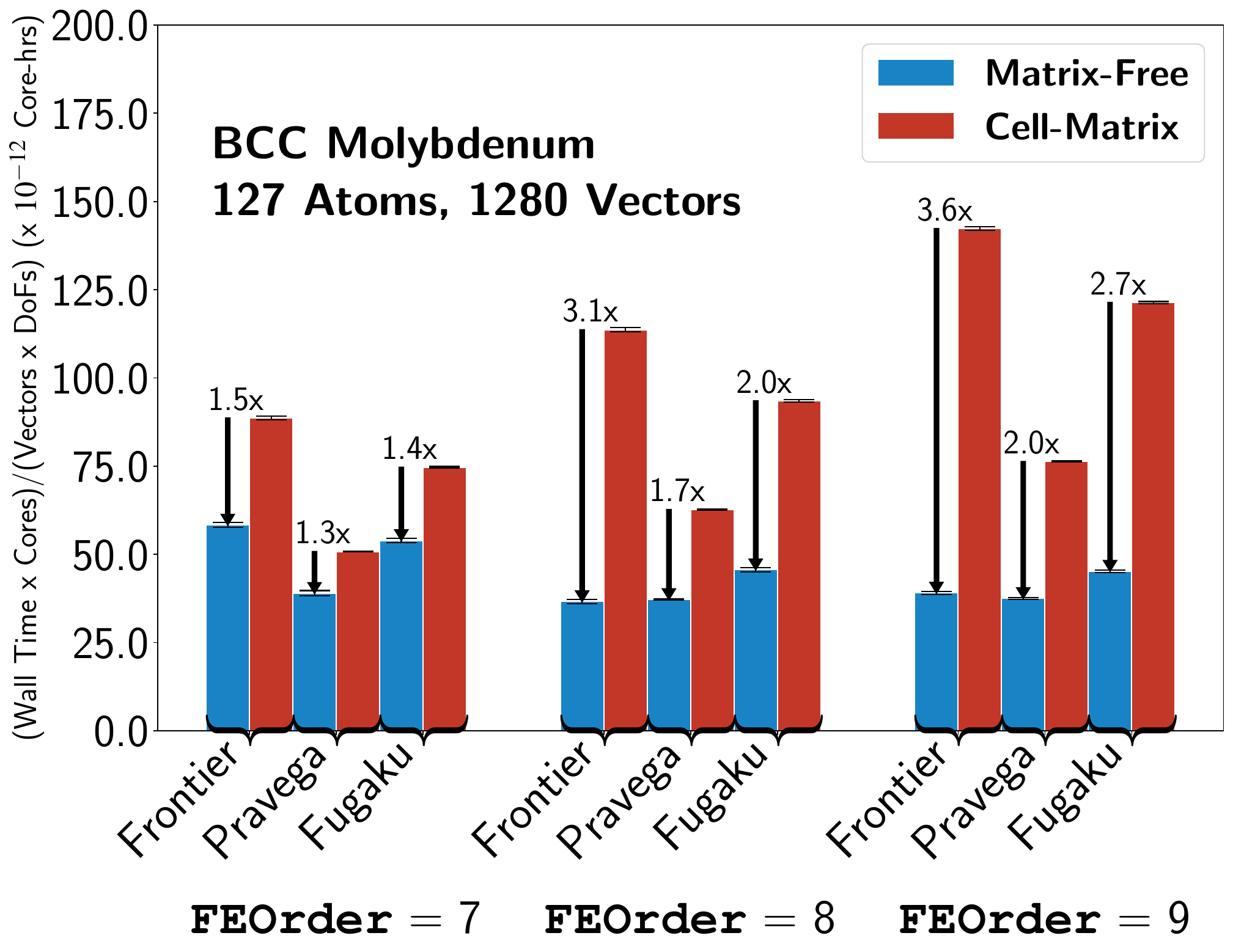}
        \label{fig:speedup_bccmo127}
    \end{subfigure}\hfill
    \begin{subfigure}[t]{.49\textwidth}
        \centering
        \includegraphics[width=\textwidth]{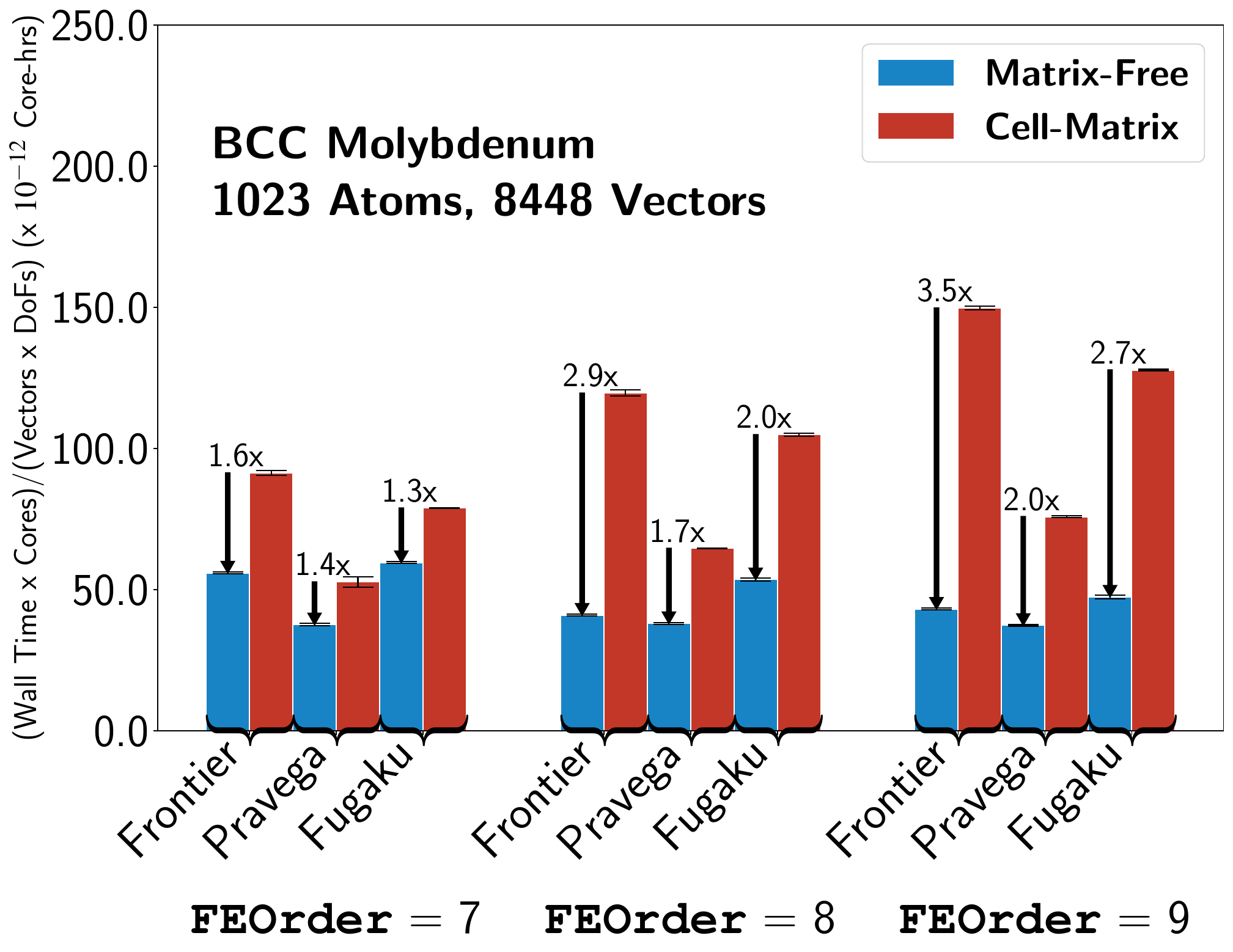}
       \label{fig:speedup_bccmo1023}
    \end{subfigure}
    \caption{Speedups of our matrix-free implementation strategies for BCC molybdenum (127 and 1023 atoms with ${\sim}4k$ DoFs/atom) periodic system with pseudopotential DFT complex-valued operator (2x2x2 k-point rule) on Frontier (Left: 2 nodes, Right: 12 nodes), Param Pravega (Left: 2 nodes, Right: 10 nodes) and Fugaku (Left: 20 nodes, Right: 200 nodes) supercomputers.}\label{fig:speedup_bccmo}
\end{figure}

Similar to \cref{fig:speedup_alnp}, the scaled time is plotted in \cref{fig:speedup_bccmo} and the performance of matrix-free is compared with the cell-matrix baseline across the material systems considered.  The corresponding speedups for the 127-atom case range from 1.5$\times$–3.6$\times$ (Frontier), 1.3$\times$–2.0$\times$ (Pravega), and 1.4$\times$–2.7$\times$ (Fugaku). For the 1023-atom system, speedups of 1.6$\times$–3.5$\times$, 1.4$\times$–2.0$\times$, and 1.3$\times$–2.7$\times$ are obtained on the respective machines Frontier, Pravega and Fugaku.
Gains similar to those of the 127-atom system are achieved for the 1023-atom system as well. Similar to the case of Al nanoparticle, the computational cost of nonlocal action of $\bH_{\text{nloc}}$ increases with the number of atoms, but our matrix-free implementation maintains the speedups over both the BCC molybdenum systems. Overall, the matrix-free algorithm maintains its performance advantage across both small and large, as well as periodic and non-periodic systems.

Importantly, the slowdown seen on Fugaku for $\texttt{FEOrder}=7$ in the non-periodic case of Al nanoparticle does not occur here, owing to the efficient  matrix-free implementation for complex-arithmetic (see \cref{sec:complexGGAOp,sec:sec4}) which reduces the amount of computation and data movement, resulting in 1.3–2.7$\times$ speedups on Fugaku. Furthermore, this complex arithmetic incurs a higher computational cost for the cell-matrix approach compared to its real arithmetic counterpart, in contrast to our matrix-free strategy. Interestingly, Pravega shows lower speedups compared to Fugaku (quite opposite to the real DFT operator case), as Fugaku's higher memory bandwidth benefits a bandwidth-bound implementation.

Finally, we note that the scaled time is consistently minimal for $\texttt{FEOrder}=8$. Moreover, the total solution time and the number of DoFs required to achieve the desired accuracy ($10^{-4}$ Ha/atom in energies and $10^{-4}$ Ha/Bohr in forces) are also the lowest for this order.

{\cblue It is also worth commenting here on how the speedups reported in \cref{fig:speedup_alnp,fig:speedup_bccmo} would behave if a quadrature order different from the default choice $n_q = n_p + 2$ were employed. The cost of the cell-matrix baseline action is essentially insensitive to $n_q$: the quadrature rule enters only the one-time construction of the cell-level matrices, after which their action costs $2n_p^6$ FLOP per cell per vector irrespective of the rule used. In contrast, the floating point operations in the matrix-free action grow with $n_q$, since all tensor contractions operate on the quadrature grid. From the operation count estimates of the performance model in \cref{sec:rooflinePerf}, lowering the quadrature order from $n_q = n_p+2$ to $n_q = n_p$ at $\texttt{FEOrder}=8$ reduces the matrix-free FLOP by about $1.8\times$, while raising it to $n_q=n_p+3$ increases the FLOP by about $1.4\times$. The corresponding change in the observed speedups is, however, considerably milder than these FLOP ratios suggest, since the matrix-free kernels are memory-bound on the architectures considered here and the quadrature-dependent data (the diagonal matrices $\bV^L$, $\bV^G_s$, $\bV^K_s$) is amortized over a batch of vectors. This behaviour is consistent with the empirical study reported in our earlier work in the context of FE discretized Helmholtz operators (see Appendix C.5 and Fig.~C.23 of \cite{PANIGRAHI2024104925}), where the batched matrix-free implementation benchmarked with the higher quadrature order $n_q = n_p+2$ retained its performance advantage over the cell-matrix baseline across scaling regimes, with speedups of 1.8$\times$--19.9$\times$ at ${\sim}43$k--45k DoFs per MPI task and 1.2$\times$--5.4$\times$ in the extreme-scaling regime of ${\sim}700$ DoFs per MPI task (for $n_v \geq 64$). We exploit precisely this quadrature dependence of matrix-free method in \cref{sec:rchfsi}, where an eigensolver tolerant to inexact matrix-vector products permits the use of $n_q=n_p$ in the subspace construction step for additional gains.\cn}

\subsubsection{\textbf{All-electron DFT operator}}\label{sec:AllElectron}

In the all-electron DFT formulation, the nonlocal pseudopotential term vanishes ($\mathbf{H}_{\text{nloc}} = 0$), and the Hamiltonian reduces to its purely local form with the action of the DFT operator on multivectors at FE cell-level encountered in each step of the iterative solver, given as follows:
\begin{align}
\bH_\text{loc}^{\left(e\right)} \bX^{\left(e,t\right)} = \left(\bT^{\left(e\right)} + \bL^{\left(e\right)} + \bnG^{\left(e\right)} - \imag \bK^{\left(e\right)} \right)\bX^{\left(e,t\right)}
\end{align}
This simplification eliminates the expensive nonlocal projector operations that arise in norm-conserving pseudopotential calculations, which scale with the number of atoms. As a result, the all-electron benchmark becomes an ideal test case for isolating the performance of the matrix-free local operator evaluation.
\begin{figure}[ht]
    \centering
    \begin{subfigure}[t]{.49\textwidth}
        \centering
        \includegraphics[width=\textwidth]{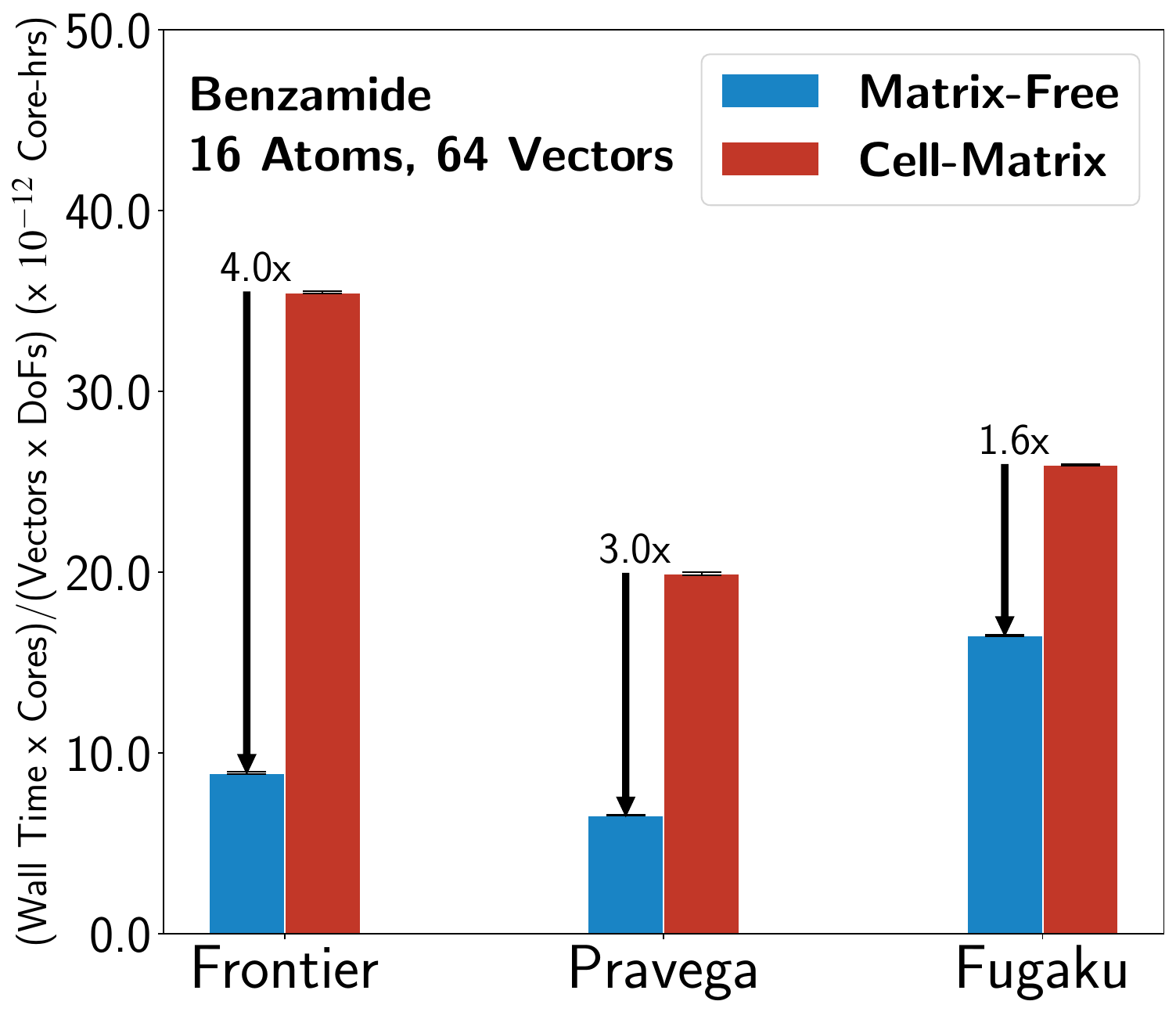}
        \label{fig:speedup_benz16}
    \end{subfigure}\hfill
    \begin{subfigure}[t]{.49\textwidth}
        \centering
        \includegraphics[width=\textwidth]{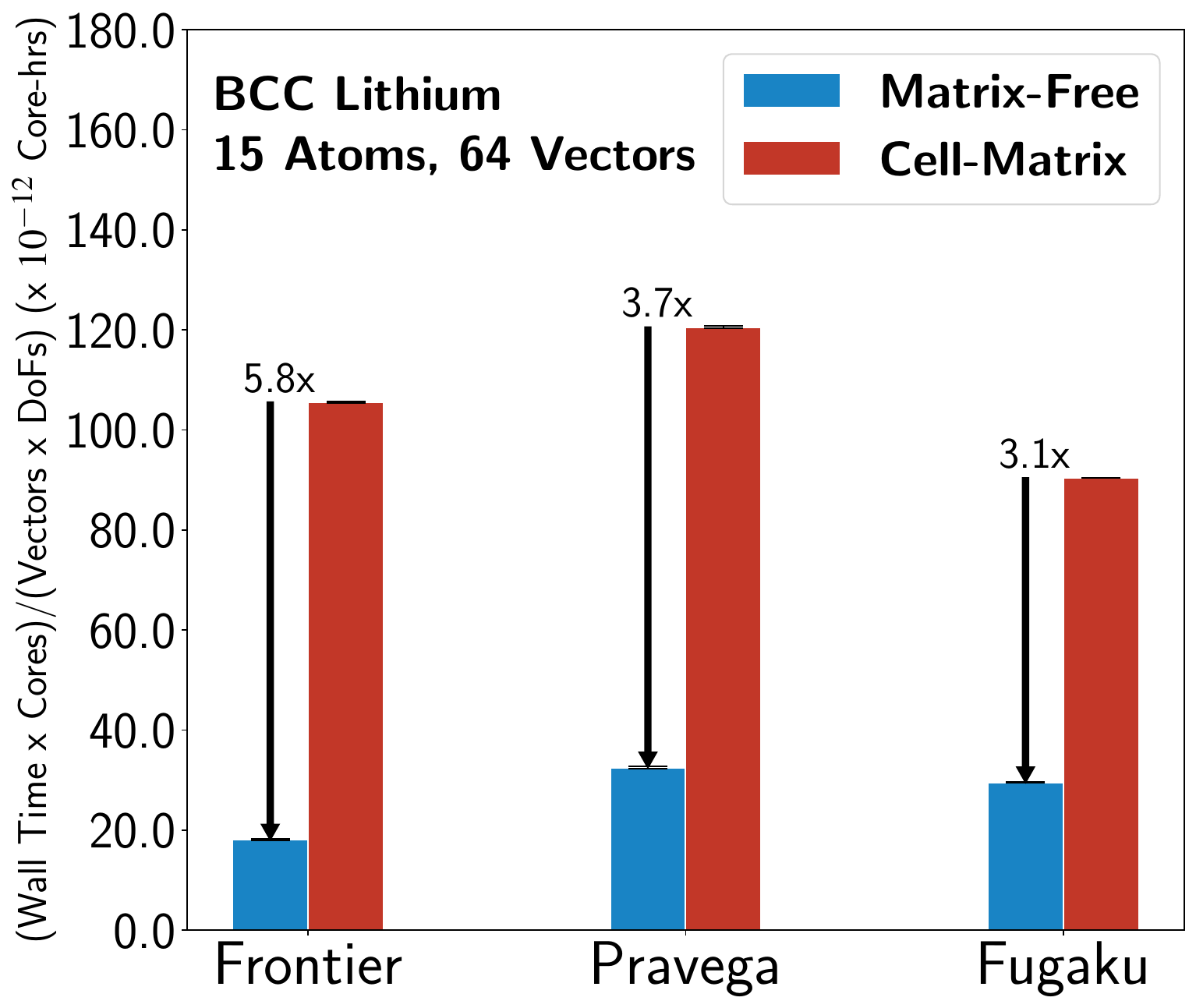}
       \label{fig:speedup_benz15}
    \end{subfigure}

    \caption{Speedups of our matrix-free implementation strategies for benzamide (real operator, 16 atoms, ${\sim}240k$ DoFs/atom) and BCC lithium (complex operator, 2x2x2 k-point rule, 15 atoms, ${\sim}140k$ DoFs/atom) with all-electron DFT operator at $\texttt{FEOrder} = 8$ on 2 nodes of Frontier, 2 nodes of Param Pravega and 20 nodes of Fugaku supercomputers respectively.}\label{fig:speedup_benzLi}
\end{figure}
We consider two representative all-electron systems to highlight both real and complex operator behavior: (i) benzamide (C$_7$H$_6$N$_2$O), a molecular system with 16 atoms treated using a non-periodic boundary condition ($\mathbf{K}^{(e)} = 0$), representing a real operator and (ii) body-centered cubic lithium $2\times 2\times 2$ supercell with monovacancy comprising 15 atoms, a metallic system treated with a  $2\times2\times2$ Monkhorst–Pack k-point sampling \cite{PhysRevB.13.5188}, representing the complex operator with periodic boundary conditions. For both systems, the \texttt{FEOrder} is chosen to be 8 as we found it to provide the most favourable performance-to-accuracy ratio. The benchmarks are conducted on Frontier, Param Pravega, and Fugaku supercomputers, using the smallest number of nodes required to accommodate each system within the available memory: 2 nodes on Frontier and Pravega, and 20 nodes on Fugaku for both benzamide and BCC lithium. \cref{fig:speedup_benzLi} presents the scaled wall time metric \textit{scaled time = (wall time $\times$ CPU cores) / (vectors $\times$ DoFs)}, used consistently throughout this work for cross-system comparison. For benzamide (the all-electron real operator case), the matrix-free implementation achieves speedups of 4.0$\times$, 3.0$\times$ and 1.6$\times$ on Frontier, Pravega, and Fugaku, respectively, relative to the baseline cell-matrix approach. The lower gains on Fugaku compared to Frontier and Pravega are similar to the non-periodic case (Al nanoparticle) reported in \cref{sec:NonPeriodic} and can be again attributed to the memory-bound behaviour observed for the case of benzamide (see \cref{fig:rooflineReal}) and worsens due to the use of only 4 MPI tasks per node on Fugaku. In contrast to benzamide, for the BCC lithium case involving the complex DFT operator, the performance benefits of the matrix-free formulation are more pronounced, with speedups of 5.8$\times$ on Frontier, 3.7$\times$ on Pravega, and 3.1$\times$ on Fugaku.  This can be attributed to the {\cblue efficient cache reuse of our matrix-free implementation in the complex-arithmetic case}. Furthermore, the increase in FLOP for the cell-matrix approach from real to complex arithmetic is higher than that of the matrix-free method, which also contributes to the observed higher speedup. In general, higher speedups relative to cell-matrix approach are observed for the cases of benzamide and BCC lithium compared to the other materials systems considered previously. This is due to the fact that the low number of vectors employed in benzamide and BCC lithium favour matrix-free methods as the small matrices are very cache-friendly, and further the reduced number of floating point operations compared to the cell-matrix approach gives huge computational gains, which is consistent with what was observed in \cite{PANIGRAHI2024104925}. The difference between matrix-free and cell-matrix methodologies is distinctly evident in all-electron $\mathbf{H}$ operator action comprising only $\mathbf{H}_{\text{loc}}$, and these results underscore the usefulness of matrix-free for reducing computations and reutilising data for complex arithmetic.

To further interpret the benchmarking results presented in this subsection, we now develop and analyze roofline performance models for the matrix-free and baseline cell-matrix implementations.

\subsection{\textbf{Roofline and Performance Model Analysis}}\label{sec:rooflinePerf}
The roofline model provides a quantitative framework to assess whether a given implementation is compute-bound or memory-bound, allowing us to identify which performance limits are set by hardware throughput and which arise from algorithmic inefficiencies. While the previous subsections demonstrated strong empirical speedups across different materials systems and architectures, the roofline analysis presented here offers a complementary architecture perspective on achieved arithmetic intensity, sustained performance, and expected scalability. {\cblue We clarify at the outset the convention adopted in this analysis. For the roofline placement of the two implementations we depart from the classical roofline of Williams et al.~\cite{10.1145/1498765.1498785}, whose arithmetic intensity is referred to the main-memory (DRAM) traffic, and instead employ the \emph{cache-aware} roofline model (CARM) of Ilic et al.~\cite{6506838}, in which the intensity is referred to the total data traffic that crosses the core--L1 boundary, that is, every floating-point datum the cores load and store. Because this traffic is counted at the core and is independent of which level of the hierarchy ultimately supplies the data, each kernel has a single arithmetic intensity and appears as a single point on the roofline. The L1, L2, L3, and DRAM lines are then not separate intensities but separate attainable-performance ceilings, each being the peak achieved when the core's request stream is served entirely from that level. The vertical position (y-coordinate) of a given approach relative to these ceilings identifies the level at which it is bandwidth-bound, or the horizontal compute roof if it is compute-bound. We adopt the cache-aware convention deliberately, for two reasons. First, the advantage of the matrix-free operator action is fundamentally one of data reuse within the cache hierarchy, and it is the cache-aware roofline, with its separate L1, L2 and L3 bandwidth ceilings, that exposes this behaviour, whereas a single DRAM ceiling cannot. Second, under the batched matrix-free evaluation used here, each MPI task owns a small subdomain whose working set is essentially resident in the last-level cache, the read-once cold load being a one-time cost amortized over the many reuses from L3, so the matrix-free operator action executes out of the last-level cache. A classical DRAM-referenced arithmetic intensity is consequently ill-defined for this cache-resident execution and does not describe it. The cache-aware roofline, referred to the core--L1 traffic the operator actually generates, is therefore both the appropriate framework for characterizing its performance and the one we can determine reliably. The same convention is applied uniformly to the matrix-free and the cell-matrix implementations, so that the two are compared on an equal footing.

We note that the analytical operation- and byte-count expressions derived below are used only to \emph{estimate} the operation counts, the data movement, and the expected speedups of the two implementations. The achieved-performance points displayed in the roofline figures (\cref{fig:rooflineReal,fig:rooflineComplex})  are obtained from hardware measurements collected on Param Pravega, on the same two-node ($96$-task) runs used for the benchmarks above, with Intel Advisor~\cite{IntelAdvisor}, whose binary instrumentation and built-in cache simulator resolve both the floating-point operations and the per-level (L1/L2/L3/DRAM) data traffic of the operator action. The achieved arithmetic intensity (x-coordinate) is then the measured core--L1 FLOP-to-byte ratio, and the achieved sustained performance (y-coordinate) is the measured FLOP divided by the measured run time. All roofline quantities are reported per node: the achieved performance is the per-task (single-core) rate aggregated over the $48$ cores of a node, and the ceilings are the corresponding measured node-aggregate peaks (the AVX-512 compute peak is $2.22$~TFLOP/s per node). The arithmetic intensity is an intensive quantity and hence independent of the core and node count, so the placement of each point relative to the roofs is identical whether expressed per core, per node, or for the full two-node run. As an independent check, these Advisor measurements were cross-validated against LIKWID hardware performance counters~\cite{psti,Gruber2022LIKWID} on a representative case, the two tools agreeing on the floating-point count of the operator action to within $15\%$.\cn}

The matrix-free implementation for the DFT problem computes the action of the finite-element discretized Hamiltonian on batches of multivectors through a sequence of tensor contractions, local element operations, and (optionally) nonlocal pseudopotential operations. We employ the even-odd decomposition \cite{Kopriva2009ImplementingEquations,Solomonoff1992ADifferentiation,Kronbichler2019FastOperators,Fischer2020ScalabilitySolvers} to reduce FLOP in our matrix-free implementation and to characterize our implementation's floating-point and memory requirements, we estimate the number of operations and bytes accessed, following the methodology outlined in Appendix A3 of \cite{Fischer2020ScalabilitySolvers} and in \cite{PANIGRAHI2024104925}. We first define
\begin{align}
    n_q^o&=\left\lfloor \frac{n_q}{2}\right\rfloor & n_q^e&=\left\lceil \frac{n_q}{2}\right\rceil & n_p^o&=\left\lfloor \frac{n_p}{2}\right\rfloor & n_p^e&=\left\lceil \frac{n_p}{2}\right\rceil
\end{align}
where $\lfloor\cdot\rfloor$ and $\lceil\cdot\rceil$ represent the floor and ceiling operations, respectively. $n_p$ represents the number of nodal points and $n_q$ represents the number of quadrature points in each spatial direction for a given FE cell.
Throughout this section, we refer to \cref{sec:sec4} and define subsequent notations wherever necessary. Here, $b^{\left(2\right)}$ denotes the batchsize of level-2 batch, $n_b^{(2)}$ denotes the number of batches in the level-2 batch, $n_b^{(1)}$ denotes the number of batches in the level-1 batch, $E$ denotes the number of FE cells, $n_{s}$ denotes the total number of projector functions in the non-local pseudopotential operator, and $m$ denotes the total number of DoFs in the FE discretized mesh.

\subsubsection{Real-valued DFT operator}
\vspace{0.1in}
\paragraph{\underline{Matrix-Free Action}}
The floating-point operation count for the matrix-free action of real DFT operator case (pseudopotential calculations) can be estimated from \cref{eqn:ggaReal} that includes contributions from the element-wise tensor contractions $\bYt_0$ to $\bYt_4$,  final evaluation of $\bY^{(e,t)}$ and the non-local pseudopotential term $\bH_\text{nloc}\bX$ as follows:\\[0.1in]
$\bYt_0: 2b^{\left(2\right)}n_b^{(2)}\left(n_p^o + n_q^o(n_p+1)\right)\left(n_p^2 + n_p n_q+n_q^2\right)n_b^{(1)} E$, \quad $\bYt_1: 6b^{\left(2\right)}n_b^{(2)}n_q^on_q^2\left(n_q+2\right)n_b^{(1)} E$, \\
$\bYt_2: 5b^{\left(2\right)}n_b^{(2)}n_q^3n_b^{(1)} E$, \quad
$\bYt_3: \left(6b^{\left(2\right)}n_b^{(2)}n_q^3\right)n_b^{(1)} E + \left(3b^{\left(2\right)}n_b^{(2)}n_q^3 + 2n_q^3 + 18b^{\left(2\right)}n_b^{(2)}n_q^3\right)n_b^{(1)} E$, \\
$\bYt_4: \left(6b^{\left(2\right)}n_b^{(2)}n_q^on_q^2\left(n_q+3\right)\right)n_b^{(1)} E + \left(2b^{\left(2\right)}n_b^{(2)}n_q^3\right)n_b^{(1)} E$, \\
$\bYt_5: 2b^{\left(2\right)}n_b^{(2)}\left(n_q^o + n_p^o(n_q+1)\right)\left(n_p^2 + n_p n_q+n_q^2\right)n_b^{(1)} E$, \quad $\bH_\text{nloc}\bX: b^{\left(2\right)}n_b^{(2)}(4n_p^3 + 1)n_{s}n_b^{(1)}$ \\[0.1in]
Subsequently, we can evaluate $\text{FLOP}$ as the following expression:
\begin{align}
\text{Total FLOP}=&\Bigr[2b^{\left(2\right)}n_b^{(2)}\left(n_p^o(n_q + 2) + n_q^o(n_p+2)\right)\left(n_p^2+n_pn_q+n_q^2\right)
+ 6b^{\left(2\right)}n_b^{(2)}n_q^on_q^2\left(2n_q+5\right) \nonumber \\
&+ 2n_q^3 + 34b^{\left(2\right)}n_b^{(2)}n_q^3\Bigr]n_b^{(1)} E + b^{\left(2\right)}n_b^{(2)}(4n_p^3 + 1)n_{s}n_b^{(1)}
\label{eqn:flopRealMF}
\end{align}
{\cblue At leading order, with $n_q\sim n_p$, this amounts to $\approx 12\,n_p^4$ FLOP per vector per cell.\cn}
{\cblue The read-once data movement of this action consists of streaming the input multivector once into cache, interpolating it to the quadrature points, applying the pointwise coefficient arrays $\bV_{s}^{G}$ and $\bV^{L}$ together with the per-cell Jacobian $\bG^{(s,d)}$, and adding the result back to the output multivector, with the input reused for the nonlocal projector action. At leading order, with $n_q\sim n_p$ and $n_v$ vectors, this read-once traffic is
\begin{align}
    \text{Bytes}_\text{MF}\approx 56\,n_p^3 \quad\text{(per vector per cell),}
\label{eqn:bytesRealMF}
\end{align}
the nonlocal-projector and quadrature-coefficient terms being subdominant.\cn}
{\cblue The read-once count of \cref{eqn:bytesRealMF} is used only to compare the data moved by the two implementations. For the cache-aware roofline placement, by contrast, we refer the arithmetic intensity (AI) to the core--L1 byte traffic the kernel actually issues, several times this read-once count since the intermediate quadrature buffers are repeatedly loaded and stored within the kernel,\cn}
\begin{align}
    \text{AI}=\text{FLOP}\Big/\text{Byte}
\end{align}
{\cblue The attainable sustained performance ($\text{SP}$) is then set by the bandwidth of the cache level that supplies this traffic,\cn}
\begin{align}
    \text{SP}=\min\left\{\text{SP}^{peak},\text{AI}\times \text{BW}^{L}\right\} \label{sp}
\end{align}
{\cblue where $\text{SP}^{peak}$ is the peak FLOP/s and $\text{BW}^{L}$ the bandwidth of the binding cache level $L\in\{\text{L1},\text{L2},\text{L3}\}$. The matrix-free action, whose working set overflows the L1 cache, is bounded by the L2/L3 bandwidth, whereas the cell-matrix action is compute-bound and attains a large fraction of $\text{SP}^{peak}$.\cn}

\paragraph{\underline{Cell-Matrix Action}} For comparison, we estimate the total number of floating point operations and bytes accessed in the case of the cell-matrix approach, the baseline used for all our benchmarks. Since the cell-matrix approach involves dense matrix-matrix products corresponding to cell-matrix and multivector matrices, we can estimate the FLOP as:
$\bH_\text{loc}\bX: 2b^{\left(0\right)}n_p^6n_b^{(0)}E$, \quad
$\bH_\text{nloc}\bX: b^{\left(0\right)}(4n_p^3 + 1)n_{s}n_b^{(0)}$
Subsequently we can evaluate total $\text{FLOP}$ as the following expression:
\begin{align}
    \text{Total FLOP} = 2b^{\left(0\right)}n_p^6n_b^{(0)}E + b^{\left(0\right)}(4n_p^3 + 1)n_{s}n_b^{(0)}
\label{eqn:flopRealCM}
\end{align}
{\cblue At leading order the dense cell-matrix product dominates, giving $\approx 2\,n_p^6$ FLOP per vector per cell.\cn}
{\cblue The cell-matrix action moves the same read-once multivector data, but additionally streams the stored cell-level Hamiltonian matrix ($8n_p^6$ bytes per cell) once per multivector. At leading order this gives
\begin{align}
    \text{Bytes}_\text{CM}\approx 56\,n_p^3 + \frac{8\,n_p^6}{n_v} \quad\text{(per vector per cell),}
\label{eqn:bytesRealCM}
\end{align}
the second term reflecting the cell-matrix storage streamed once per block of $n_v$ vectors. 

\vspace{0.1in}

The leading-order behaviour of the operation and byte counts derived above is instructive, and we discuss it in more detail below. With $n_q \sim n_p$, the matrix-free action involves approximately $12n_p^4$ FLOP per vector per cell against the $56n_p^3$ read-once bytes of \cref{eqn:bytesRealMF}, whereas the cell-matrix action involves $2n_p^6$ FLOP against the $56n_p^3 + 8n_p^6/n_v$ bytes of \cref{eqn:bytesRealCM}. The matrix-free method thus performs $\sim n_p^2/6$ fewer floating point operations.}\cn

\paragraph{\underline{Comparison of Matrix-Free and Cell-Matrix Action}}
\cblue 
\cref{fig:rooflineReal} shows the achieved sustained performance of the matrix-free and cell-matrix implementations on the cache-aware roofline for the real-valued DFT operator action (\cref{eqn:HX}). On the architectures considered here, the two implementations occupy opposite regimes of the cache-aware roofline.  The matrix-free points lie in the bandwidth-bound region, below the L1 ceiling and close to the L2/L3 bandwidth ceilings from which the working set is streamed, for both the Al nanoparticle (147 atoms, $\texttt{FEOrder} = 8$, $N_v = 256$) and the benzamide (16 atoms, $\texttt{FEOrder} = 8$, $N_v = 64$) benchmarks on Param Pravega. The cell-matrix points lie instead in the compute-bound region, sustaining roughly $57\%$--$60\%$ of the AVX-512 floating-point peak due to large dense matrix products being executed through optimized BLAS. 

The matrix-free action attains both a low sustained performance and a low arithmetic intensity compared to cell-matrix action. Its data movement is non-contiguous, it incurs runtime MPI communication, and the extraction and assembly steps ($\bQ^{(i_b,e,t)}$ and ${\bQ^{(i_b,e,t)}}^T$) of the multilevel batched layout dominate its cost (see \cite{PANIGRAHI2024104925}), all of which keep the action bandwidth-bound. Its operation count is also much smaller than that of the cell-matrix approach, which lowers the achieved rate further. The time to solution is nonetheless the lower of the two, since that operation count is smaller by a factor of about $n_p^2/6$ while the data moved is no larger than in the cell-matrix approach, and is in fact smaller, because the stored cell matrices are never streamed. The contractions on small one-dimensional matrices reach a smaller fraction of the floating-point peak than the large dense products underlying the cell-matrix approach, so the higher sustained FLOP/s of the latter is expended on operations that the matrix-free approach does not need to perform. 

Further, the DFT operator action with $n_q = n_p$ (the quadrature rule used in the operator benchmarks), the estimated cache-aware (core--L1) arithmetic intensity of the matrix-free action is approximately $0.37$, $0.36$ and $0.43$ FLOP/byte at $\texttt{FEOrder}=7,8,9$. 
{\cblue Since the time for a single operator action is $t=\text{FLOP}/\text{SP}$, the speedup of the matrix-free over the cell-matrix action, $s=t_\text{CM}/t_\text{MF}$, follows from \cref{sp} as
\begin{align}
    s = \frac{\text{FLOP}_\text{CM}}{\text{FLOP}_\text{MF}}\times\frac{\text{SP}_\text{MF}}{\text{SP}_\text{CM}},\qquad \text{SP}_\text{MF}=\text{AI}\times\text{BW}^{\text{L2}},\quad \text{SP}_\text{CM}=f\,\text{SP}^{peak}, \label{eqn:speedupCARM}
\end{align}
the matrix-free action being bandwidth-bound at the L2 ceiling and the cell-matrix action sustaining a fraction $f$ of the compute peak. In evaluating the above \cref{eqn:speedupCARM}, the operation-count ratio $\text{FLOP}_\text{CM}/\text{FLOP}_\text{MF}$ and the matrix-free arithmetic intensity $\text{AI}$ entering $\text{SP}_\text{MF}$ are both taken from the analytical operation- and byte-count model of this section, while $\text{BW}^{\text{L2}}$, $\text{SP}^{peak}$ and the cell-matrix fraction $f$ are the measured machine quantities. For the real operator at $\texttt{FEOrder}=8$ on Param Pravega, for instance, $\text{SP}_\text{MF}=\text{AI}\times\text{BW}^{\text{L2}}\approx 0.36\times 11.3\approx 4$~GFLOP/s per core is only about $15\%$ of the cell-matrix rate $\text{SP}_\text{CM}=f\,\text{SP}^{peak}\approx 0.6\times 46.3\approx 28$~GFLOP/s per core. So, combined with the leading-order operation-count reduction $n_p^2/6 = 13.5$ at $\texttt{FEOrder}=8$, evaluating \cref{eqn:speedupCARM} predicts a $2.0\times$ speedup\cn} for the matrix-free approach over the cell-matrix baseline for the Al nanoparticle system (147 atoms, $\texttt{FEOrder} = 8$, $N_v = 256$), matching the measured value of 2.0$\times$ (see \cref{fig:rooflineReal}). Similarly, for the benzamide benchmark (16 atoms, $\texttt{FEOrder} = 8$, $N_v = 64$), the model predicts a {\cblue 2.1}$\times$ computational advantage, while the observed speedup is 3.0$\times$ (see \cref{fig:rooflineReal}). This difference between estimated and observed speedups reflects the approximations in the model, chiefly the operation-count ratio and the fraction $f$ of the compute peak sustained by the cell-matrix action. Nonetheless, it provides a useful estimate of the expected speedups.

{\cblue 
The efficient utilisation and reuse of terms, along with the proposed multilevel batched layout, hardware-tuned batchsizes, and the even–odd decomposition method, provide a significant computational advantage over the cell-matrix approach across supercomputing architectures.

\begin{figure}[ht]
    \centering
    \includegraphics[width=\textwidth]{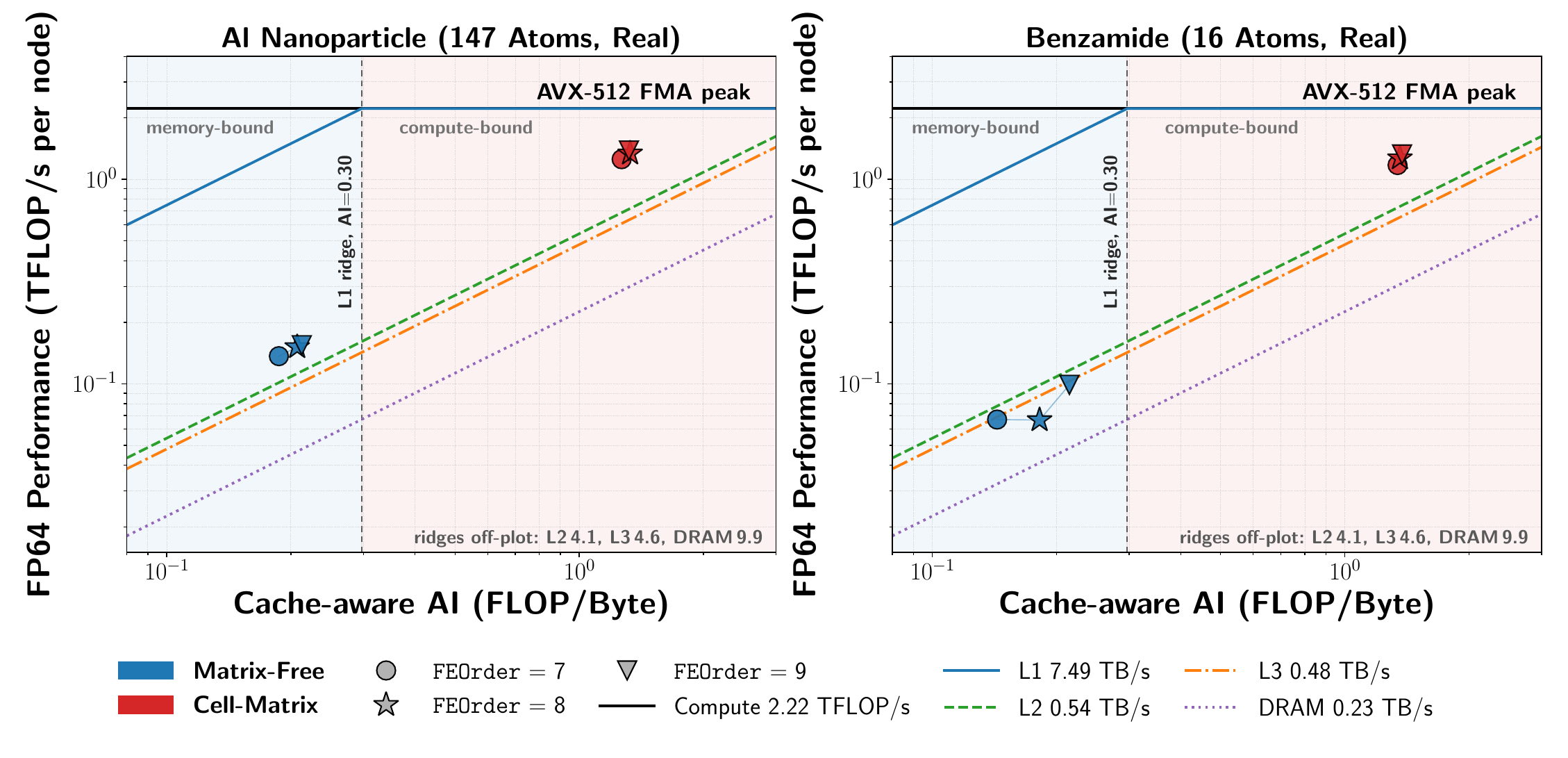}
    \caption{{\cblue Cache-aware (CARM) roofline analysis of the matrix-free (blue) and cell-matrix (red) implementations for the Al nanoparticle (147 atoms) and benzamide (16 atoms) systems, real-valued DFT operator, on Param Pravega. Method is encoded by colour and the finite-element interpolating polynomial order by marker shape (circle $\texttt{FEOrder} = 7$, star $\texttt{FEOrder} = 8$, triangle $\texttt{FEOrder} = 9$). The x-axis is the measured core--L1 (cache-aware) arithmetic intensity, the y-axis is the sustained performance per node (aggregated over the 48 cores). The horizontal line is the measured AVX-512 compute peak and the diagonals are the measured L1/L2/L3/DRAM bandwidth ceilings (per node). The marked L1 ridge (machine balance, $\text{AI}=0.30$) is the balance point of the L1 ceiling, the lower-bandwidth L2 and L3 ceilings balancing at correspondingly higher intensities, so a point lying to the right of the L1 ridge may still be bandwidth-bound at the cache level that actually supplies its data. The faint blue and red shading marks the memory-bound (left of the L1 ridge) and compute-bound (right) regions, respectively. Each kernel is a single point at its core-measured cache-aware intensity. The matrix-free action is bandwidth-bound, sitting on the L2/L3 ceilings, while the cell-matrix action is compute-bound.}} \label{fig:rooflineReal}
\end{figure}
\FloatBarrier

\subsubsection{Complex-valued DFT Operator}
\vspace{0.1in}
\paragraph{\underline{Matrix-Free Action}} Complex DFT operators, encountered in periodic and all-electron metallic systems, introduce an additional imaginary component and for the matrix-free complex case, the total operation and memory counts described below follow directly from those of the real case with some modifications accounting for additional operations in our complex implementation\cn. The operation counts can then be estimated from \cref{eqn:ggaCmplx} as follows: \\[0.1in]
$\bYt_0: 2b^{\left(2\right)}n_b^{(2)}\left(n_p^o + n_q^o(n_p+1)\right)\left(n_p^2 + n_p n_q+n_q^2\right)n_b^{(1)} E$, \quad $\bYt_1: 6b^{\left(2\right)}n_b^{(2)}n_q^on_q^2\left(n_q+2\right)n_b^{(1)} E$, \\
$\bYt_2: 5b^{\left(2\right)}n_b^{(2)}n_q^3n_b^{(1)} E$, \quad
$\bYt_3: 5b^{\left(2\right)}n_b^{(2)}n_q^3n_b^{(1)} E$, \quad $\bYt_4: \left(6b^{\left(2\right)}n_b^{(2)}n_q^3\right)n_b^{(1)} E
\\ + \left(3b^{\left(2\right)}n_b^{(2)}n_q^3 + 2n_q^3 + 18b^{\left(2\right)}n_b^{(2)}n_q^3\right)n_b^{(1)} E$, \quad
$\bYt_5: \left(6b^{\left(2\right)}n_b^{(2)}n_q^on_q^2\left(n_q+3\right)\right)n_b^{(1)} E + \left(3b^{\left(2\right)}n_b^{(2)}n_q^3\right)n_b^{(1)} E$, \\
$\bYt_6: 2b^{\left(2\right)}n_b^{(2)}\left(n_q^o + n_p^o(n_q+1)\right)\left(n_p^2 + n_p n_q+n_q^2\right)n_b^{(1)} E$, \quad $\bH_\text{nloc}\bX: b^{\left(2\right)}n_b^{(2)}(8n_p^3 + 3)n_{s}n_b^{(1)}$ \\[0.1in]
Subsequently, we can evaluate $\text{FLOP}$ as follows:
\begin{align}
\text{Total FLOP} &= 2b^{\left(2\right)}n_b^{(2)}\left(n_p^o(n_q + 2) + n_q^o(n_p+2)\right)\left(n_p^2+n_pn_q+n_q^2\right)n_b^{(1)} E
+ 6b^{\left(2\right)}n_b^{(2)}n_q^on_q^2\left(2n_q+5\right)n_b^{(1)} E \nonumber \\ &+ 2n_q^3n_b^{(1)} E + 40b^{\left(2\right)}n_b^{(2)}n_q^3n_b^{(1)} E + b^{\left(2\right)}n_b^{(2)}(8n_p^3 + 3)n_{s}n_b^{(1)}
\label{eqn:flopComplexMF}
\end{align}
{\cblue Its leading-order tensor-contraction structure mirrors that of the real matrix-free action (\cref{eqn:flopRealMF}), with the complex arithmetic and the $-\imag\bK$ term contributing additional operations.\cn}
{\cblue The read-once data movement of the complex action follows that of the real case with complex ($16$-byte) scalars, so at leading order it is roughly double,
\begin{align}
    \text{Bytes}^{\mathbb{C}}_\text{MF}\approx 112\,n_p^3 \quad\text{(per vector per cell).}
\label{eqn:bytesComplexMF}
\end{align}\cn}

\paragraph{\underline{Cell-Matrix Action:}}
Following a similar approach, for the cell-matrix method we can estimate the floating point operations for complex valued DFT operator accounting for additional complex multiply-adds as
$\bH_\text{loc}\bX: 6b^{\left(0\right)}n_p^6n_b^{(0)}E$, \quad
$\bH_\text{nloc}\bX: 3b^{\left(0\right)}(4n_p^3 + 1)n_{s}n_b^{(0)}$. Subsequently we can evaluate
\begin{align}
    \text{Total FLOP} = 6b^{\left(0\right)}n_p^6n_b^{(0)}E + 3b^{\left(0\right)}(4n_p^3 + 1)n_{s}n_b^{(0)}
\label{eqn:flopComplexCM}
\end{align}
{\cblue At leading order this is $\approx 6\,n_p^6$ FLOP per vector per cell, three times the real cell-matrix count.\cn}
{\cblue As in the real case, the complex cell-matrix action additionally streams the stored cell matrix ($16n_p^6$ bytes per cell) once per multivector, giving at leading order
\begin{align}
\text{Bytes}^{\mathbb{C}}_\text{CM}\approx 112\,n_p^3 + \frac{16\,n_p^6}{n_v} \quad\text{(per vector per cell).}
\label{eqn:bytesComplexCM}
\end{align}\cn}

\paragraph{\underline{Comparison of Matrix-Free and Cell-Matrix Action}}
{\cblue From \cref{fig:rooflineComplex}, the matrix-free points for the complex-valued operator again lie in the bandwidth-bound region of the cache-aware roofline, close to the L2/L3 bandwidth ceilings, for both the BCC molybdenum (127 atoms, $\texttt{FEOrder} = 8$, $N_v = 1280$) and the BCC lithium (15 atoms, $\texttt{FEOrder} = 8$, $N_v = 64$) benchmarks on Param Pravega, while the cell-matrix points are compute-bound, sustaining roughly $78\%$--$79\%$ of the AVX-512 floating-point peak. 

The complex DFT operator action scales up both the operation count and the data movement relative to the real case (for the cell-matrix action the operation count triples while the data movement roughly doubles). Because the two scale by similar factors, the cache-aware arithmetic intensity of the complex operator stays comparable to that of the real operator (measured matrix-free intensities of $0.18$--$0.39$ FLOP/byte against $0.14$--$0.21$ for the real systems), rather than markedly higher.\cn} {\cblue Applying \cref{eqn:speedupCARM} to the complex operator, again with the operation-count ratio and arithmetic intensity taken from the analytical model (\cref{eqn:flopComplexMF,eqn:bytesComplexMF,eqn:flopComplexCM,eqn:bytesComplexCM}),\cn} we estimate a speedup of {\cblue 1.5}$\times$ for the BCC molybdenum (127 atoms) periodic benchmark on Param Pravega ($\texttt{FEOrder}=8$, $N_v=1280$), compared to the observed 1.7$\times$. For the BCC lithium (15 atoms, $\texttt{FEOrder}=8$, $N_v=64$) all-electron system, the model estimates {\cblue 3.2}$\times$, while the measured value is around 3.7$\times$. {\cblue The complex DFT operator thus reproduces the roofline picture observed in real-valued operator, at higher total operation and byte counts.\cn}

{\cblue We close this subsection with a comparison of the memory footprints of the two approaches. The cell-matrix approach stores the FE cell-level Hamiltonian matrices, requiring $8cn_p^6E$ bytes ($c=1$ for real and $c=2$ for complex arithmetic), whereas the matrix-free approach stores only the diagonal quadrature-point data of \cref{eqn:refact} along with the one-dimensional matrices $\bN^{1D}$ and $\widetilde{\bD}^{1D}$ (about two orders of magnitude smaller in terms of memory at the FE polynomial orders of interest). For smaller problems involving only a modest number of vectors, such as the all-electron benzamide and BCC lithium systems, this operator storage is the dominant memory cost, so eliminating it allows the matrix-free variant to fit the same problem on fewer nodes, which is especially valuable on memory-constrained machines such as Fugaku (32 GB/node). For the large pseudopotential benchmarks, by contrast, the node-level multivector storage common to both approaches dominates the total memory, so the node counts are set by the eigensolver working memory rather than the operator storage. But identical node counts were used for both implementations in all benchmarks for fair comparision.\cn} 

\begin{figure}[ht]
    \centering
    \includegraphics[width=\textwidth]{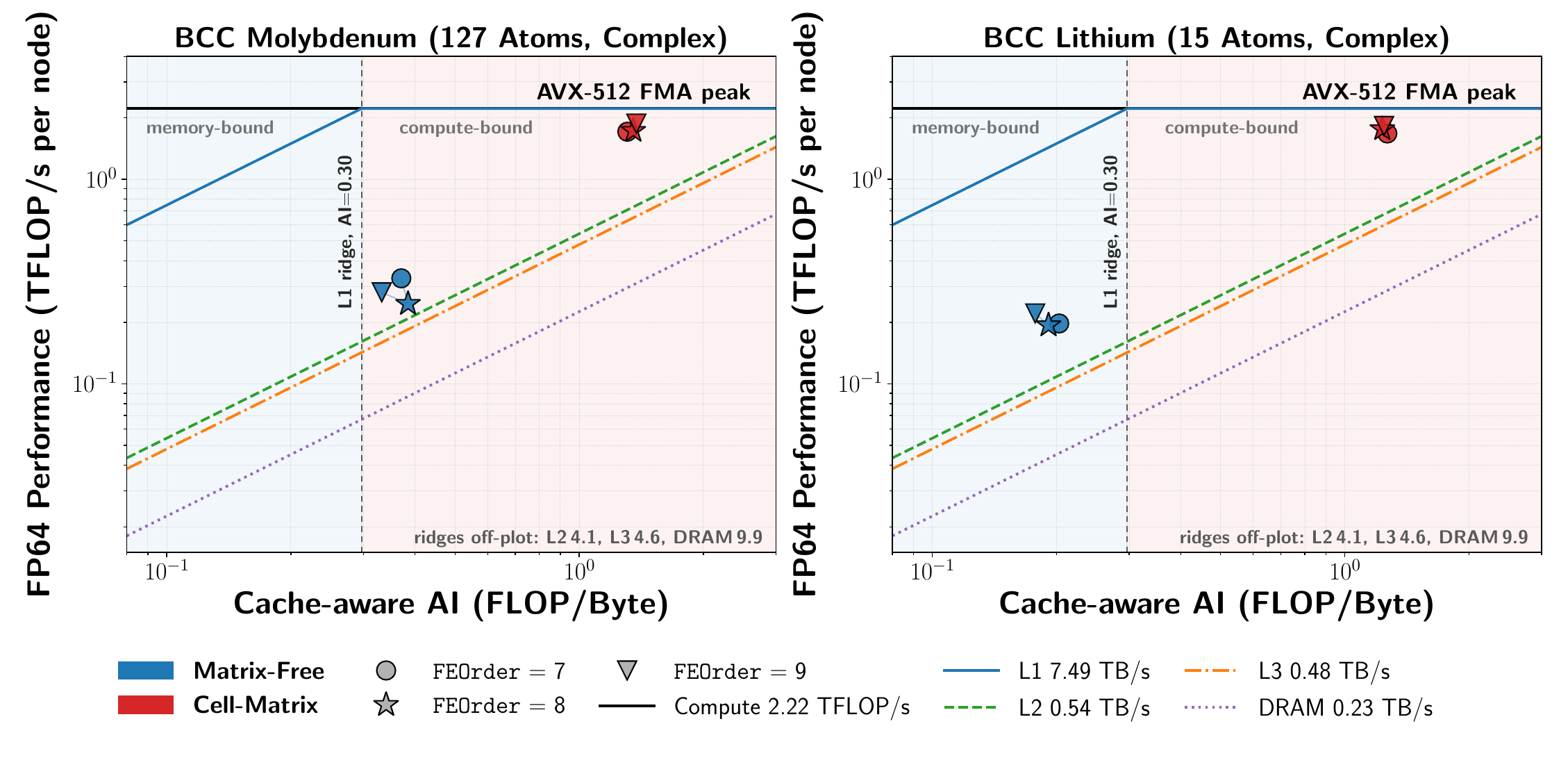}
    \caption{{\cblue Cache-aware (CARM) roofline analysis of the matrix-free (blue) and cell-matrix (red) implementations for the BCC molybdenum (127 atoms) and BCC lithium (15 atoms) systems, complex-valued DFT operator, on Param Pravega. Encoding, axes and roofs are as in \cref{fig:rooflineReal}: method by colour, $\texttt{FEOrder}$ by marker shape (circle $\texttt{FEOrder} = 7$, star $\texttt{FEOrder} = 8$, triangle $\texttt{FEOrder} = 9$). The x-axis is the measured core--L1 (cache-aware) arithmetic intensity, the y-axis the sustained performance per node. The ceilings are the measured node-aggregate compute peak and L1/L2/L3/DRAM bandwidths, with the L1 ridge marking the L1 balance point and the faint blue/red shading marking the memory- and compute-bound regions (the per-level ridges and the shading being as described in \cref{fig:rooflineReal}). As in the real case, the matrix-free action is bandwidth-bound and the cell-matrix action is compute-bound.}} \label{fig:rooflineComplex}
\end{figure}
\FloatBarrier

\subsection{\textbf{Kohn--Sham DFT eigensolver leveraging matrix-free action using mixed precision}}\label{sec:rchfsi}
We now present an important benchmark involving the solution of the finite-element discretized Kohn--Sham DFT eigenvalue problem (DFT-EVP), which uses the proposed matrix-free implementation to evaluate matrix-multivector products that arise during the iterative procedure used to solve the DFT-EVP. Consequently, we note that the FE discretized DFT eigenproblem as described in \cref{ssec:ksdftfe}, can be expressed as $\bH \hat{\bu}_n = \epsilon^{h}_n \bM \hat{\bu}_n$ (for notational convenience here, the superscript $\bk$ has been dropped). The DFT Hamiltonian operator as discussed in \cref{eq:Hloc}, $\bH = \bT + \bL + \bnG - \imag\bK$ can be real ($\bK = \mathbf{0}$) or complex valued depending on whether one is solving non-periodic or periodic DFT calculations, respectively. Further, the matrix $\bM$ is the FE overlap matrix that arises due to the non-orthogonality of finite-element basis functions. We employ the recently developed residual-based Chebyshev filtered subspace iteration strategy (R-ChFSI) \cite{KODALI2026110239}, a variant of Chebyshev filtered subspace iteration (ChFSI) \cite{Zhou2006ParallelAcceleration,Motamarri2020DFT-FECalculations} to solve the DFT generalised Hermitian eigenproblem as described in \cref{alg:RChFSiGhep}. The algorithm for R-ChFSI involves a subspace construction step employing a Chebyshev polynomial filter of $\bM^{-1}\bH$ that magnifies the wanted spectrum, generating a subspace rich in desired eigenvectors, followed by a Rayleigh-Ritz step that involves subspace diagonalization of the projected FE discretized generalized eigenproblem onto this filtered space. In particular, the subspace construction step in R-ChFSI is designed to accommodate inexact matrix-vector products while maintaining robust convergence properties \cite{KODALI2026110239}. This aspect can be exploited in the subspace construction step of R-ChFSI to employ a diagonal approximation of $\bM^{-1}$ using a Gauss-Lobatto quadrature rule coincident with the nodal points \cite{Das2022DFT-FEDiscretization} alongside FP32 arithmetic in the matrix-multivector products for improving computational efficiency.

For accurate evaluations of integrals arising in $\bL$ and $\bnG$ for the DFT operator $\bH_{\text{loc}}$, we usually employ a quadrature rule with number of points $n_q = \texttt{FEOrder} + 3 = n_p+2$, however this might increase the cost of matrix-free action on multivectors compared to the case where $n_q=n_p$ and can in principle reduce the speedups over the cell-matrix baseline. To this end, we leverage the fact that the subspace construction step in R-ChFSI is tolerant to inexact matrix-vector multiplications, and consequently, we employ a lower quadrature rule $n_q=n_p=\texttt{FEOrder}+1$ for evaluating tensor contractions during the matrix-free action of the DFT operator on multivectors in the subspace construction step using a Chebyshev polynomial filter of suitable polynomial degree. Our numerical experiments demonstrate that this reduced quadrature rule, utilising FP32 lower-precision arithmetic, does not compromise the achievable accuracy in our DFT calculations due to the use of R-ChFSI rather than the usual ChFSI.

We employ our cell-matrix baseline, as described earlier, to compute matrix-multivector products during the R-ChFSI procedure and conduct comparative studies with our matrix-free implementation on multi-node CPUs, as described subsequently. The finite-element meshes with $\texttt{FEOrder}=8$ are employed for the benchmark studies reported here, which yield the shortest time to solution (in core-hours) with the least number of degrees of freedom to achieve the desired accuracy in energy and forces, as reported in \cref{sec:DFToperator}. The Chebyshev polynomial degree employed in the benchmark calculations is around 30.

{\setlength{\algomargin}{0.5em}
\begin{algorithm}[ht]
\caption{Subspace Iteration accelerated using Chebyshev polynomial of degree $p$ (R-ChFSI)}
\label{alg:RChFSiGhep}
\begin{algorithmic}
\STATE{\emph{Initial Guess}:  Let $\bX^{(0)} = \begin{bmatrix}
    \bx_1^{(0)} & \bx_2^{(0)} &\dots&\bx_n^{(0)}\end{bmatrix}$ be the initial guess of the eigenvectors ($\{\bu_j\}$).}
    \WHILE{$\norm{\bH\bx^{(i+1)}_j-\epsilon_j^{(i+1)}\bM\bx^{(i+1)}_j} \geq \tau$}
    \STATE{\emph{Chebyshev Filtered Subspace Construction}: Construct $\bY_p^{(i)}=C_p(\bH)\bX^{(i)}$ using \cref{alg:rchebFilt}}
    \STATE{\emph{Rayleigh-Ritz step}: Solve the $n\times n$ generalized EVP: $({\bY_p^{(i)}}^\dagger\bH\bY_p^{(i)})\bE=({\bY_p^{(i)}}^\dagger\bM\bY_p^{(i)})\bE\bLam$; set $\bX^{(i+1)}=\bY_p^{(i)}\bE$, $\bLam^{(i+1)}=\bLam$.}
    \ENDWHILE
\end{algorithmic}
\end{algorithm}
}

{\setlength{\algomargin}{0.75em}
\begin{algorithm}
\caption{R-ChFSI filtering procedure for generalized Hermitian eigenvalue problem $\bH \bX = \bM \bX \bLam$}
\label{alg:rchebFilt}

    \KwIn{$\bH$ and $\bM$ matrices; approximate inverse of $\bM$ denoted $\bD^{-1}$ (diagonal approximation due to GLL quadrature rule); Chebyshev polynomial order $p$; bounds of the eigenspectrum $\lambda_{\min}$ and $\lambda_{\max}$; upper bound of the wanted spectrum $\lambda_{T}$; initial guess of eigenvectors $\bX^{(i)}$ and eigenvalues $\bLam^{(i)}$}

    \KwTemp{$\bX$, $\bY$, $\bR_{\bX}$, $\bR_{\bY}$, $\bLam_{\bX}$ and $\bLam_{\bY}$}

    \KwResult{The filtered subspace $\bY_p^{(i)}$}

\begin{algorithmic}
\STATE{$e \gets \tfrac{\lambda_{\max}-\lambda_{T}}{2}$; $c \gets \tfrac{\lambda_{\max}+\lambda_{T}}{2}$; $\sigma \gets \tfrac{e}{\lambda_{\min}-c}$; $\sigma_1 \gets \sigma$; $\gamma \gets \tfrac{2}{\sigma_1}$}
\STATE{$\bX \gets \bX^{(i)}$; $\bY \gets \bH\bX^{(i)}-\bM\bX^{(i)}\bLam^{(i)}$}
\STATE{$\bR_{\bX}\gets 0$; $\bR_{\bY}\gets \frac{\sigma_{1}}{e}\bY$}
\STATE{$\bLam_{\bX} \gets \bI$; $\bLam_{\bY} \gets \frac{\sigma_{1}}{e}\left(\bLam^{(i)}-c\bI\right)$}
\FOR{$k\gets 2$ to $p$}
\STATE{$\sigma_2\gets \frac{1}{\gamma-\sigma}$}
\STATE{$\bR_{\bX}\gets\tfrac{2\sigma_2}{e}[(\bH\bD^{-1}-c\bI)\bR_{\bY}+\bY\bLam_{\bY}]-\sigma\sigma_2\bR_{\bX}$}
\STATE{$\bLam_{\bX}\gets \frac{2\sigma_2}{e}\bLam_{\bY}(\bLam^{(i)}-c\bI)-\sigma\sigma_2\bLam_{\bX}$}
\STATE{swap$\left(\bR_{\bX},\bR_{\bY}\right)$; swap$\left(\bLam_{\bX},\bLam_{\bY}\right)$; $\sigma = \sigma_2$}
\ENDFOR
\STATE{$\bX \gets\bD^{-1}\bR_{\bY}+\bX\bLam_{\bY}$}
\RETURN \upshape $\bX$
\end{algorithmic}
\end{algorithm}
}

\subsubsection{Matrix-free subspace construction step using FP32 arithmetic}
In this subsection, we first benchmark the performance of the Chebyshev polynomial filtering step (subspace construction step) in the R-ChFSI eigensolver (\cref{alg:rchebFilt}), employing the matrix-free action of the DFT operator, relative to the cell-matrix baseline, for both FP64 and FP32 arithmetic. To accommodate FP32 arithmetic in our matrix-free implementation, we extend the same framework in \cref{sec:multiBatch} from FP64 doubles to FP32 floats by doubling the batch size at each level of batch, namely $b^{\left(2\right)}$ and $b^{\left(1\right)}$. Hence, the proposed multilevel batched layout is easily amenable to different floating point representation. We choose the non-periodic system Al nanoparticle (561 atoms, 1024 eigenvectors) and the periodic system BCC molybdenum (1023 atoms, 8448 eigenvectors) for these studies, and the matrix-free speedups over the cell-matrix baseline are plotted in \cref{fig:speedup_fp32}. The results demonstrate the advantage of using FP32 precision to evaluate matrix-free multivector products. We notice the speedups due to switching to FP32 are more for the case of the complex DFT operator than the real DFT operator. {\cblue This follows from the larger arithmetic and data volume of the complex operator (roughly twice the data movement of the real case). Halving the precision then reduces the streamed data proportionally more, and the doubled batch size packs the real and imaginary components efficiently into the SIMD lanes, yielding higher computational gains.\cn}

\begin{figure}[ht]
    \centering
    \begin{subfigure}[t]{.49\textwidth}
        \centering
        \includegraphics[width=\textwidth]{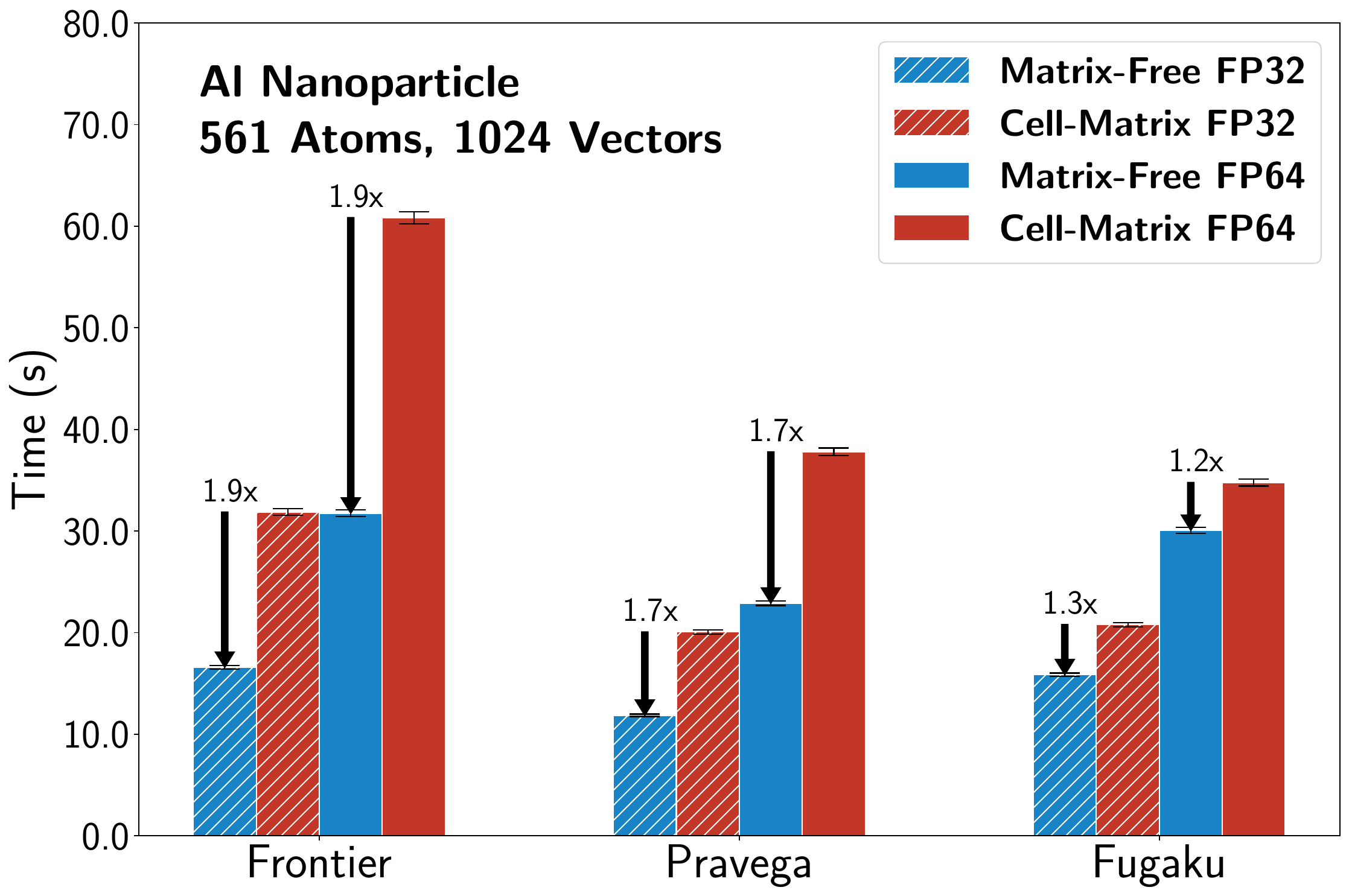}
        \label{fig:speedup_fp32_Al561}
    \end{subfigure}\hfill
    \begin{subfigure}[t]{.49\textwidth}
        \centering
        \includegraphics[width=\textwidth]{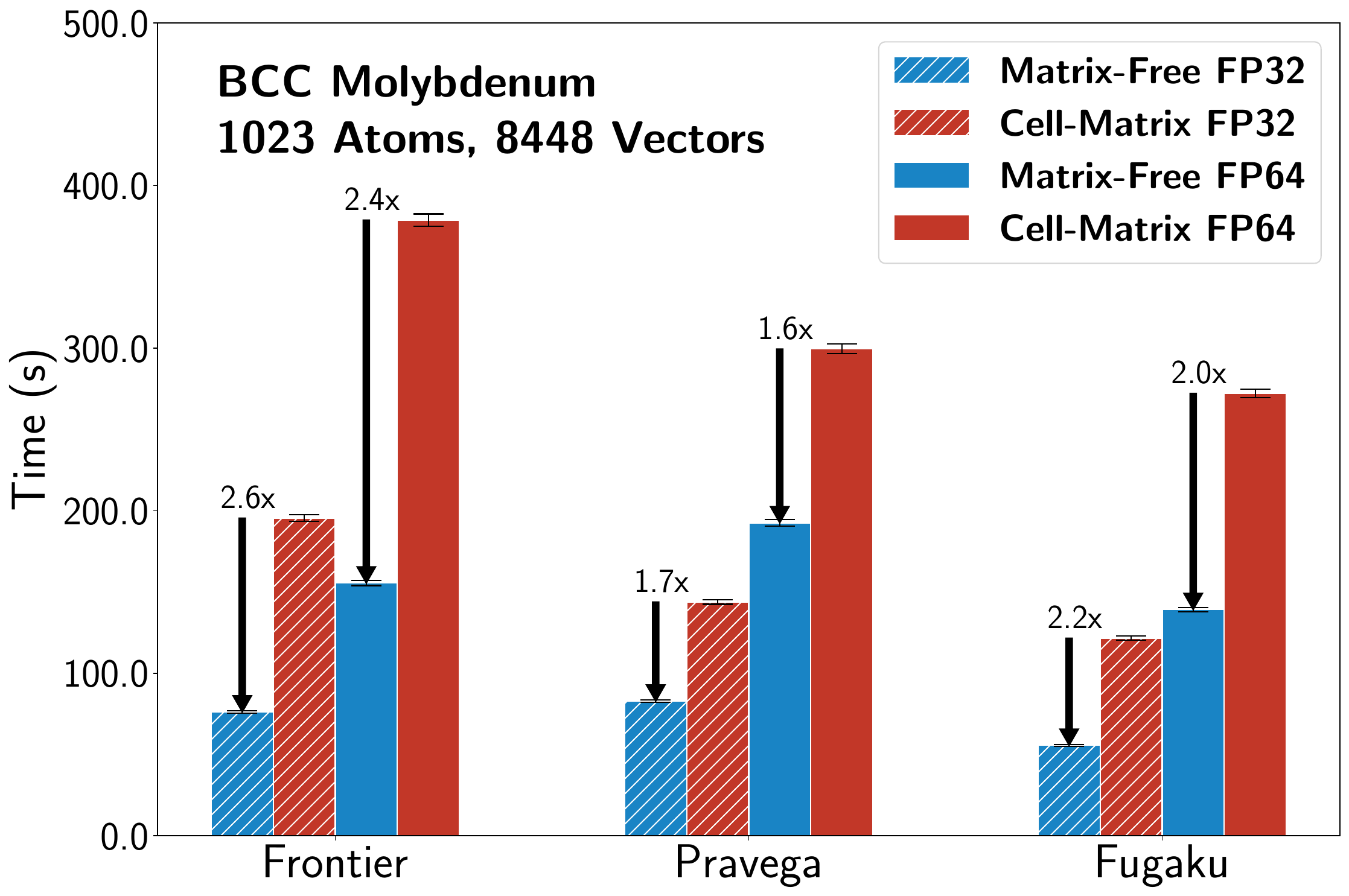}
       \label{fig:speedup_fp32_BCCMo1023}
    \end{subfigure}

    \caption{Speedups of our proposed matrix-free implementation strategies using FP32 and FP64 precision in the Chebyshev filtering step simulating Al nanoparticle 561 atoms (Real-valued DFT operator) and BCC molybdenum 1023 atoms (Complex-valued DFT operator) with $\texttt{FEOrder} = 8$ on 4 and 12 nodes of Frontier, 4 and 10 nodes of Param Pravega and 70 and 200 nodes of Fugaku supercomputers respectively.}\label{fig:speedup_fp32}
\end{figure}
\begin{figure}[ht]
    \centering
    \begin{subfigure}[t]{.49\textwidth}
        \centering
        \includegraphics[width=\textwidth]{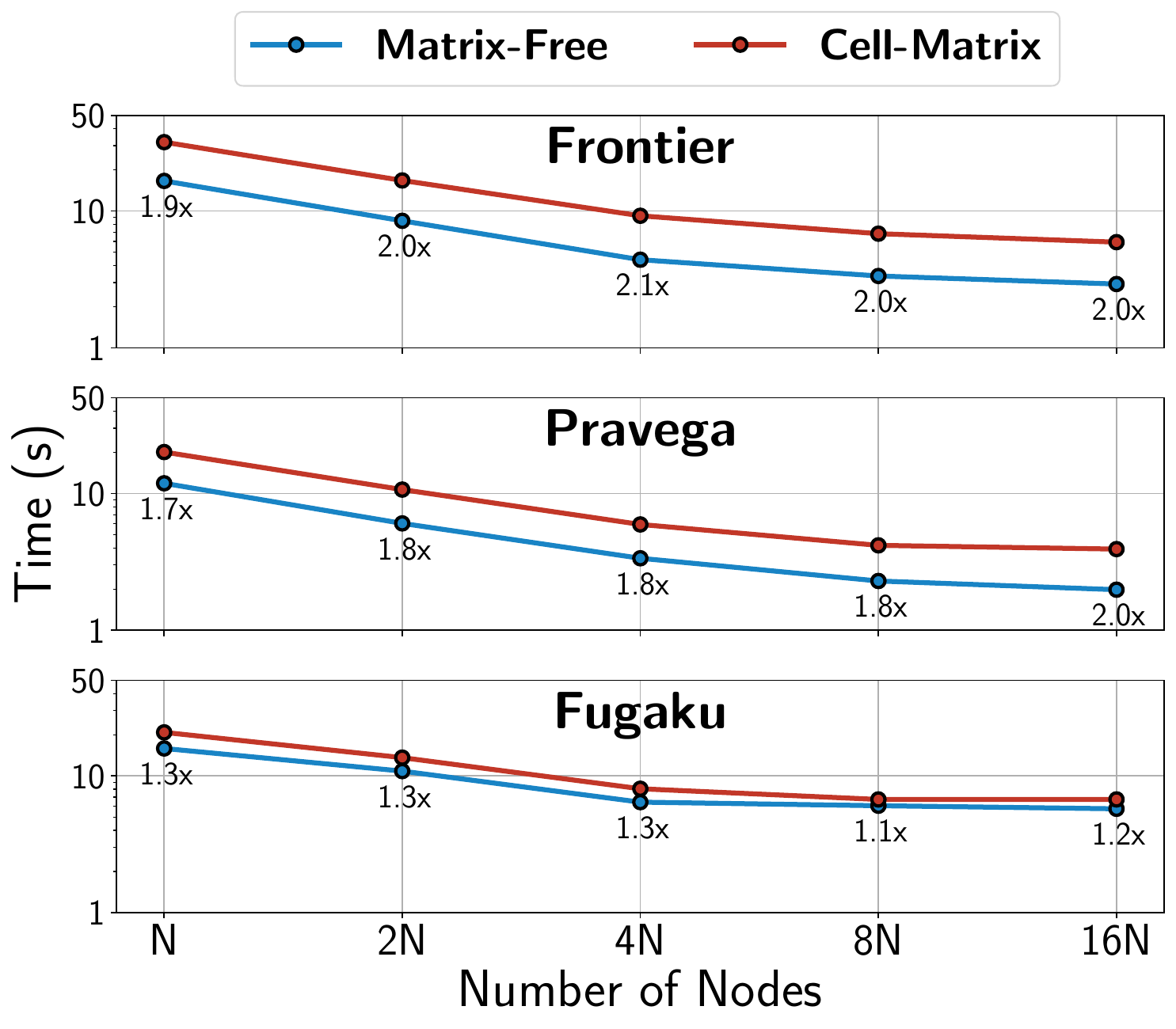}
        \label{fig:scaling_Al561}
    \end{subfigure}\hfill
    \begin{subfigure}[t]{.49\textwidth}
        \centering
        \includegraphics[width=\textwidth]{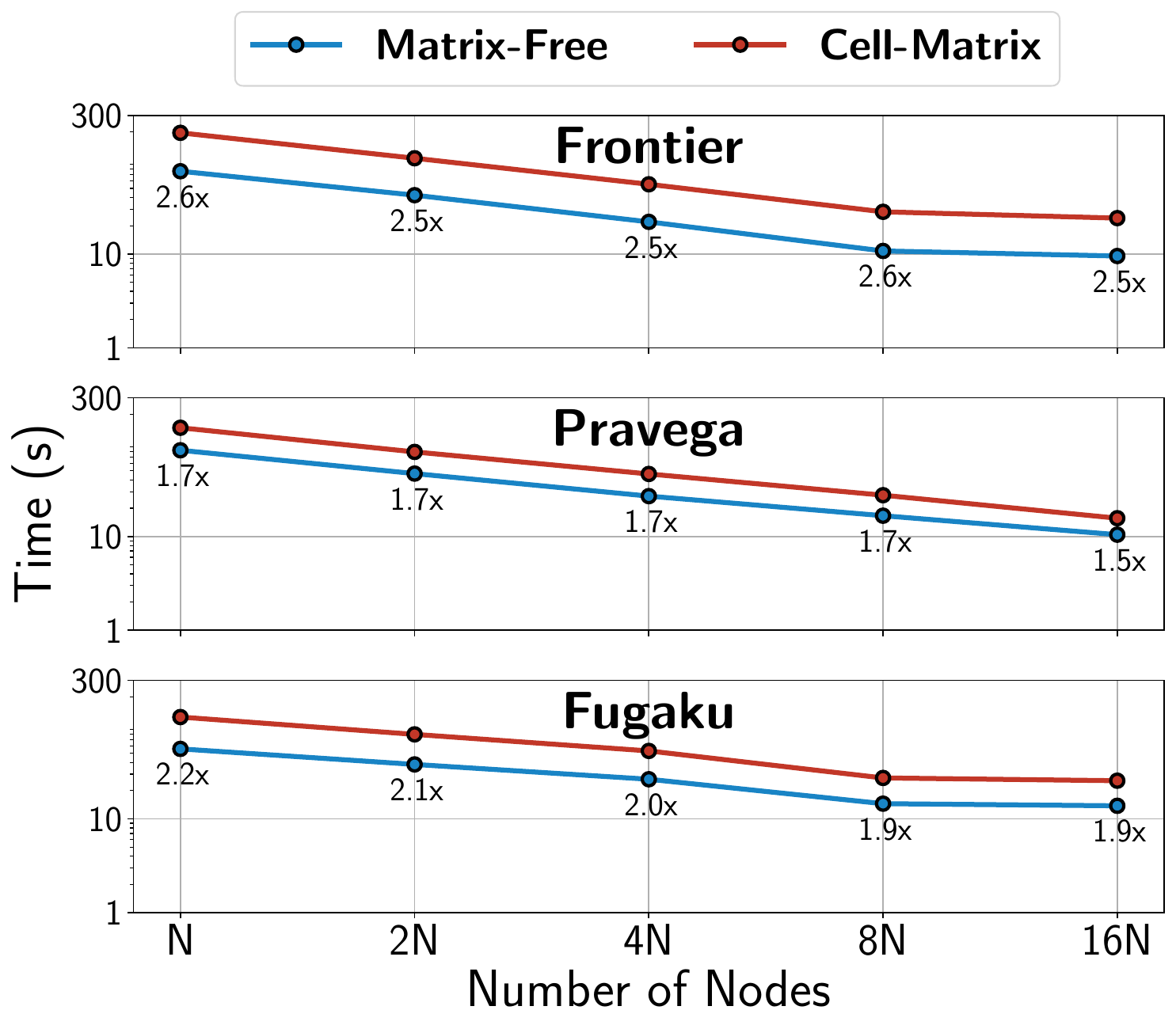}
       \label{fig:scaling_BCCMo1023}
    \end{subfigure}
    \caption{{\cblue Strong-scaling study\cn} comparing the proposed matrix-free implementation with cell-matrix baseline for subspace construction step of Al nanoparticle 561 atoms (left) and BCC molybdenum 1023 atoms (right) with $\texttt{FEOrder} = 8$, FP32 precision on Frontier (Left: N = 4, Right: N = 12), Param Pravega (Left: N = 4, Right: N = 10) and Fugaku (Left: N = 70, Right: N = 200) supercomputers.}\label{fig:scalingAlBCCMo}
\end{figure}
In \cref{fig:scalingAlBCCMo}, we present the {\cblue strong-scaling} behaviour of the Chebyshev filtering step (subspace construction step) in FP32 arithmetic of the R-ChFSI eigensolver on various MPI tasks, comparing the proposed matrix-free implementation to the cell-matrix baseline on three supercomputers. {\cblue\ Here the material system (and hence the problem size) is held fixed while the number of nodes is increased from the minimum node count $N$ that fits the problem in memory up to $16N$\cn}. Our implementation has a clear performance advantage over the cell-matrix implementation across various MPI tasks. In the case of Al nanoparticle 561 atoms, we achieve a speedup of 1.9$\times$--2.1$\times$ (Frontier), 1.7$\times$--2.0$\times$ (Pravega) and 1.1$\times$--1.3$\times$ (Fugaku) over the cell-matrix implementation. These results indicate that our implementation retains similar speedups as the number of nodes increases across supercomputers. Further, even at extreme scale regime for Frontier (64 nodes, 3584 cores, 900 DoFs/core) and Pravega (64 nodes, 3072 cores, 1000 DoFs/core), relative to the cell-matrix baseline, we observe a 2.0$\times$ speedup and on Fugaku (1120 nodes, 4480 cores, 700 DoFs/core), we observe 1.2$\times$ speedup. In the case of BCC molybdenum 1023 atoms, we achieve a speedup of 2.5$\times$--2.6$\times$ (Frontier), 1.5$\times$--1.7$\times$ (Pravega) and 1.9$\times$--2.2$\times$ (Fugaku) over the cell-matrix implementation. Hence, as observed for Al nanoparticles, our implementation achieves similar speedups on BCC molybdenum as the number of nodes increases across supercomputers. At extreme scales, we observe speedups of 2.5$\times$ on Frontier (192 nodes, 10752 core, 200 DoFs/core), 1.5$\times$ on Pravega (160 nodes, 7680 core, 280 DoFs/core) and 1.9$\times$ on Fugaku (3200 nodes, 12800 core, 170 DoFs/core). Thus, our matrix-free implementation demonstrates excellent scalability across supercomputing architectures even at extreme scaling regime.
\FloatBarrier

\subsubsection{Full eigensolver leveraging matrix-free action}

\begin{figure}[ht]
    \centering
    \begin{subfigure}[t]{.49\textwidth}
        \centering
        \includegraphics[width=\textwidth]{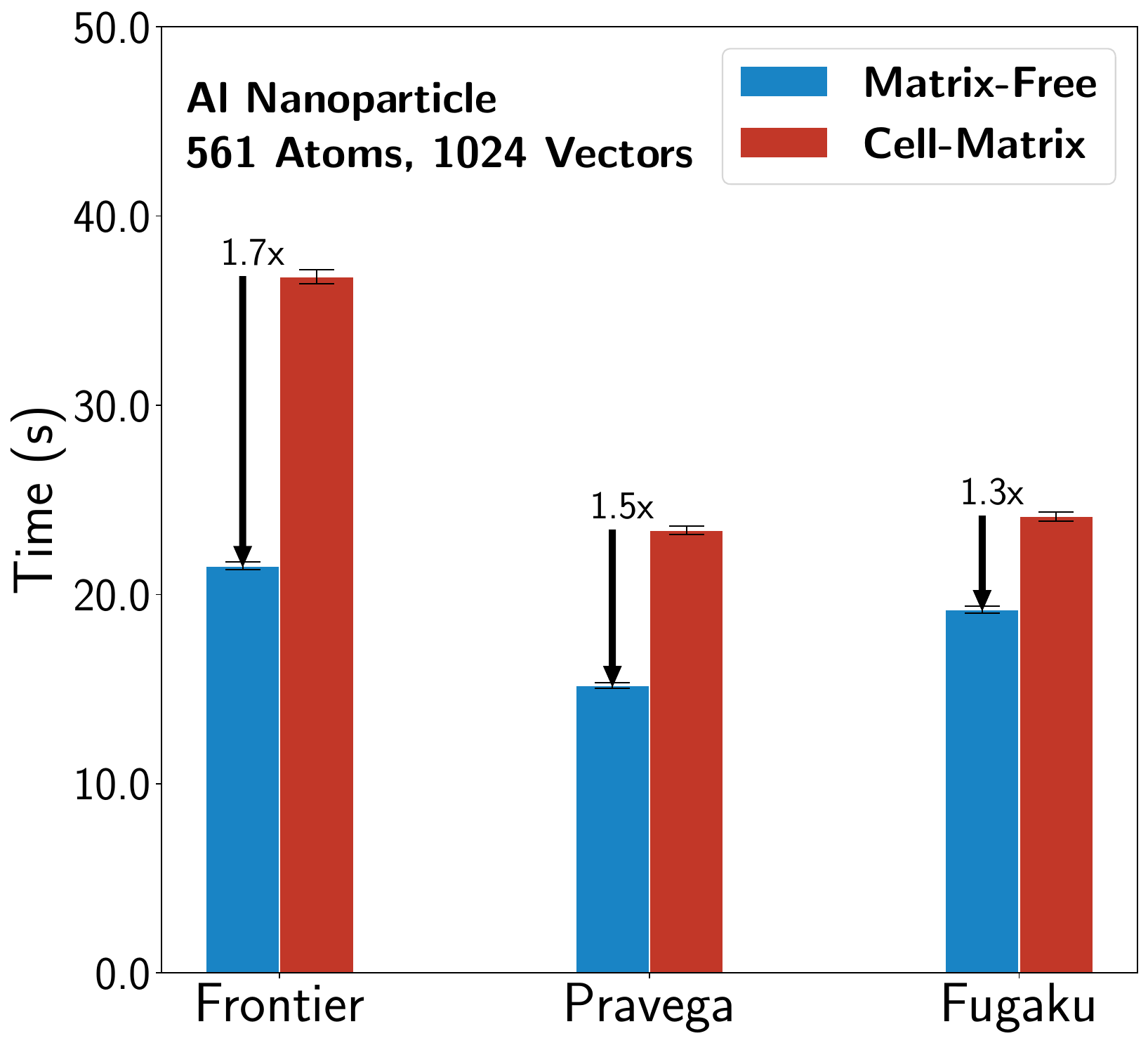}
        \label{fig:eigenRR_Al561}
    \end{subfigure}\hfill
    \begin{subfigure}[t]{.49\textwidth}
        \centering
        \includegraphics[width=\textwidth]{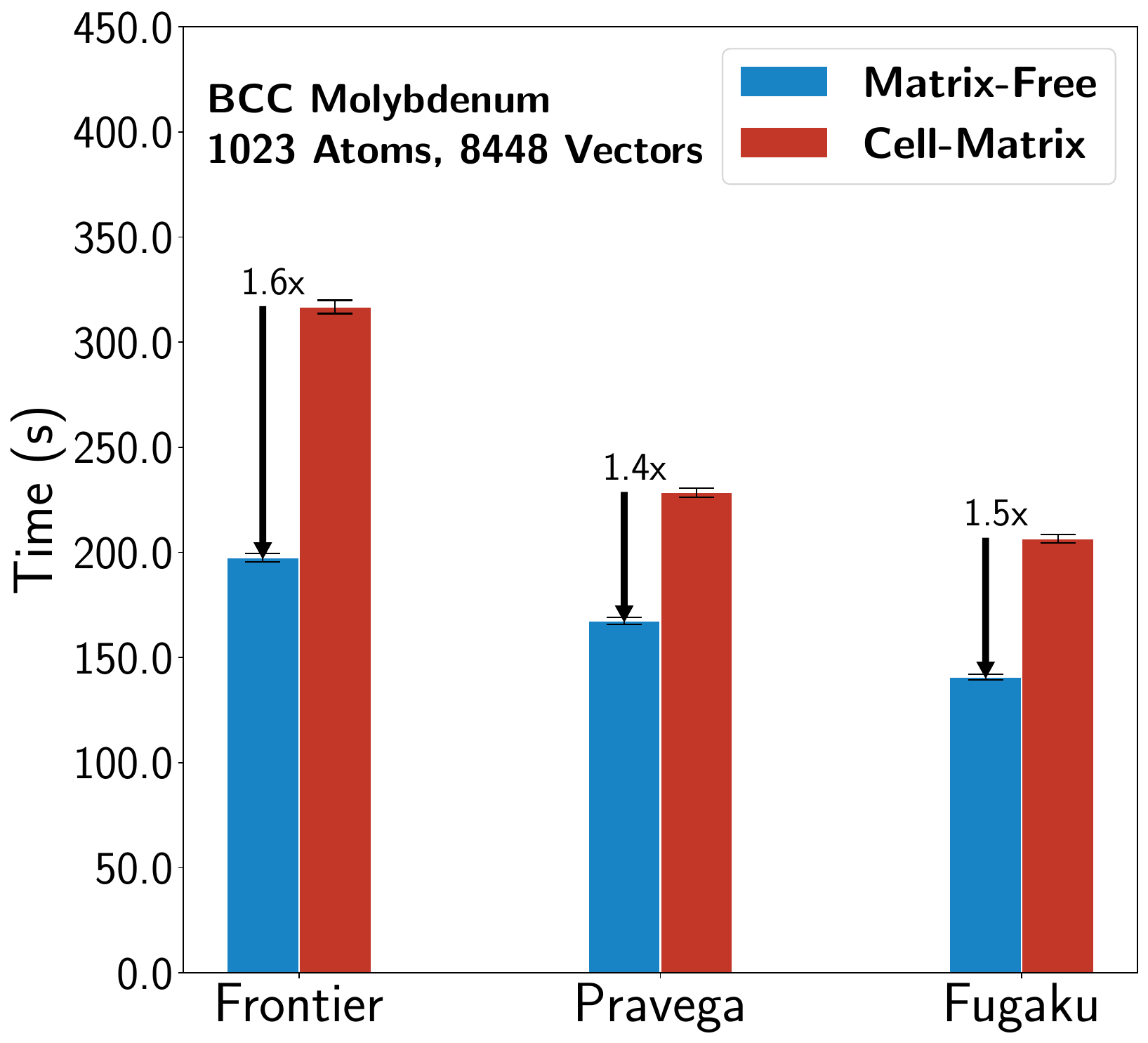}
       \label{fig:eigenRR_BCCMo1023}
    \end{subfigure}
    \caption{Speedups of our proposed matrix-free implementation with cell-matrix baseline for the full eigensolver (subspace construction step and subspace projection step) for Al nanoparticle 561 atoms (left) and BCC molybdenum 1023 atoms (right)  with $\texttt{FEOrder} = 8$, FP32 precision on Frontier (Left: 4 nodes, Right: 12 nodes), Param Pravega (Left: 4 nodes, Right: 10 nodes) and Fugaku (Left: 70 nodes, Right: 200 nodes) supercomputers.}\label{fig:eigenRRAlBCCMo}
\end{figure}
\begin{figure}[ht]
    \centering
    \begin{subfigure}[t]{.49\textwidth}
        \centering
        \includegraphics[width=\textwidth]{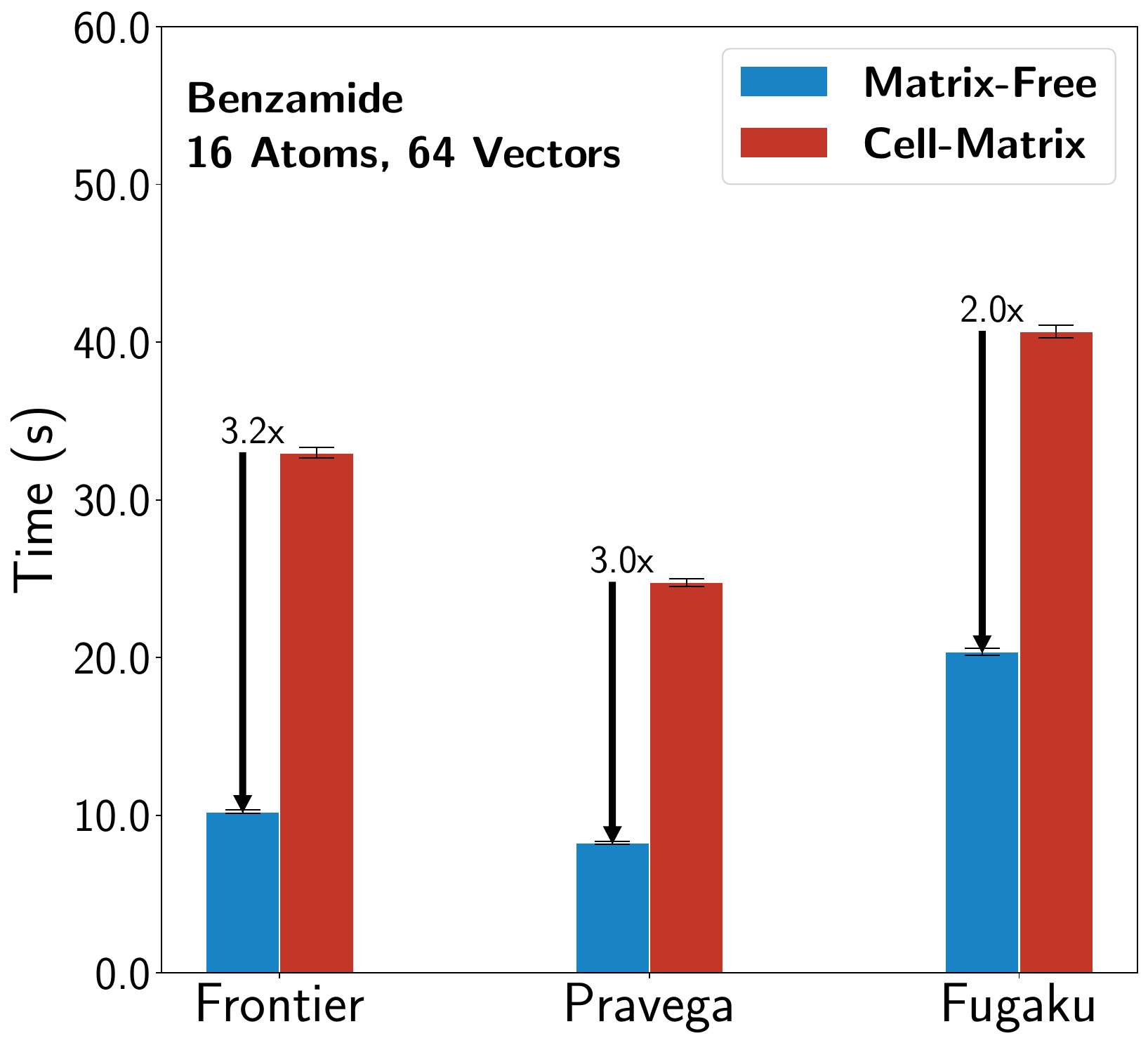}
        \label{fig:eigenRR_Benz16}
    \end{subfigure}\hfill
    \begin{subfigure}[t]{.49\textwidth}
        \centering
        \includegraphics[width=\textwidth]{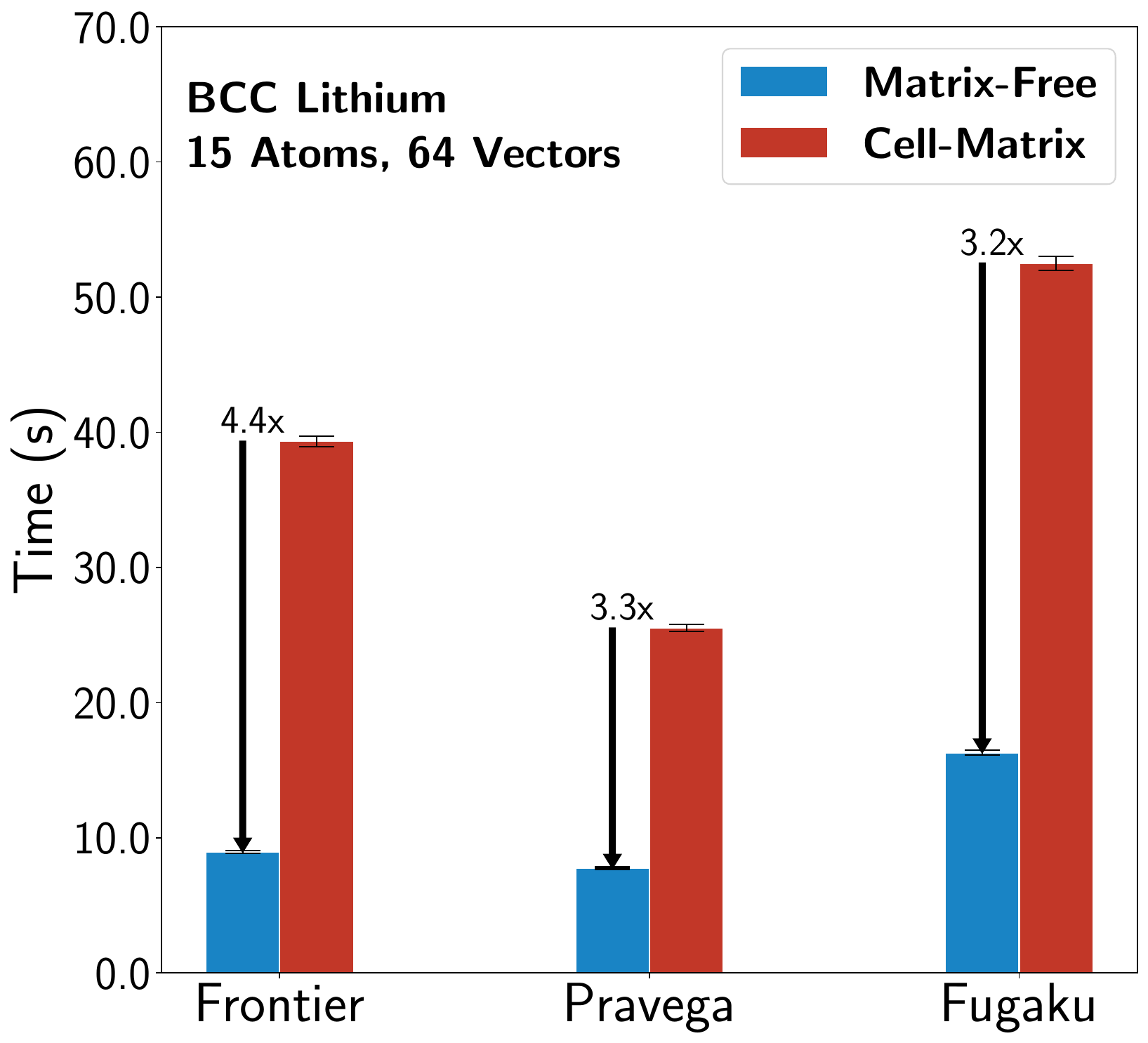}
       \label{fig:eigenRR_BCCLi15}
    \end{subfigure}
    \caption{Speedups of our proposed matrix-free implementation with cell-matrix baseline for full eigensolver (subspace construction step and subspace projection step) for benzamide 16 atoms (left) and BCC lithium 15 atoms (right)  with $\texttt{FEOrder} = 8$ and FP32 precision on Frontier (2 nodes each), Param Pravega (2 nodes each) and Fugaku (20 nodes each) supercomputers.}\label{fig:eigenRRBenzBCCLi}
\end{figure}

While we showcased the speedups over the cell-matrix baselines for the subspace construction (Chebyshev filtering step) of the R-ChFSI eigensolver algorithm employed here, we now present the speedups for the entire eigensolver that includes the Rayleigh–Ritz step (subspace projection and subspace diagonalisation) in \cref{fig:eigenRRAlBCCMo}. We observe a slight drop in the speedup due to the additional cost of this Rayleigh–Ritz step. {\cblue These results indicate the overall gains achievable with the matrix-free approach for solving the DFT eigenproblem.} In particular, for the Al nanoparticle 561 atoms system, our proposed matrix-free method achieves speedups of 1.7$\times$ on Frontier, 1.5$\times$ on Pravega and 1.3$\times$ on Fugaku. Similarly, for the BCC molybdenum 1023-atom system, our proposed matrix-free method achieves speedups of 1.6$\times$ on Frontier, 1.4$\times$ on Pravega, and 1.5$\times$ on Fugaku. In the all-electron cases, for benzamide 16-atom system our proposed matrix-free method achieves speedups of 3.2$\times$ on Frontier, 3.0$\times$ on Pravega and 2.0$\times$ on Fugaku. And for BCC lithium 15-atom system, our proposed matrix-free method achieves speedups of 4.4$\times$ on Frontier, 3.3$\times$ on Pravega and 3.2$\times$ on Fugaku. The higher speedups observed in the all-electron case are expected, as already observed in \cref{fig:speedup_benzLi}. The low number of vectors favours matrix-free methods, as the small matrices are very cache-friendly, and the lower number of floating-point operations compared to cell matrices yields significant computational gains.
\FloatBarrier

\section{Conclusion and future work}\label{sec:sec6}
This work presents a comprehensive matrix-free finite-element framework for accelerating Kohn--Sham density functional theory calculations on modern distributed CPU architectures. A significant computational challenge posed by \emph{ab initio} simulations is the repeated application of the FE discretized Kohn--Sham DFT operator to hundreds or thousands of trial vectors within iterative eigensolvers. As material system sizes and FE polynomial orders increase, traditional approaches based on global sparse matrix-vector multiplications or FE cell-level matrix-vector multiplications become increasingly limited by memory bandwidth, irregular data access, and high memory footprint. These challenges are particularly acute for high-order finite elements used in DFT, where finite-element matrices grow rapidly in size, and the poor cache reuse of global sparse matrix-vector multiplications becomes insufficient to saturate the hardware capabilities of present-day CPU architectures. The current state-of-the-art cell-matrix approach mitigates some of these limitations by exploiting optimized BLAS routines and achieving high throughput through multithreaded parallelism. However, even this strategy ultimately encounters performance bottlenecks for large material systems and higher-order finite-elements due to the need to store sizable FE cell-level Hamiltonian matrices and node-level multivectors, which impose substantial pressure on memory capacity, bandwidth and efficient cache utilization. 

In this work, we develop matrix-free algorithms tailored to the structure of the FE discretized Kohn--Sham DFT operator and demonstrate how they can be integrated efficiently within multivector-based iterative eigensolvers. The methodology is built around the observation that high-order FE bases for hexahedral finite-elements admit a tensor-product structure, allowing the local part of discretized DFT operator action to be expressed through sequences of structured tensor contractions rather than explicit multiplication by 3D finite-element matrices. By evaluating these contractions on the fly using quadrature-based representations of the operator coefficients, the method eliminates the need to construct or store FE cell-matrices and dramatically reduces memory movement. The resulting formulation efficiently handles all components of the Kohn--Sham operator: the kinetic energy term, local potential, gradient-density contributions from GGA exchange-correlation, periodic Bloch-shift terms, and the separable nonlocal pseudopotentials, while supporting both real and complex arithmetic. Importantly, the framework is applicable to pseudopotential and all-electron DFT calculations, accommodating periodic, semi-periodic, and non-periodic boundary conditions. A key algorithmic contribution of this work is the development of a multilevel batched data layout for efficient operator evaluation on CPU architectures with vector instruction sets such as AVX2, AVX-512, and SVE. By partitioning the multivector into two levels of batches, the implementation aligns tensor contractions with SIMD widths and ensures that intermediate data remain in cache across contraction sequences. This layout not only improves cache reuse but also allows for a unified treatment of real and complex-valued operators by storing real and imaginary components in separate batches. Complex matrix-free operations are thus reduced to combinations of real contractions, lowering the total number of floating-point operations. Additional optimizations, including even-odd decomposition to exploit the symmetry of shape functions and shape function gradients, reuse of intermediate tensors, uniform quadrature rules across tensor contractions to simplify data management, and nonblocking MPI communication for boundary exchanges, collectively contribute to high performance and strong parallel scalability while preserving accuracy.

The performance benchmarks conducted in this work span representative material systems and operator types, including (i) real-valued pseudopotential calculations for aluminum nanoparticles, (ii) complex-valued periodic pseudopotential calculations for BCC molybdenum with a vacancy, and (iii) all-electron DFT calculations for benzamide (real-valued) and BCC lithium (complex-valued). Across these systems, the matrix-free approach consistently outperforms the state-of-the-art FE cell-matrix implementation in DFT-FE. For pseudopotential systems, speedups range from 1.5--3.6$\times$ on Frontier (AMD x86-64 CPUs), 1.3--2.4$\times$ on Param Pravega (Intel x86-64 CPUs), and {\cblue 0.9--2.7$\times$ on Fugaku (Fujitsu ARM CPUs), depending on FE order and operator structure, with the lone value below unity arising for the real-valued operator at $\texttt{FEOrder}=7$ on Fugaku\cn}. For all-electron DFT simulations involving a lesser number of multivectors than pseudopotential calculations, the performance benefits are even more pronounced, with speedups up to 5.8$\times$ at higher FE orders. These gains reflect the ability of the matrix-free kernels to more effectively utilize memory bandwidth and vector units while avoiding the memory traffic associated with loading dense FE matrices. {\cblue The cache-aware roofline analyses reported in this study indicate that our matrix-free action on multivectors is memory-bandwidth-bound, streaming its working set from the L2/L3 caches, whereas the cell-matrix baseline is compute-bound. The matrix-free action therefore sustains a lower fraction of the floating-point peak and a lower arithmetic intensity than the cell-matrix baseline.\cn} Still, a significant computational advantage over the cell-matrix approach across supercomputing architectures is observed, attributed to lower arithmetic complexity, increased cache reuse, reduced memory footprint, and favourable access patterns in the proposed batched layout. The analyses also clarify differences in our matrix-free implementation across the three architectures: Frontier and Pravega benefit from high memory per core and wide SIMD units, while Fugaku’s lower memory per core ratio and node-level restriction to four MPI tasks slightly diminish the achievable speedups. Nevertheless, in all cases, the matrix-free implementation consistently provides a substantial performance advantage. An additional strength of the proposed approach lies in its compatibility with iterative eigensolvers that are resilient to controlled approximations in operator application on multivectors. In particular, when combined with the residual-based Chebyshev-filtered subspace iteration (R-ChFSI) eigensolver, the matrix-free kernels can be evaluated at reduced quadrature orders and lower floating-point precision without degrading the accuracy of total energies or atomic forces computed from DFT calculations. As demonstrated in the benchmarking studies, this synergy between R-ChFSI eigensolver and matrix-free operator evaluation further reduces the time-to-solution and enhances scalability for large-scale DFT simulations.

{\cblue In conclusion, the results of this work show that matrix-free algorithms can substantially reduce the time to solution of finite-element based \emph{ab initio} electronic structure calculations on current and emerging CPU architectures, without compromising accuracy or parallel scalability.\cn} {\cblue We remark that the benefits of the proposed methodology are not tied to the distributed multi-node setting. The gains arise from reduced floating point operations, smaller cache-resident working sets and reduced data movement within a node, and therefore carry over to single-node workstations with large core counts.} Looking ahead, extending the matrix-free framework {\cblue developed here for the full Kohn--Sham DFT operator\cn} to heterogeneous and GPU-accelerated architectures represents a natural next step. The tensor contractions underlying the operator evaluation map well to GPU hardware, provided data layout and thread-parallelism are carefully tuned. {\cblue Beyond DFT, the algorithmic ingredients developed here, namely the multilevel batched layout, the unified treatment of real and complex arithmetic, and the batched handling of constraints and inter-process communication, carry over to other real-space PDE discretizations in computational mechanics and materials science that require the action of FE operators on a large number of vectors.\cn}

\section{Acknowledgments}
The authors gratefully acknowledge the seed grant from Indian Institute of Science (IISc) and the SERB Startup Research Grant from the Department of Science and Technology (DST), India (Grant Number:SRG/2020/002194) for the purchase of a hybrid CPU-GPU  cluster, which provided computational resources for this work. The research utilised the NSM supercomputer PARAM Pravega at the Indian Institute of Science. This research also used Frontier resources of the Oak Ridge Leadership Computing Facility (OLCF) at the Oak Ridge National Laboratory (ORNL), supported by the Office of Science of the U.S. Department of Energy (DoE) under Contract No. DE-AC05-00OR22725. This work was also supported by MEXT as ``Program for Promoting Research on the Supercomputer Fugaku" (Project ID: hp240529) and used computational resources of supercomputer Fugaku provided by the ``RIKEN Center for Computational Science" through the HPCI System Research Project (HPCI ID: hpci010876). G.P. acknowledges financial support in the form of the Prime Minister's Research Fellowship (PMRF) from the Ministry of Education (MoE), India. P.M. acknowledges the Google India Research Award 2023 for financial support during the course of this work.

\section{Declaration of Generative AI and AI-assisted technologies in the writing process}
During the preparation of this work the authors used ChatGPT, Gemini and Writefull in order to proofread. After using this tool/service, the authors reviewed and edited the content as needed and take full responsibility for the content of the publication.

\bibliographystyle{elsarticle-num-names}
\bibliography{references}

\end{document}